\documentclass[12pt]{article}
\usepackage{amsmath}
\usepackage{enumerate}
\usepackage{amsfonts}
\usepackage{graphicx}
\usepackage{latexsym}
\usepackage{verbatim}
\begin{document}
\def\be{\begin{equation}}
\def\bea{\begin{eqnarray}}
\def\ee{\end{equation}}
\def\eea{\end{eqnarray}}
\newcommand{\ba}{\begin{array}}
\newcommand{\ea}{\end{array}}
\def\d{\partial}
\def\eps{\varepsilon}
\def\la{\lambda}
\def\b{\bigskip}
\def\nn{\nonumber \\}
\def\p{\partial}
\def\t{\tilde}
\def\h{{1\over 2}}
\def\cH{\mathcal{H}}
\def\mb{{\bar m}}
\def\nb{{\bar n}}
\def\ch{\,\mbox{ch}}
\def\sh{\,\mbox{sh}}

\setcounter{tocdepth}{2}

\makeatletter
\def\blfootnote{\xdef\@thefnmark{}\@footnotetext}  
\makeatother

\begin{center}
{\LARGE Killing(--Yano) Tensors in String Theory}
\\
\vspace{18mm}
{\bf   Yuri Chervonyi  and   Oleg Lunin}
\vspace{14mm}

Department of Physics,\\ University at Albany (SUNY),\\ Albany, NY 12222, USA\\ 

\vskip 10 mm

\blfootnote{ichervonyi@albany.edu,~olunin@albany.edu}

\end{center}

\begin{abstract}

We construct the Killing(--Yano) tensors for a large class of charged black holes in higher dimensions and study general properties of such tensors, in particular, their behavior under string dualities. Killing(--Yano) tensors encode the symmetries beyond isometries, which lead to insights into dynamics of particles and fields on a given geometry by providing a set of conserved quantities. By analyzing the eigenvalues of the Killing tensor, we provide a prescription for constructing several conserved quantities starting from a single object, and we demonstrate that Killing tensors in higher dimensions are always associated with ellipsoidal coordinates. We also determine the transformations of the Killing(--Yano) tensors under string dualities, and find the unique modification of the Killing--Yano equation consistent with these symmetries. These results are used to construct the explicit form of the Killing(--Yano) tensors for the Myers--Perry black hole in arbitrary number of dimensions and for its charged version.
\b

\end{abstract}

\newpage

\addtocontents{toc}{\vskip-5pt}
\addtocontents{toc}{\protect\enlargethispage{\baselineskip}}
{\footnotesize
\tableofcontents
\addtocontents{toc}{\vskip-5pt}
}

\newpage

\section{Introduction and summary}\label{SectionIntro}
\renewcommand{\theequation}{1.\arabic{equation}}
\setcounter{equation}{0}

Symmetries of dynamical equations have always played very important role in string theory. Conformal symmetry of the worldsheet led to Polyakov's reformulation of the theory \cite{Polyakov}, making it amenable to quantization, and provided powerful tools for performing calculations \cite{BPZ}. Study of string dualities \cite{duality} led to great insights into dynamics of string theory at strong coupling and to formulation of the gauge/gravity duality \cite{mald}. More recently discovery of hidden symmetries of equations for a classical string led to the discovery of integrability \cite{BenaRoibPolch,MinZarembo}, which stimulated a great progress in understanding of string dynamics and gauge/gravity duality (see \cite{IntReview} for the review and list of references). To gain additional insights into properties of quantum gravity and strong interactions it is very important to look for new examples of integrable string backgrounds. Since at low energies strings behave as point--like particles, integrable structures must give rise to hidden symmetries of supergravity, which will be investigated in this article. 

Integrability of classical strings on certain backgrounds is guaranteed by an infinite number of conserved quantities which can be extracted from reformulating the dynamical equations as a linear Lax pair \cite{LaxPair}. Unfortunately, there is no algorithmic procedure for constructing such pairs, and they have to be guessed. 
Interestingly, there exists a procedure for demonstrating that a particular background does not have a Lax pair, and it has been applied in \cite{Pando,StepTseyt} to rule out several promising candidates, such as strings on a conifold and on asymptotically--flat geometry produced by D3 branes. Unfortunately, this procedure for ruling out integrability is rather complicated, and it has to be applied on a case--by--case basis, so in \cite{ChL} we used a different approach based on the study of geodesics. Since at low energies strings behave as point particles, integrability must survive as a hidden symmetry of such objects, and this gives a very coarse necessary condition for integrability, which can be tested for large classes of backgrounds. Interestingly, this condition was sufficient for ruling out integrability on all known supersymmetric  geometries produced by D--branes, with an exception of AdS$_p\times$S$^q$ and a couple of other examples \cite{ChL}. Of course, to analyze the integrability of geodesics one has to start with explicit solutions, and the nontrivial integrable deformations of  AdS$_p\times$S$^q$ \cite{Beta,Eta} had to be constructed using special techniques rather than obtained as members of known families\footnote{Analysis of \cite{ChL} focused only on geometries supported by the Ramond--Ramond fluxes, which allowed us to analyze very large families. The `isolated points' discussed \cite{Beta,Eta} contained mixed fluxes, and they would have survived the analysis of \cite{ChL} had it been performed. Integrability of strings on the beta--deformed backgrounds \cite{Beta} has been discussed in \cite{BetaIntegr}.}. This article is a continuation of the  program initiated in \cite{ChL}: it extends the earlier results to geometries without supersymmetry, and, more importantly, it uncovers the hidden symmetries underlying integrability of geodesics. In spite of this continuity, this paper does not require familiarity with \cite{ChL}.

Study of geodesics has a long history in general relativity, and the most powerful methods are based on the analysis of the Hamilton--Jacobi (HJ) equation. It is well-know that such equation separates if the background contains cyclic (ignorable) directions, but sometimes separation happens even between non--cyclic coordinates. The simplest example of such `accidental separation' comes from the three--dimensional flat space in spherical coordinates: the polar angle $\theta$ separates in the HJ equation, although the metric depends on this coordinate. In this case the separation can be attributed to the SU(2) symmetries of the sphere, but similar argument cannot be applied to the Kerr black hole, which has only U(1)$\times$U(1) isometry, although the $\theta$ coordinate still separates.  The technical aspects of this separation will be reviewed in section \ref{SecSeparKT}, and here we just recall that the separation is associated with a hidden symmetry encoded in the Killing tensor (KT) \cite{Carter,Penrose}. The same tensor also leads to separation of the Klein--Gordon equation even beyond the eikonal approximation. The Kerr metric also gives rise to separable Dirac equation, this is guaranteed by an additional symmetry encoded in the Killing--Yano tensor (KYT) \cite{Yano}.  Over the last four decades Killing(--Yano) tensors have been found for other geometries both in general relativity \cite{GRyano} and in string theory \cite{StrYano}, and in this article we will construct KYT for a large class geometries in arbitrary numbers of dimensions, which contains most of the known examples as special cases. 

Killing(--Yano) tensors encode all continuous symmetries of solutions in general relativity, but string theory also has discrete symmetries associated with dualities, which can be promoted to a continuous group of 
{solution-generating transformations} in supergravity. This leads to a very natural question: what happens  with Killing(--Yano) tensors under action by this group? Answering this question is one of the main goals of this paper. A slightly different question was answered in the article \cite{KYdual}, which identified the subset of duality transformation leaving the Killing--Yano tensor invariant. As we will see, in general both Killing and Killing--Yano tensors are changed by the dualities, even the 
{\it equation for the KYT is modified}. However, for the special cases discussed in  \cite{KYdual} our results agree with that paper. In this article we focus on dualities in the NS--NS sector since our preliminary study of the Ramond--Ramond backgrounds indicates that T duality applied to such geometries may change the rank of the KYT and even produce Killing--Yano tensors of mixed rank. A very brief discussion of this point is given in section \ref{SectionModifiedKYT}. 

\bigskip

\noindent
This paper has the following organization.

In sections \ref{SectionKTKYT} and \ref{SecKYTranks} we review some well-known properties of Killing(--Yano) tensors, and in section \ref{SecSeparKT} we rewrite them in a slightly unusual form which becomes crucial for the subsequent discussion. Usually one uses the Killing tensor to produce a conserved quantity which leads to separation of the HJ and Klein--Gordon equations, and only one such quantity can be constructed from a given Killing tensor. In section  \ref{SecSeparKT} we argue that if one looks further and studies the {\it eigenvalues} of the Killing tensor, then a single KT can lead to a {\it family} of conserved quantities since the detailed analysis of eigenvalues allows one to construct a family of Killing tensors from a single representative using an {\it algebraic} procedure (i.e., without solving differential equations). As a bi--product of this analysis we also demonstrate that separation caused by nontrivial Killing tensors in any number of dimensions can only happen in (degenerate) ellipsoidal coordinates, this generalizes the earlier result of \cite{ChL} to non--supersymmetric geometries. In section \ref{SecKYTranks} we also show that the eigenvectors of the Killing tensors lead to simple expressions for the Killing--Yano tensors when the latter exist.

After developing this general technology we apply it in section \ref{SecMyersPerry} to write the Killing--Yano and Killing tensors for the Myers--Perry black holes \cite{MyersPerry} in arbitrary number of dimensions with arbitrary number of rotations. In section \ref{SecMPF1} this construction is extended to charged solutions built from Myers--Perry geometries by application of the solution--generating dualities, and relatively simple explicit expressions for the Killing(--Yano) tensors are derived.

The general effects of string dualities on Killing(--Yano) tensors are discussed in section \ref{SecKillingsDualities}, where it is demonstrated that Killing vectors (KV) and Killing tensors survive under dualities if certain conditions on the Kalb--Ramond field are satisfied, and the resulting transformations for the KV and KT are derived\footnote{For Killing vectors, a very nice interpretation of the transformation law in terms of the Double Field Theory \cite{DFT}  is discussed in section \ref{SecKVDuality}, but unfortunately a natural embedding of KT and KYT in this formalism is still missing.}. For the Killing--Yano tensors the situation is rather different: while dualities generically destroy the standard KYT, they preserve the modified version of the KYT equation, which is derived in section \ref{SectionModifiedKYT}. We demonstrate that such duality--invariant modification is unique and derive the transformation laws for the Killing--Yano tensor.  Several examples of the modified KY tensors are discussed in section \ref{SecExamplesF1NS5}. 

While studying massless particles, one encounters Conformal Killing(--Yano) tensors (CKT and CKYT), and their behavior under string dualities has some unusual aspects. The conformal objects are discussed throughout the paper along with their standard counterparts. 
Some technical details are presented in appendices.


\section{Killing(--Yano) tensors in higher dimensions}
\label{SectionKTKYTGeneral}
\subsection{Killing tensors and Killing--Yano tensors}
\label{SectionKTKYT}
\renewcommand{\theequation}{2.\arabic{equation}}
\setcounter{equation}{0}

Symmetries play very important role in physics, and symmetries of geometries are encoded in Killing vectors and Killing tensors. In this section we will review some well--known properties of these objects and establish the notation which will be used in the rest of the paper. 

We begin with recalling that  the Killing vector (KV) is defined as a vector field $V$ which leaves the metric invariant. In other words, the Lie derivative of the metric along $V$ must vanish:
\bea\label{KVDefA}
{\cal L}_V g_{MN}=0,
\eea
Relation (\ref{KVDefA}) can be rewritten as 
\bea
{\cal L}_V g_{MN}=V^P\d_P g_{MN}+\d_M V^P g_{PN}+\d_N V^P g_{MP}=
\nabla_M V_N+\nabla_N V_M=0,
\eea
and it implies that the metric does not change under an infinitesimal transformation
\bea 
x'^M=x^M+\epsilon V^M.
\eea
Since Killing vectors encode symmetries, they are always associated with conserved quantities. Specifically, the expression
\bea
I=V_M\frac{dx^M}{ds}
\eea 
is conserved along any geodesic.

The correspondence between Killing vectors and integrals of motion is not one--to-one: some conserved quantities are not associated with KV. However, it was shown by Penrose and Walker \cite{Penrose} that any integral of motion that depends on momentum comes either from a Killing vector or from a rank--two Killing tensor as
\bea
I=K_{MN}\frac{dx^M}{ds}\frac{dx^N}{ds},
\eea 
where $K_{MN}$ satisfies a linear equation
\bea\label{KTeqnDef}
\nabla_{M}K_{NP}+\nabla_{N}K_{MP}+\nabla_{P}K_{MN}=0.
\eea
To determine whether the integrals of motion survive in quantum theory as well, one should analyze 
separability of the Klein--Gordon equation, and as shown in \cite{KMhelm}, the relevant conserved quantity must be associated with eigenvalues of the differential operator 
\bea
{\hat K}\equiv \frac{1}{\sqrt{-g}}\d_M\left[\sqrt{-g}K^{MN}\d_N\right]+k(x)
\eea
with some function $k(x)$. As demonstrated in \cite{Moon,KMhelm}, operator ${\hat K}$ commutes with $\nabla_M\nabla^M$ if and only if $K^{MN}$ satisfies equation (\ref{KTeqnDef}) and one more condition which will not be discussed here.

In general, presence of the Killing tensor does not imply separability of the Dirac equation, this requires existence of an anti--symmetric Killing--Yano tensor (KYT) $Y_{MN}$ which satisfies the defining equation
\cite{Yano}
\bea\label{KYTdef}
\nabla_M Y_{NP}+\nabla_N Y_{MP}=0.
\eea
This equation can be generalized to tensors of arbitrary rank as \cite{Tachibana} 
\bea\label{DefKYT}
\nabla_{(M}Y_{N)P_1\dots P_{k-1}}=0,\quad Y_{P_1\dots P_k}=Y_{[P_1\dots P_k]}.
\eea
In four dimensions KYT of rank $k>2$ can be dualized into vectors and scalars, but in string theory one encounters interesting solutions of (\ref{DefKYT}), which will be discussed  throughout this paper. It is also possible to define Killing tensors of rank $k>2$ as solutions of the equation \cite{Penrose}
\bea
\nabla_{(M_1}K_{M_2\dots M_{k+1})}=0,
\eea
but such objects will not play any role in our discussion.

Any KYT gives rise to a Killing tensor of rank two via the relation
\bea\label{KTfromKYTodrin}
{K}_{MN}={Y}_M{}^{A_1\dots A_{k-1}}{Y}_{NA_1\dots A_{k-1}}.
\eea
This equation has a simple interpretation: separability of the Dirac equation implies one for the Klein--Gordon equation in the same coordinates. In section \ref{SecKTandHJ} we will present a detailed analysis of Killing tensors and outline a procedure for ``extracting the square root'' from them which allows one to construct the Killing--Yano tensors, if they exist.

So far we discussed the integrals of motion for massive particles, but some additional symmetries might arise in the massless case. For example, while the metric
\bea
ds^2=dr^2+r^2d\phi^2
\eea
is not invariant under rescaling of $r$ coordinate, massless particles are not sensitive to such rescaling, so while 
\bea
{\cal V}=r\d_r
\eea
is not a Killing vector, it does lead to conserved quantities for {\it massless} particles. Such {\it conformal Killing vectors} (CKV) satisfy equation
\bea
\nabla_M {\cal V}_N+\nabla_N {\cal V}_M=vg_{MN},
\eea
where $v$ is an arbitrary functions of all coordinates. If $v$ is a constant, then the corresponding CKV is called homothetic \cite{Homothetic}, and such vectors will play an important role in the analysis presented in section \ref{SubsectionCKV}.

The conformal Killing(--Yano) tensors (CKT and CKYT) are defined as solutions of equations
\bea\label{CKTdef}
&&\nabla_{(M_1}{\cal K}_{M_2...M_{k+1})}=W_{(M_1...M_{k-1}}g_{M_{k}M_{k+1})},\\
&&\nabla_{(M_1}{\cal Y}_{M_2)...M_{k+1}}=g_{M_1M_2}Z_{M_3...M_{k+1}}+
\sum_{i=3}^{k+1}(-1)^ig_{M_i(M_1}Z_{M_2)...{M_{i-1}}M_{i+1}...M_{k+1}}.\nonumber
\eea
with coordinate--dependent tensors $W$ and $Z$. Notice that under rescaling of the metric, CKV, CKT and CKYT transform in a simple way\footnote{The relevant transformations are derived in Appendix \ref{AppConfResc}.}, so they survive S duality and transition from the string to the Einstein frame. Ordinary Killing vectors have the same feature, as long as we impose a reasonable restriction on the dilaton:
\bea
{\cal L}_V e^{2\Phi}=V^M\d_M e^{2\Phi}=0.
\eea
On the other hand, the ordinary KT and KYT are usually destroyed by coordinate--dependent rescaling of the metric, so they exist only in one frame. Conformal transformations of the KT and KYT  are discussed in Appendix \ref{AppConfResc}.

We will mostly focus on rank--2 KT and CKT, and they can be constructed by squaring KYT or CKYT:
\bea\label{KTfromKYT}
{\cal K}_{MN}={\cal Y}_M{}^{A_1\dots A_{k-1}}{\cal Y}_{NA_1\dots A_{k-1}}, \quad 
W_{M}=2{\cal Y}_{MA_1\dots A_{k-1}}Z^{A_1\dots A_{k-1}}.
\eea
For rank-1 and rank--2 (C)KYT this construction is well-known, and direct computation shows that it works for all $k$. 

Conformal Killing tensors ${\cal K}_{MN}$ with $W_M=-\nabla_M \phi$ have a special property: they can be extended to the standard KT $K_{MN}$ by
\be\label{KTfromCKT}
K_{MN}={\cal K}_{MV}+\phi g_{mn}.
\ee
To see this one can take a covariant derivative of \eqref{KTfromCKT} and symmetrize the result:
\bea
\nabla_{(M}K_{NP)}=\nabla_{(M} {\cal K}_{NP)}+\nabla_{(M} \phi g_{NP)}=0.
\eea
This construction will be illustrated in section \ref{SecKYTranks} by comparing KT and CKT for  rotating black holes.

\subsection{Killing tensors and the Hamilton--Jacobi equation}
\label{SecKTandHJ}
\label{SecSeparKT}

Solutions of the equation for the KT,
\bea\label{KTeqn4}
\nabla_P K_{MN}+\nabla_M K_{NP}+\nabla_N K_{PM}=0
\eea
form a linear space, in particular, a `trivial subspace' is spanned by combinations of the metric and Killing vectors,
\bea\label{Ktriv}
K^{triv}_{MN}=e_0 g_{MN}+\sum_{i,j} e_{ij}V^{(i)}_MV^{(j)}_N,
\eea
with constant coefficients $e_0$, $e_{ij}$. In this subsection we will establish a one--to--one correspondence between {\it nontrivial} Killing tensors and separation of variables in the Hamilton--Jacobi equation
\bea\label{HJone}
g^{MN}\d_M S\d_N S+\mu^2=0.
\eea

\subsubsection{Killing tensors from the Hamilton--Jacobi equation}

There are several notions of separability for equation (\ref{HJone}), and we focus on the standard one by assuming that 
\bea\label{SeparSimple}
S=S(x_1,\dots x_k)+S(x_{k+1}\dots x_n).
\eea
This assumption can be generalized to R--separability as
\bea\label{SeparCompl}
S=S(x_1,\dots x_k)+S(x_{k+1}\dots x_n)+S_0(x_1\dots x_n),
\eea
where $S_0(x_1\dots x_n)$ is a {\it known} function of its arguments\footnote{The counterpart of (\ref{SeparCompl}) for the Schr{\"o}dinger equation is $$\Psi=X(x_1\dots x_k)Y(x_{k+1}\dots x_n)\Psi_0(x_1\dots x_n)$$ with \textit{known} function $\Psi_0$. For non-trivial $\Psi_0$ this is known as R--separation \cite{KMhelm}.} \cite{MorseFesh}. However, this generalization will not play any role in our discussion.

Equation (\ref{HJone}) separates as (\ref{SeparSimple}) if and only if three conditions are satisfied:
\begin{enumerate}[(a)]
\item{Coordinates $x^M$ can be divided into cyclic coordinates $z$ and two other groups, which will be denoted by $x$ and $y$. The metric does not depend on coordinates $z$.}
\item{There exists a separation function $f$, such that
\bea\label{gEigenTens}
 g^{MN}=\frac{1}{f}\left(X^{MN}+Y^{MN} \right),\quad
 \d_x Y^{MN}=\d_y X^{MN}=0,\ X^{y^i M}=0,\ Y^{x^iM}=0.
\eea
}
\item{Function $f$ can be decomposed as
\bea\label{Eqn6Other}
f=f_x-f_y,\qquad \d_y f_x=0,\quad \d_x f_y=0,\quad \d_zf_x=\d_z f_y=0.
\eea
}
\end{enumerate} 
Conditions (a)--(c) allow us to rewrite equation (\ref{HJone}) as
\bea
X^{MN}\d_M S\d_N S+\mu^2 f_x=-Y^{MN}\d_M S\d_N S+\mu^2 f_y,
\eea
where the left--hand side depends only on $x$, and the right--hand side depends only on $y$. This implies that 
\bea\label{SeparInt}
I\equiv\left[X^{MN}-f_xg^{MN}\right]\d_M S\d_N S
\eea
must be an integral of motion, and as such it must be associated with a Killing tensor:
\bea\label{IntKT}
I=K^{MN}\d_M S\d_N S.
\eea
We conclude that separation of variables (a)--(c) is associated with Killing tensor
\bea\label{KeigenTens}
K^{MN}=X^{MN}-\frac{f_x}{f}\left( X^{MN}+Y^{MN} \right)=-\frac{f_yX^{MN}+f_xY^{MN}}{f}.
\eea
If condition (c) is not satisfied, then equation (\ref{HJone}) separates only for $\mu=0$, and the associated {\it conformal} Killing tensor is 
\bea
K^{MN}=X^{MN}.
\eea

After reviewing the standard procedure for extracting the Killing tensor from separation of variables \cite{Carter, Penrose}, we discuss the inverse problem: recovery of separation from a given Killing tensor.


\subsubsection{Separation of variables from Killing tensor}
\label{SecSeparKTsub}

Every Killing tensor gives rise to an integral of motion via (\ref{IntKT}), and such constant must be associated with separation of variables as in (\ref{SeparInt}). While the separation functions $(f_x,f_y)$ and the corresponding tensors $(X^{MN},Y^{MN})$ are encoded in the Killing tensor, extracting them requires further analysis, and as we will demonstrate,
 this analysis may lead to an entire family of the Killing tensors which can be constructed {\it algebraically} from one representative. Schematically our results can be represented as
\bea\label{diagram}
\begin{array}{c}
\mbox{Eigenvalues}\\ \mbox{of KT}
\end{array}
\quad\Rightarrow\quad \mbox{separation} \quad\Rightarrow\quad 
\begin{array}{c}
m\mbox{--parameter}\\
\mbox{family of KTs}
\end{array}\, \quad\Leftrightarrow\quad
\begin{array}{c}
m\,\mbox{conserved}\\\mbox{charges}
\end{array}
\eea

To justify the usefulness of eigenvalues we recall equations (\ref{gEigenTens}) and (\ref{KeigenTens}):
\bea
g^{MN}=\frac{1}{f}\left(X^{MN}+Y^{MN} \right),\qquad
K^{MN}=-\frac{f_yX^{MN}+f_xY^{MN}}{f}
\eea
and consider an eigenvalue problem:
\bea\label{Keigen}
K^{MN}Z_N=\Lambda g^{MN}Z_N.
\eea
Assuming that metric has at least one non--cyclic direction\footnote{This assumption is violated only for flat space in Cartesian coordinates.} $x$ and that there is at least one component $K^{xN}\ne 0$, the $M=x$ component of (\ref{Keigen}) becomes
\bea
-\frac{f_y}{f}X^{xN}Z_N=\Lambda \frac{1}{f}\,X^{xN}Z_N\quad\Rightarrow\quad
\Lambda=-f_y.
\eea
In other words, some eigenvalues of the Killing tensor give the separation functions, and corresponding eigenvectors can be used to recover the relevant tensors $(X^{MN},Y^{MN})$. The cyclic coordinates complicate this construction, so they should be ignored to recover the separation function and added back in the end. Specifically, we propose the following procedure for extracting the separation function from the Killing tensor:
\begin{enumerate}[(1)]
\item Find the eigenvalues and eigenvectors of the KT:
\bea\label{KTanGmain}
K_{MN}=\sum_a \Lambda_a e^{(a)}_M e^{(a)}_N,\qquad 
g_{MN}=\sum_a  e^{(a)}_M e^{(a)}_N.
\eea
Notice that some eigenvalues may vanish of be degenerate.
\item Build the projectors\footnote{To avoid cumbersome formulas, we focus on non--degenerate eigenvalues. In general the left hand side of (\ref{ProjLam}) should refer to an eigenvalue $\Lambda$ and the right--hand side should contain summation over all $a$ with $\Lambda_a=\Lambda$. Since degeneracy clutters notation without introducing new effects, we use (\ref{ProjLam}).} 
\bea\label{ProjLam}
P^{(a)}_{MN}=e^{(a)}_M e^{(a)}_N.\nonumber
\eea
Projector $P$ will be called cyclic if
\bea\label{CyclProj}
\sum_N {[P^{(a)}]_M}^N\d_N \Lambda_b=0 \quad\mbox{for all}\quad (a,b).
\eea
If all projectors are cyclic, the Killing tensor can be built from Killing vectors and the metric.
\item Remove all directions associated with cyclic projectors and construct the reduced metric and Killing tensor:
\bea\label{KandGred}
K^{red}_{MN}=\left[{\sum_a} \Lambda_a e^{(a)}_M e^{(a)}_N\right]_{red},\qquad 
g^{red}_{MN}=\left[{\sum_a}  e^{(a)}_M e^{(a)}_N\right]_{red}.\nonumber
\eea
Non--cyclic components of equation (\ref{KTeqn4}) imply that $K^{red}_{MN}$ is a Killing tensor for  
$g^{red}_{MN}$. Nontrivial $K^{red}_{MN}$ and $g^{red}_{MN}$ imply that Killing tensor cannot be constructed from the Killing vectors and the metric. 
\item Separation of variables implies that 
\bea
\sum_M e^{(a)}_M dx^M=\sqrt{g_a}dx^a,\qquad \d_j\d_k\ln g_m=0\
\mbox{for different}\ (i,j,k).
\eea
Then analysis of the Killing equations shows that {\it generically} the reduced metric and Killing tensor must have the form
\bea\label{KillLaMain}
&&ds^2_{red}=\sum_k g_k (dx_k)^2,\quad 
K_{red}=\sum_k \Lambda_k g_k (dx_k)^2,\nn
&&g_k=h_k(x_k)\prod_{j\ne k} [x_k-x_j],\quad \Lambda_j=\d_j\Lambda,
\eea
where $\Lambda(x_1\dots x_n)$ is a linear polynomial in every $(x_1\dots x_n)$ symmetric under interchange of every pair of arguments. 
\item
Separation of variables in the reduced metric is accomplished by multiplying the reduced HJ equation by
\bea\label{RhoKmain}
\rho_k=\prod_{j\ne k} [x_k-x_j].
\eea
Then the reduced HJ equation can be written as
\bea\label{SeparHJmain}
\frac{1}{h_k}(\d_k S)^2=\sum_{p=0}^{n-1} (x_k)^p I^{(k)}_p(x_1\dots x_{k-1},x_{k+1}\dots x_n),
\eea
which implies that all $I^{(k)}_p$ must be constant\footnote{Integrals of motion $I^{(k)}_p$ are closely related to the separation constants which arise from breaking the HJ equation into pieces using St{\"a}ckel determinant. A detailed discussion of the St{\"a}ckel's method can be found in chapter 5 of \cite{MorseFesh}.}. This construction separates variable $x_k$, and other coordinates can be separated in the same fashion 
\item
After coordinates $(x_1\dots x_n)$ have been constructed, cyclic directions can be added back, and upon multiplication by (\ref{RhoKmain}) the complete $d$--dimensional HJ equation takes the form (\ref{SeparHJmain}). This follows from the fact that $K$ from (\ref{KTanGmain}) was a Killing tensor for the $d$--dimensional metric.
\item
A given Killing tensor corresponds to a particular function $\Lambda$ in (\ref{KillLaMain}), and a family of Killing tensors for the reduced metric can be constructed by keeping the same coordinates and introducing an arbitrary polynomial $\Lambda$.
\end{enumerate}
Steps (1)--(7) outline our construction, and the details and justification are presented in the Appendix \ref{AppEllips}. A different approach to separation functions and Killing tensors was developed in \cite{KalMil1}, and our results are consistent with theirs. 

Expressions (\ref{KillLaMain}) generalize Jacobi's ellipsoidal coordinates \cite{Jacobi} to curved space, and we derived them assuming that the dependence on $(x_1\dots x_n)$ is generic. Specifically we assumed that $g_1$ depends on all $n$ coordinates. It is also possible to have some degenerate cases where some $x_j$ does not appears in $g_1$, but such solutions can be obtained by taking some singular limits of the ellipsoidal coordinates. In the appendix \ref{AllFlatEllips} we review such singular limits for the ellipsoidal coordinates in flat three--dimensional space.

\bigskip

To summarize, in this subsection we clarified the relation between Killing tensors and separation of variables.  
It is well--known that separation of variables leads to a Killing tensor, which is associated with a conserved quantity \cite{Carter, Penrose}, but in higher dimensions, where the metric can depend on three or more variables and may admit more than one nontrivial Killing tensor, the correspondence is more interesting. As illustrated in the diagram (\ref{diagram}), a single separation of variables may give rise to a family of Killing tensors, and the entire family can be constructed from a single member by studying its eigenvalues. In section \ref{SecMyersPerry} our construction will be applied to an important example of the Myers--Perry black hole, and in section \ref{SecMPF1} it will be extended to the charged version of that solution. But first we discuss the additional symmetry structures which appear when the geometry admits a Killing--Yano tensor.

\subsection{Killing--Yano tensors of various ranks}
\label{SecKYTranks}

While Killing--Yano tensors (KYT) of rank two are well-known from general relativity in four dimensions, the objects with higher rank are less familiar, so in this subsection we will present several examples of such Killing--Yano tensors and discuss their relation to Killing tensors.

Recall that the Killing--Yano tensors are defined as solutions of equation (\ref{DefKYT})
\bea\label{DefKYT1}
\nabla_{(M}Y_{N)P_1\dots P_{k-1}}=0,\quad Y_{P_1\dots P_k}=Y_{[P_1\dots P_k]}.
\eea
As reviewed in section \ref{SectionKTKYT}, any Killing--Yano tensor leads to a Killing tensor via (\ref{KTfromKYTodrin}). For example, any $d$--dimensional space admits a trivial KYT of rank $d$, which is defined as a volume form, and it squares to the metric. Nontrivial KYT may square to the metric as well, as illustrated by our first example: a space that has a factorized form
\bea
ds^2=g_{mn}(x)dx^mdx^n+h_{\mu\nu}(y)dy^\mu dy^\nu,
\eea
where two subspaces have the same dimensionality $n$. Then volume forms on $x$ and $y$ spaces give rise to a family of Killing--Yano tensors:
\bea\label{AdSKYT}
&&Y=c_1 \mbox{Vol}_{g}+c_2 \mbox{Vol}_{h}\quad\Rightarrow \nn
&&K_{MN}dX^MdX^N=(n-1)!\left[c_1^2 g_{mn}(x)dx^mdx^n+c_2^2 h_{\mu\nu}(y)dy^\mu dy^\nu\right].
\eea
It is clear that a non--trivial KY tensor can square to the metric as long as $c_1^2=c_2^2$. For generic values of constants $c_1$ and $c_2$ Killing tensor has two distinct eigenvalues, and each of them has 
degeneracy $n$.

A large class of geometries admitting Killing--Yano tensors comes from rotating black 
holes\footnote{Another interesting class of geometries admitting Killing--Yano tensors comes from putting 
D--branes on singular points of Calabi--Yau manifolds. Killing--Yano tensors for Sasaki-Einstein manifolds appearing in this construction have been recently constructed in \cite{BabVis}.}, and in the next section we will construct the KYTs for black holes with arbitrary number of rotations. Before performing this general analysis we review the situation for the well--known example of the Kerr black hole \cite{Kerr} and extract important lessons from it. The non--trivial Killing tensor for the Kerr geometry was constructed by Carter \cite{Carter}, and we begin with rewriting the metric in convenient frames defined as eigenvectors of that KT:
\bea\label{Kerr4D}
ds^2&=&-e_t^2+e_r^2+e_\theta^2+e_\phi^2,\nonumber\\
e_{t}&=&\frac{\sqrt{\Delta}}{\rho}(dt-as_\theta^2 d\phi),\quad 
e_{\phi}=\frac{s_\theta}{\rho}\left[(r^2+a^2)d\phi-adt\right],\quad
e_{r}=\frac{\rho}{\sqrt{\Delta}}dr,\quad e_{\bf\theta}=\rho d\theta,\nonumber\\
\Delta&=&r^2+a^2-2mr,\quad \rho^2=r^2+a^2c_\theta^2,\quad
c_\theta=\cos\theta,\quad s_\theta=\sin\theta.
\eea
Then expressions for the Killing and Killing--Yano tensors become very compact:
\bea\label{Kerr4DKT}
K&=&r^2\left[e_\phi^2+e_\theta^2\right]+(ac_\theta)^2\left[e_t^2-e_r^2\right],\quad
Y=r e_\theta\wedge e_\phi+(ac_\theta) e_r\wedge e_t.
\eea
We observe that the eigenvalues of $K$ ($r^2$ and $-(ac_\theta)^2$) appear in pairs, and $Y$ is constructed from these eigenvalues and corresponding eigenvectors in a simple way. As we will see in the next section, this double degeneracy persists in all even dimensions. Notice that the separating function defined in the previous subsection is equal to the difference of eigenvalues, and in the present case equation (\ref{Eqn6Other}) becomes
\bea
f_x=r^2,\quad f_y=-(ac_\theta)^2,\quad f=r^2+(ac_\theta)^2.
\eea

In odd dimensions the situation is different\footnote{Since the number of eigenvalues is odd, the double degeneracy is not possible. To avoid unnecessary complications, we write (\ref{5DKerrNtr}) for one rotation, more general case will be discussed in the next section.}, and to get some insights, we look at a rotating black hole in five dimensions \cite{MyersPerry}. Solving equations for the Killing--Yano tensor, constructing the corresponding KT, and defining the frames as its eigenvalues, we find
\bea\label{5DKerrNtr}
ds^2&=&-e_t^2+e_r^2+e_\theta^2+e_\phi^2+e_\psi^2,\nonumber\\
K&=&r^2\left[e_\phi^2+e_\theta^2\right]+(ac_\theta)^2\left[e_t^2-e_r^2\right]+
[r^2-(ac_\theta)^2]e_\psi^2,\\
Y&=&\left[r e_\theta\wedge e_\phi+(ac_\theta) e_r\wedge e_t\right]\wedge e_\psi.\nonumber
\eea
The frames are defined by
\bea\label{5DKerrNtrFrm}
e_{t}&=&\frac{\sqrt{\Delta}}{\rho}(dt-as_\theta^2 d\phi),\quad 
e_{\phi}=\frac{s_\theta}{\rho}\left[(r^2+a^2)d\phi-adt\right],\nonumber\\
e_{r}&=&\frac{\rho}{\sqrt{\Delta}}dr,\quad e_{\bf\theta}=\rho d\theta,\quad 
e_\psi=rc_\theta d\psi,\\
\Delta&=&r^2+a^2-M,\quad \rho^2=r^2+a^2c_\theta^2.\nonumber
\eea
Notice that eigenvalues of $K$ come in two pairs and one special value corresponding to $e_\psi$. In the next section we will demonstrate that this pattern persists in all odd dimensions with arbitrary number of rotations. As expected from (\ref{Eqn6Other}), the separating function $f$ is equal to the difference of two non--cyclic eigenvalues 
\bea
f_x=r^2,\quad f_y=-(ac_\theta)^2,\quad f=r^2+(ac_\theta)^2,
\eea
but now the Killing tensor has an additional eigenvector $e_\psi$ associated with cyclic coordinates, and the corresponding eigenvalue is 
\bea
\Lambda_\psi=f_x+f_y=r^2-(ac_\theta)^2.
\eea
Analysis of section \ref{SecKTandHJ} did not put any restrictions on cyclic eigenvectors and eigenvalues. 

In addition to the standard KYT, rotating black holes may admit a conformal KYT, which satisfies  equations (\ref{CKTdef}) and gives rise to a conformal KT (CKT) via (\ref{KTfromKYT}). In particular, the CKYT and CKT for the Kerr metric \eqref{Kerr4D} are 
\bea\label{CKYTKerr4d}
\mathcal{Y}&=&re_r\wedge e_t-(ac_\theta)e_\theta\wedge e_\phi,\quad 
Z=dt-\frac{2mr}{\rho\sqrt{\Delta}}e_t,\nn
\mathcal{K}&=&r^2[e_t^2-e_r^2]+(ac_\theta)^2[e_\theta^2+e_\phi^2],\quad 
W=-d[r^2-a^2c_\theta^2],
\eea
and for the rotating black hole in five dimensions (\ref{5DKerrNtrFrm}) they are given by
\bea\label{CKYTrank2Kerr5d}
\label{CKYTKerr5d}
\mathcal{Y}&=&r e_r\wedge e_t-(ac_\theta)e_\theta\wedge e_\phi,\quad
Z=dt-\frac{M}{\rho\sqrt{\Delta}}e_t,\nn
\mathcal{K}&=&r^2[e_t^2-e_r^2]+(ac_\theta)^2[e_\theta^2+e_\phi^2],\quad
W=-d[r^2-a^2c_\theta^2].
\eea
Notice that vectors $W$ appearing in (\ref{CKYTKerr4d}) and (\ref{CKYTKerr5d}) are written as gradients of scalar functions, which means that they give rise to standard Killing tensors via (\ref{KTfromCKT}). Direct calculations show that application of 
(\ref{KTfromCKT}) to (\ref{CKYTKerr4d}) and (\ref{CKYTKerr5d}) leads to the Killing tensors given in (\ref{Kerr4DKT}) and (\ref{5DKerrNtr}). Conformal KYT 
(\ref{CKYTKerr4d}) and (\ref{CKYTKerr5d}) will play an important role in the general analysis presented in section \ref{SecKillingsDualities}. 

\section{Example: Killing--Yano tensors for the Myers--Perry black hole}
\label{SecMyersPerry}
\renewcommand{\theequation}{3.\arabic{equation}}
\setcounter{equation}{0}

In this section we construct a family of Killing--(Yano) tensors for the Myers--Perry black hole using the techniques introduced in section \ref{SecSeparKT}. The cases of odd and even dimensions have to be treated differently, so we begin with MP solution in even dimensions ($d=2n+2$) \cite{MyersPerry,Myers}:
\bea\label{MPeven}
ds^2&=&-dt^2+\frac{mr}{FR}\Big(dt+\sum_{i=1}^n a_i\mu_i^2 d\phi_i\Big)^2+\frac{FR dr^2}{R-mr}
+\sum_{i=1}^n(r^2+a_i^2)\Big(d\mu_i^2+\mu_i^2 d\phi_i^2\Big)\nn
&&+r^2d\alpha^2.
\eea
Here variables $(\mu_i,\alpha)$ are subject to constraint
\bea
\alpha^2+\sum_{i=1}^n\mu_i^2=1,
\eea
and functions $F$, $R$ are defined by 
\bea\label{MPdefFR}
F=1-\sum_{k=1}^n\frac{a_k^2\mu_k^2}{r^2+a_k^2},\quad R=\prod_{k=1}^n (r^2+a_k^2).
\eea

To find the KYT for the geometry (\ref{MPeven}) we observe that the square of the KYT gives a KT with some components along non--cyclic coordinates, so following the general procedure outlined in section 
\ref{SecSeparKT}, we begin with looking at the non--cyclic part of the metric:
\bea\label{MPevNC}
ds_{NC}^2=\frac{FR dr^2}{R-mr}+\sum_{i=1}^n(r^2+a_i^2)d\mu_i^2+r^2\Big(d[1-\sum_{i=1}^n\mu_i^2]^{1/2}\Big)^2.
\eea
As demonstrated in section \ref{SecSeparKTsub}, in the appropriate frames the Killing tensor and geometry (\ref{MPevNC}) must have the form\footnote{In this section we have to distinguish between $e^a=e^a_M dx^M$ and 
$e_a=e_a^M\d_M$, so the frame indices are written in the appropriate places. In the rest of the paper we abuse notation and write $e_a=e^a_M dx^M$ to simplify formulas.}
\bea\label{ay2}
K_{mn}dx^m dx^n=\sum_m \Lambda_m (e^m)^2,\quad ds_{NC}^2=\sum_m (e^m)^2,
\eea
where 
\bea\label{ay2a}
e^m=h_m(x_m)\left[\prod_{k\ne m}[x_m-x_k]\right]dx^m,\quad 
\Lambda_m=\d_m \Lambda(x_0\dots x_{n}),
\eea
and $\Lambda$ is a symmetric polynomial linear in every argument. To determine the new coordinates $(x_1\dots x_{n+1})$ in terms of $(r,\mu_1\dots\mu_n)$ we begin with $m=0$ case when metric (\ref{MPevNC}) becomes flat and the relation between $(x_0\dots x_{n+1})$ and $(r,\mu_1\dots\mu_n)$ is given in terms of well--known ellipsoidal coordinates \cite{MorseFesh}:
\bea\label{EplsdEven}
x_0=r^2,\quad (a_i\mu_i)^2=\frac{1}{c_i^2}\prod_{k=1}^n (a_i^2+x_k),\quad
c_i^2=\prod_{k\ne i}(a_i^2-a_k^2).
\eea
Note that here the variables are arranged in the following order
\bea
r^2>0>x_1>-a_1^2>x_2>-a_2^2>\cdots>x_n>-a_n^2.
\eea
It turns out that mass does not spoil this relation, and in terms of $(x_0\dots x_{n})$ metric (\ref{MPevNC}) takes the form (\ref{ay2})--(\ref{ay2a}):
\bea\label{FrameX}
e^r=\frac{dr}{\sqrt{R-mr}}\sqrt{\prod_{k}(r^2-x_k)}
,\qquad e^{x_i}=\frac{1}{2}dx_i\sqrt{\frac{(r^2-x_i)}{-x_i H_i}\prod_{k\ne i}(x_i-x_k)}.
\eea
From now on Latin indices take values $(1\dots n)$, and we also define convenient quantities $d_i$, $H_i$ and rewrite $FR$ in terms of the new coordinates:
\bea\label{MPnotation}
d_i=\prod_{k\ne i}(x_i-x_k),\quad H_i=\prod_k(x_i+a_k^2),\quad
FR=\prod_k(r^2-x_k).
\eea

So far we have ignored the cyclic coordinates since components of the Killing tensor in these directions contain an ambiguity of adding an arbitrary combination of Killing vectors:
\bea\label{AddKV}
K^{MN}\rightarrow K^{MN}+\sum_{a,b}c_{ab}V^M_a V^N_b,\quad
V_0=\d_t,\quad V_i=\d_{\phi_i}.
\eea
Once the proper non--cyclic coordinates $(x_0\dots x_{n})$ are found, we can determine the remaining components of the Killing tensor by studying the separation of variables associated with it. Specifically, we 
look at the Hamilton--Jacobi equation associated with (\ref{MPeven}) and write it in coordinates $(x_0\dots x_{n})$:
\bea\label{qq1}
\sum_i\frac{4H_i (-x_i)}{(r^2-x_i)d_i}(\d_i S)^2+\frac{R-mr}{FR}(\d_r S)^2+g^{ab}\d_a S\d_b S=-\mu^2.
\eea
To separate $r$ coordinate, we have to multiply the last relation by 
\bea\label{RhoR}
\rho_r=R F=\prod_k (r^2-x_k)
\eea 
and introduce integrals of motion $I_k$ as coefficients in front of various powers of $r$. Then we will find
\bea\label{IkIntgr}
(R-mr)(\d_r S)^2=\sum_{k=0}^n I_k r^{2k}.
\eea
Notice that one Killing tensor leads to several integrals of motion, while the standard prescription \cite{Carter, Penrose} allows us to construct only one:
\bea
I=K^{MN}\d_M S\d_N S.
\eea
The `extra' conserved quantities came as the result of our analysis of eigenvalues: the coordinates $(r^2,x_1\dots x_n)$ define a {\it family} of the Killing tensors parameterized by the polynomial $\Lambda$,
and the coordinates can be extracted from any special solution. Then starting with any member of the family and analyzing its eigenvalues, we can recover other Killing tensors by changing coefficients in $\Lambda$, as summarized by (\ref{diagram}). 

Extraction of the explicit expressions for $I_k$ is straightforward, but we will be interested in a different aspect of (\ref{IkIntgr}). To extend the relations (\ref{ay2}) beyond non--cyclic variables, we should identify the relevant cyclic frames, in particular, they should form pairs with $e_r$ and $e_{x_i}$\footnote{This follows  from the existence of the Killing--Yano tensor, as discussed below.}. To extract the partner of $e_r$, we set $(r^2-x_i)\rightarrow 0$ in (\ref{IkIntgr})\footnote{This is a very formal manipulation: although we set  $(r^2-x_i)\rightarrow 0$ for all $i$, we assume that $x_i-x_j\ne 0$. The goal of this operation is to remove all $x$--dependent terms from (\ref{IkIntgr}). We also recall that (\ref{IkIntgr}) comes from multiplying  (\ref{qq1}) by (\ref{RhoR}).}, then the right--hand side coming from (\ref{qq1}) contains only one frame:
\bea
e_t\propto \d_t-\sum_i \frac{a_i}{r^2+a_i^2}\d_{\phi_i}
\eea
Raising the index and normalizing this frame, we find
\bea\label{FrameT}
e^t=\sqrt{\frac{R-mr}{FR}}\left[dt+\sum_i\frac{G_i}{a_ic_i^2}d\phi_i\right],\quad
G_i\equiv \prod_k(x_k+a^2_i)
\eea
To extract the remaining frames, we write a counterpart of (\ref{IkIntgr}) by multiplying (\ref{qq1}) by 
\bea\label{RhoI}
\rho_i=(r^2-x_i)\prod_{k\ne i}(x_i-x_k)=(r^2-x_i)d_i.
\eea
This gives
\bea
\sum_i4H_i x_i(\d_i S)^2=\sum_{k=0}^n I^{(i)}_k (x_i)^{2k}.\nonumber
\eea
As before, we formally replace $(r^2-x_i)$ and $(x_j-x_i)$ by zero to extract 
\bea\label{FrameI}
e_i\propto \d_t-\sum_k\frac{a_k}{a_k^2+x_i}\d_{\phi_i}\quad\Rightarrow\quad
e^i=\sqrt{\frac{H_i}{d_i(r^2-x_i)}}\left[dt+\sum_k\frac{G_k(r^2+a_k^2)}{a_kc_k^2(x_i+a_k^2)}d\phi_k\right].
\eea
For future reference we summarize the frames and notation associated with Myers--Perry black hole in even dimensions\footnote{See (\ref{MPdefFR}), (\ref{EplsdEven}), (\ref{FrameX}), (\ref{MPnotation}), (\ref{FrameT}), (\ref{FrameI}).} 
\bea\label{AllFramesMP}
e^t&=&\sqrt{\frac{R-mr}{FR}}\left[dt+\sum_k\frac{G_k}{a_kc_k^2}d\phi_k\right],\quad
e^r=\sqrt{\frac{FR}{{R-mr}}}dr,
\nn
e^i&=&\sqrt{\frac{H_i}{d_i(r^2-x_i)}}\left[dt+\sum_k\frac{G_k(r^2+a_k^2)}{a_kc_k^2(x_i+a_k^2)}d\phi_k\right],\quad
e^{x_i}=\sqrt{-\frac{(r^2-x_i)d_i}{4 x_i H_i}}dx_i\nn
e_t&=&-\sqrt{\frac{R^2}{FR(R-mr)}}\left[\d_t-
\sum_k\frac{a_k}{r^2+a_k^2}\d_{\phi_k}\right],\quad 
e_r=\sqrt{\frac{R-mr}{FR}}\d_r,\nn
e_i&=&-\sqrt{\frac{H_i}{d_i(r^2-x_i)}}\left[\d_t-\sum_k\frac{a_k}{x_i+a_k^2}\d_{\phi_k}
\right],\quad e_{x_i}=\sqrt{-\frac{4x_iH_i}{d_i(r^2-x_i)}}\d_{x_i}\\
d_i&=&\prod_{k\ne i}(x_i-x_k),\quad H_i=\prod_k(x_i+a_k^2),\quad
G_i= \prod_k(x_k+a^2_i),
\nn
R&=&\prod_{k} (r^2+a_k^2),\quad
FR=\prod_k(r^2-x_k),\quad c_i^2=\prod_{k\ne i}(a_i^2-a_k^2).\nonumber
\eea
In terms of frames (\ref{AllFramesMP}) the metric and the Killing tensor become
\bea\label{SymmKT}
&&ds^2=-(e^t)^2+(e^r)^2+\sum_k [(e^{x_k})^2+(e^k)^2],\nn
&&K_{MN}dx^Mdx^N=\Lambda_r[-(e^t)^2+(e^r)^2]+\sum_k \Lambda_k [(e^{x_k})^2+(e^k)^2].
\eea
Here $\Lambda_r$ and $\Lambda_k$ are symmetric polynomials, as guaranteed by the general construction of section \ref{SecSeparKT}. The most general KT is obtained by adding Killing vectors (see (\ref{AddKV})) and the metric to the last expression, and this leads to modification of eigenvalues. We are primarily interested in KT that comes from squaring a Killing--Yano tensor, this requires a double degeneracy in the eigenvalues, so (\ref{SymmKT}) is the most natural choice. 

The simplest KYT  is the volume form,
\bea\label{KYTnn}
Y^{(2n)}=e^t\wedge e^r\wedge \prod_k \left[e^{x_k}\wedge e^k\right],
\eea
and its square gives a trivial KT with $\Lambda_r=\Lambda_k=1$ in (\ref{SymmKT}). 
Experience with KYT for the Kerr metric suggests that there is also a KYT of rank $2(n-1)$ and it should have the form
\bea\label{Ydown2}
Y^{(2n-2)}=\la_r \prod_k \left[e^{x_k}\wedge e^k\right]+
e^t\wedge e^r\left[\sum_i \la_i \prod_{k\ne i} \left[e^{x_k}\wedge e^k\right]\right].
\eea
In the four--dimensional Kerr metric we had
\bea\label{KYTLaKerr}
\la_r=\sqrt{r^2},\quad \la_1=\sqrt{-x_1},\quad \Lambda_r=x_1,\quad \Lambda_1=r^2,
\eea
and generalization to higher dimensions is straightforward\footnote{The sign difference between \eqref{KYTLaKerr} and \eqref{LaDwn2} is explained by different conventions for Kerr BH (where we use $ \sqrt{a^2}=a$) and Myers-Perry BH (where $\sqrt{a_k^2}=a_k$) and the relation $a_1=-a$.}:
\bea\label{LaDwn2}
\lambda_r=\sqrt{r^2},\quad \lambda_k=-\sqrt{-x_k},\quad \Lambda_r=\sum_k x_k,\quad
\Lambda_i=r^2+\sum_{k\ne i}x_k.
\eea
Direct calculation shows that (\ref{Ydown2}) with (\ref{LaDwn2}) solves the equation for the KYT. A clear pattern appears: 
\begin{quote}
{\it To construct a KYT of rank $2(n-k)$ one should start with (\ref{KYTnn}) and symmetrically remove $k$ pairs using the rule
\bea\label{ReplaceKYT}
e^t\wedge e^r\rightarrow \sqrt{r^2},\quad e^{x_i}\wedge e^i\rightarrow -\sqrt{-x_i}.
\eea
Then the square of this KYT is the KT (\ref{SymmKT}) with}
\bea\label{ReplaceKYT1}
\Lambda_r=\d_{x_0}\Lambda,\quad \Lambda_i=\d_i\Lambda,\quad \Lambda=x_0x_1\dots x_k+perm,\quad x_0=r^2.
\eea 
\end{quote}
For example, for $k=2$ this procedure gives
\bea
Y^{(2n-4)}&=&\la_r \sum_j \la_j\prod_{k\ne j} \left[e^{x_k}\wedge e^k\right]+
e^t\wedge e^r\left[\sum_{j<m} \la_j\la_m \wedge \prod_{k\ne j,m} \left[e^{x_k}\wedge e^k\right]\right],
\\
\lambda_r&=&\sqrt{r^2},\quad \lambda_k=-\sqrt{-x_k},\quad \Lambda_r=\sum_{k<m} x_kx_m,\quad
\Lambda_i=r^2\sum_k x_k+\sum_{j<k}x_jx_k.\nonumber
\eea
Rather than proving the procedure (\ref{ReplaceKYT}) we connect it to a very nice discussion of \cite{Kub1,Kub2,KubThes,Kub3}, where it was shown that a family of KYT can be constructed starting from 
\bea\label{Kub1}
h=\sum_i a_i\mu_id\mu_i\wedge\Big[a_i dt+(r^2+a_i^2)d\phi_i\Big]+rdr\wedge \Big[dt+\sum_i a_i\mu_i^2d\phi_i\Big]
\eea
by applying an operation
\bea\label{Kub2}
Y^{2(n-k)}=\star\left[\wedge h^k\right].
\eea
While our equations (\ref{ReplaceKYT}), (\ref{ReplaceKYT1}) give simpler expressions for the KYT and KT due to the use of convenient frames, they reduce to the construction (\ref{Kub1})--(\ref{Kub2}) once (\ref{Kub1}) is rewritten in the frames  \eqref{AllFramesMP}:
\bea\label{HInFrame}
h=re^r\wedge e^t+\sum_i\sqrt{-x_i}e^{x_i}\wedge e^i.
\eea
Construction (\ref{Kub2})--(\ref{HInFrame})  is proven in Appendix \ref{AppPCKYT}, and here we just outline the steps:
\begin{enumerate}
\item Expression (\ref{HInFrame}) gives a conformal Killing--Yano tensor (CKYT) for the Myers--Perry black hole, and the two--form $h$ is closed.
\item The product ${\cal Y}=[\wedge h^k]$ has the same properties as $h$ (i.e., it is a closed CKYT).
\item A Hodge dual of any closed CKYT is a KYT.
\end{enumerate}
Justifications of these statements are scattered throughout the literature 
\cite{Carig,Kub1,KubThes}, and Appendix \ref{AppPCKYT} provides streamlined derivations. 
Construction (\ref{Kub2})--(\ref{HInFrame}) of the KYT will be extended to a charged black hole in section \ref{SecMPF1}.

\bigskip

We conclude this section by a brief discussion of the Myers--Perry black hole in odd dimensions. Instead of starting with (\ref{MPeven}) one should begin with
\bea\label{MPodd}
ds^2=-dt^2+\frac{mr^2}{FR}\Big(dt+\sum_{i=1}^n a_i\mu_i^2 d\phi_i\Big)^2+\frac{FR dr^2}{R-mr^2}
+\sum_{i=1}^n(r^2+a_i^2)\Big(d\mu_i^2+\mu_i^2 d\phi_i^2\Big),
\eea
then repetition of the previous analysis leads to the counterpart of (\ref{AllFramesMP}):
\bea\label{AllFramesMPOdd}
e^t&=&
\sqrt{\frac{R-mr^2}{FR}}\left[dt+\sum_k\frac{a_kG_k}{c_k^2}d\phi_k\right],\quad
e^r=\sqrt{\frac{FR}{R-mr^2}}dr,
\nn
e^i&=&\sqrt{-\frac{H_i}{x_id_i(r^2-x_i)}}\left[dt+
\sum_k\frac{G_ka_k(r^2+a_k^2)}{c_k^2(x_i+a_k^2)}d\phi_k\right],\quad
e^{x_i}=\sqrt{\frac{d_i(r^2-x_i)}{4H_i}}dx_i,
\nonumber\\
e_t&=&-\sqrt{\frac{R^2}{FR(R-mr^2)}}\left[ \d_t
-\sum_k\frac{a_k}{r^2+a_k^2}\d_{\phi_k}\right],\quad e_r=\sqrt{\frac{R-mr^2}{FR}}\d_r,\\
e_i&=&-\sqrt{-\frac{H_i}{x_id_i(r^2-x_i)}}\left[\d_t-\sum_k\frac{a_k}{x_i+a_k^2}\d_{\phi_k}
\right],\quad e_{x_i}=\sqrt{\frac{4H_i}{d_i(r^2-x_i)}}\d_{x_i},\nonumber
\eea
and to one more frame that was not present in the even--dimensional case:
\bea
e^\psi=\sqrt{\frac{\prod a_i^2}{r^2\prod (-x_k)}}\left[dt +
\sum_k \frac{G_k(r^2+a_k^2)}{c^2_k a_k} d\phi_k \right],~ 
e_\psi=-\sqrt{\frac{\prod a_i^2}{r^2\prod (-x_k)}}\left[\d_t-\sum_k\frac{1}{a_k}\d_{\phi_k}
\right].
\eea
Notice that one of the relations (\ref{EplsdEven}) between Myers--Perry and ellipsoidal coordinates is modified\footnote{Notice that in contrast to the even-dimensional case, where $\mu_i$ were not constrained, now there is a relation $\sum \mu_i^2=1$, and, as a consequence, there only $n-1$ coordinates $x_i$.}:
\bea\label{EplsdOdd}
\mu_i^2=\frac{1}{c_i^2}\prod_{k=1}^{n-1} (a_i^2+x_k).
\eea
This leads to a new expression for
\bea\label{FROdd}
FR=r^2\prod_k(r^2-x_k)
\eea
and we still have the remaining relations
\bea\label{OtherOdd}
d_i&=&\prod_{k\ne i}(x_i-x_k),\quad H_i=\prod_k^n(x_i+a_k^2),\quad
G_i= \prod_k^{n-1}(x_k+a^2_i),
\nn
R&=&\prod_{k}^n (r^2+a_k^2),\quad c_i^2=\prod_{k\ne i}(a_i^2-a_k^2).
\eea
Note a very special form of the relative coefficients in frames $e_a$: they depend only on $r$ in 
$e_t$, only on $x_i$ in $e_i$, and they are constant in $e_\psi$.

The Killing--Yano tensors are still given by construction (\ref{Kub2}) with
\bea
h=re^r\wedge e^t+\sum_i\sqrt{-x_i}e^{x_i}\wedge e^i.
\eea
The separation factors are 
\bea
\rho_r=r^2\prod_j^{n-1}(r^2-x_j),\quad \rho_i=x_i(r^2-x_i)\prod_{k\ne i}[x_i-x_k].
\eea
This reduces to (\ref{RhoR}), (\ref{RhoI}) if we introduce $x_{n}\equiv 0$.


\section{Killing(--Yano) tensors and string dualities}
\label{SecKillingsDualities}
\renewcommand{\theequation}{4.\arabic{equation}}
\setcounter{equation}{0}

In this section we will analyze transformations of various tensors under string dualities. Specifically, we will focus on T dualities along $U(1)$ isometries and assume that Killing--(Yano) tensors do not depend on coordinates parameterizing the isometries. We will also consider larger classes of U duality transformations. Our results are summarized below:
\begin{itemize}
\item Generically, the Killing vectors depending on the direction of T duality are destroyed (as we will show in section \ref{SubsectionzKV}), and Killing vectors with trivial dependence on the duality direction survive the duality, as long as original fluxes respect the symmetry associated with Killing vectors (see section \ref{SubsectionKV}).
\item Conformal Killing vectors are destroyed by the T duality with an exception of the homothetic CKV. The latter acquire nontrivial dependence upon the duality direction in 
the dual geometry (see section \ref{SubsectionCKV}).
\item KT equation remains the same, but there are constraints on the $B$ field and the dilaton (\ref{Hcond2fields}), \eqref{ConstrBKT},  (see section \ref{SecKTdual}).
\item Extension of T duality to the CKT is possible only for special solutions, and some examples are presented in Appendix \ref{AppCKT}.
\item KYT equation is modified by terms containing the Kalb--Ramond field \eqref{KYTBfield}, and there is an additional constraint \eqref{RestBKYTGuess}  (or, more generally, (\ref{ConstrBcompBdy})) on this field (see section \ref{SectionModifiedKYT}). 
\item Extension of T duality to CKYT is possible only for special solutions.
\end{itemize}
We will now discuss all theses properties in detail.


\subsection{Killing vectors and T duality}
\label{SecKVDuality}

In this subsection we will analyze the transformations of the Killing vectors under combinations of T dualities and reparametrizations. The most natural formalism for such study is provided by the Double Field Theory (DFT) \cite{DFT}, which is reviewed in Appendix \ref{AppDFT}, and a very simple interpretation of our results in terms of this approach is presented in the end of section \ref{SubsectionKV}.

We will begin with a pure metric 
\be\label{DimRedKV}
ds^2=e^C[dz+A_idx^i]^2+\hat g_{ij}dx^idx^j, \quad B_{MN}=0
\ee
that admits two Killing vectors, $Z=\d_z$ and $V=V^M\d_M$, and study the transformation of vector $V$ under T duality along $z$ direction. We will look at three situations and the results are summarized as follows:
\begin{enumerate}[(a)]
\item The $z$--independent vectors $V$ (i.e., vectors commuting with $Z$) have counterparts after T duality, and the transformation law is derived in section \ref{SubsectionKV}.
\item The $z$--dependent vectors $V$ (i.e., vectors with $[V,Z]\ne 0$) may be destroyed by the duality transformation, and in general the numbers of such vectors before and after T duality do not match. Some examples are discussed in section \ref{SubsectionzKV}.
\item Conformal Killing Vectors of the original geometry are destroyed by T duality unless one introduces $z$--dependence in the dual frame. This construction is discussed in section \ref{SubsectionCKV}.
\end{enumerate}
In case (a) we will find an additional constraint on the Kalb--Ramond field after duality:
\bea\label{HfieldCond}
H_{MNP}V^P=\nabla_M W_N-\nabla_N W_M,\quad\mbox{with arbitrary}\quad W_N,
\eea 
and we will demonstrate that any geometry that has a Killing vector $V$ satisfying (\ref{HfieldCond}) can be dualized in a direction commuting with $V$ without destroying the Killing vector. We will also show that condition (\ref{HfieldCond}) arises naturally from the equation for a Killing vector in DFT.  

\subsubsection{Killing vectors commuting with T duality direction}
\label{SubsectionKV}

Let us first assume that geometry (\ref{DimRedKV}) solves Einstein's equations without $B$ field, and that it admits a Killing vector $V$:
\bea\label{Apr5}
\nabla_M V_N+\nabla_N V_M=0
\eea
which commutes with $Z=\d_z$. In Appendix \ref{AppDimRed1} we perform dimensional reduction of this equation in geometry (\ref{DimRedKV}) before and after T duality in $z$ direction. Using tildes to denote the quantities after T duality, we find various components of (\ref{Apr5}) and its dual counterpart:
\bea\label{KVAllEqs}
\begin{array}{|r|c|c|}
\hline
&\nabla_M V_N+\nabla_M V_N=0&\Big.\nabla_M\tilde V_N+\nabla_N \tilde V_M=0\\
\hline
zz:&\d_rC V^r=0&\Big.\d_rC \tilde V^r=0\\
mz:&F_{mr}V^r=\d_m(e^{-C}V_z)&\d_m(e^{C}\tilde V_z)=0\\
mn:&\hat\nabla^mV^n+\hat\nabla^nV^m=0&\Big.\hat\nabla^m\tilde V^n+\hat\nabla^n\tilde V^m=0\\
\hline
\end{array}
\eea
Here $\hat\nabla$ denotes the covariant derivative corresponding to metric ${\hat g}_{ij}$. 

Comparison of two columns on (\ref{KVAllEqs}) leads to the transformation law
\bea\label{May15}
\tilde V^r=V^r, \quad \tilde V^z\equiv e^C\tilde V_z=\mbox{const}.
\eea
Relation (\ref{May15}) ensures that the Killing equations after T duality are satisfied, but the $(mz)$ component of the original equation imposes a constraint on the new $B$ field:
\bea\label{BConstrComp}
{\tilde B}_{mz}=A_m\quad \Rightarrow\quad {\tilde H}_{zmp}V^p=\d_m(e^{-C}V_z).
\eea
Notice that this is the only relation in the dual frame that contains the original $V_z$. 

The implications of the constraint (\ref{BConstrComp}) are analyzed in Appendix \ref{AppDRKV}, where it is shown that a pair of relations
\bea\label{BConstrCovar}
&&\nabla_M V_N+\nabla_M V_N=0,\nonumber\\
&&H_{MNP}V^P=\nabla_M W_N-\nabla_N W_M
\eea
is preserved by T duality as long as one imposes the the transformation 
\bea\label{May5a}\label{KVtransf}
&&{\tilde V}^a=V^a,\quad {\tilde W}_z= -e^{-C}V_z,\quad
{\tilde V}_z= -e^{-C}W_z,\nonumber\\
&&{\tilde W}_n=W_n-{\tilde A}_n e^{-C}V_z-A_n W_z+\d_n f,
\eea
with arbitrary function $f$. Although we motivated (\ref{BConstrCovar}) by starting with a pure metric, the map (\ref{KVtransf}) leaves (\ref{BConstrCovar}) invariant for arbitrary configurations of the $B$ field before and after the duality.

The system (\ref{BConstrCovar}) is the unique extension of the equation for Killing vector consistent with T duality, and in Appendix \ref{AppDFT} we show that (\ref{BConstrCovar}) can be written as a single equation for a Killing vector on an extended space used in the Double Field Theory (DFT). 
Specifically, if the metric and the $B$ field are combined in a single matrix (\ref{DefDFT})\footnote{In equations (\ref{DefDFTmain})--(\ref{LieDFTmain}) and in Appendix 
\ref{AppDFT} we deviate from the notation used throughout this paper and denote the spacetime indices by lower--case letters, while reserving the capital ones to label the ``double space". This notation is standard in the DFT literature.}
\bea\label{DefDFTmain}
\cH_{IJ}=\begin{pmatrix}g^{ij}& -g^{ik}B_{kj}\\B_{ik}g^{kj}&g_{ij}-B_{ik}g^{kl}B_{lj} \end{pmatrix},
\eea
then equations (\ref{BConstrCovar}) appear as different components of a single equation for $\xi^P$:
\be\label{LieDFTmain}
{L}_{\xi}\mathcal{H}_{MN}\equiv\xi^P\d_P\cH_{MN}+(\d_M\xi^P-\d^P\xi_M)\cH_{PN}+(\d_N\xi^P-\d^P\xi_N)\cH_{MP}=0
\ee
Here $\xi^I=(\tilde\lambda_i,\lambda^i)$ is the generalized gauge parameter, where $\tilde\lambda_i$ corresponds to the gauge transformation of the Kalb--Ramond field $B_{ij}$ and  $\lambda^i$ generates diffeomorphisms. Equation (\ref{LieDFTmain}), which involves the generalized Lie derivative in double space $L_\xi$, implies that the system (\ref{BConstrCovar}) is covariant under combinations of diffeomorphisms and T--dualities.


\subsubsection{Killing vectors with $z$ dependence}
\label{SubsectionzKV}

In the previous subsection we assumed that components of the Killing vector $V$ did not depend on the direction of T duality\footnote{In covariant form this condition is written as $[Z,V]=0$.} and demonstrated that components of the Killing vector transform in a simple way (\ref{KVtransf}). Here we will use several examples to argue that situation for the $z$--dependent Killing vectors is rather different: even the number of such vectors can be changed by application of T duality. 

We begin with the simplest example of a pure metric
\be
ds^2=f(dz^2+dy^2)+g_{mn}dx^m dx^n
\ee
which admits a Killing vector corresponding to rotations in the $(y,z)$ plane:
\be\label{OrigKV}
V=y\d_z-z\d_y.
\ee
Performing the T duality along $z$ direction and  
solving equations for the Killing vector in the dual configuration,
\bea
ds^2=\frac{dz^2}{f}+f dy^2+g_{mn}dx^m dx^n,
\eea
we find that there are only two KVs with nontrivial $(y,z)$ components:
\bea
V=c_1\d_y+c_2\d_z
\eea
unless $f=const$, where there is also a counterpart of (\ref{OrigKV}):
\bea
V=f^2y\d_z-z\d_y.
\eea
We conclude that the $z$--dependent Killing vector (\ref{OrigKV}) disappears unless $f$ is equal to constant. 

The same phenomenon can be seen in a more interesting geometry produced by smeared fundamental strings \cite{FundString}:
\bea\label{Apr5a}
ds^2&=&H^{-1}(dz^2-dt^2)+dr^2+r^2d\Omega_p^2+\sum_{k=1}^{7-p} dx_kdx_k, \nn
B&=&(H^{-1}-1)dt\wedge dz, \quad e^{2\Phi}=H^{-1}, \quad H=1+\frac{Q}{r^{p-1}}.
\eea
The most general Killing vector with $(z,t)$ components has the form
\bea
V=c_1\d_t+c_2\d_z+c_3(t \d_z+z \d_t).
\eea
T duality along $z$ direction leads to a metric produced by a plane wave, which has only two independent Killing vectors with components in $(t,z)$ directions:
\bea
V=c_1\d_t+c_2\d_z.
\eea
Once again, $z$--dependent Killing vector disappears after T duality.
In section \ref{SectionModifiedKYT} we will encounter a similar situation with Killing--Yano tensors (KYT): at first sight they seem to be destroyed by T duality. To cure this problem we will modify the equation for KYT by adding an extra term containing the Kalb--Ramond field. This solution would not work in the present case: since the geometry dual to (\ref{Apr5a}) does not contain matter fields, the original equation (\ref{Apr5}) is the unique relation consistent with invariance under diffeomorphisms. 

To summarize, we conclude that $z$--dependent Killing vectors can appear and disappear under T dualities, so they don't have well--defined transformation properties. We expect the situation to be at least as bad for the Killing(--Yano) tensors, so in sections \ref{SecKTdual} and \ref{SectionModifiedKYT} we will focus only on $z$--independent objects. However, 
$z$--dependence can lead to very interesting effects for conformal Killing vectors, which will be discussed now.

\subsubsection{Conformal Killing Vectors and T duality.}
\label{SubsectionCKV}

Conformal Killing vectors (CKV) do not leave the metric invariant, but rather they lead to rescalings by a conformal factor. Such vectors satisfy differential equation
\be
\nabla_M {\cal V}_N+\nabla_N {\cal V}_M=g_{MN}v,
\ee
with some function $v$. Dimensional reduction of this equation gives the counterpart of 
(\ref{KVAllEqs})\footnote{Recall that we are starting with a pure metric, so there are no $g_{zm}$ components after duality. Reductions (\ref{CKVAllEqs}) and (\ref{CKVAllEqsZ}) follow directly from Appendix \ref{AppDimRed1}.} :
\bea\label{CKVAllEqs}
\begin{array}{|c|c|c|}
\hline
&\nabla_M {\cal V}_N+\nabla_N {\cal V}_M=g_{MN}v
&\Big.\tilde\nabla_M \tilde {\cal V}_N+\tilde\nabla_N\tilde {\cal V}_M=\tilde g_{MN}\tilde v\\
\hline
zz&\h\d_re^C {\cal V}^r=e^Cv&\Big.\h\d_re^{-C} \tilde {\cal V}^r=e^{-C}\tilde v\\
mz&F_{mr}{\cal V}^r=\d_m(e^{-C}{\cal V}_z)&\d_m(e^{C}\tilde {\cal V}_z)=0\\
mn&\tilde\nabla^m{\cal V}^n+\tilde\nabla^n{\cal V}^m=g^{mn}v&
\Big.\hat\nabla^m\tilde {\cal V}^n+\hat\nabla^n\tilde {\cal V}^m=g^{mn}\tilde v\\
\hline
\end{array}
\eea
Imposing the relation ${\cal V}^n=\tilde{\cal V}^n$, we conclude that $v={\tilde v}$, then $(zz)$ components lead to contradiction unless $C$ is a constant or $v$ is equal to zero. To cure this problem, we allow $z$ dependence in the conformal Killing tensor after duality and replace (\ref{CKVAllEqs}) by\footnote{Notice that introduction of $z$ dependence after duality puts the initial and final system on a different footing. Similar situation is encountered in the non--Abelian T duality \cite{NabTdual}, but there an analog of $z$--dependence is introduced for the dynamical fields, while here we are looking at the Killing vectors.}
\bea\label{CKVAllEqsZ}
\begin{array}{|c|c|c|}
\hline
&\nabla_M {\cal V}_N+\nabla_N {\cal V}_M=g_{MN}v
&\Big.\tilde\nabla_M \tilde {\cal V}_N+\tilde\nabla_N\tilde {\cal V}_M=\tilde g_{MN}\tilde v\\
\hline
zz&\h\d_re^C {\cal V}^r=e^Cv&
\Big.\d_z{\tilde{\cal V}}_z+\h\d_re^{-C} \tilde {\cal V}^r=e^{-C}\tilde v\\
mz&F_{mr}{\cal V}^r=\d_m(e^{-C}{\cal V}_z)&\d_m(e^{C}\tilde {\cal V}_z)+\d_z {\tilde{\cal V}}_m=0\\
mn&\hat\nabla^m{\cal V}^n+\hat\nabla^n{\cal V}^m=g^{mn}v&
\Big.\hat\nabla^m\tilde {\cal V}^n+\hat\nabla^n\tilde {\cal V}^m=g^{mn}\tilde v\\
\hline
\end{array}
\eea
Once again setting
\bea
\tilde {\cal V}^n={\cal V}^n, \quad \tilde v=v,
\eea
we find a system of equations for ${\tilde{\cal V}}^z$:
\bea
 {\cal V}^r\d_r C=2v,\quad \d_z {\tilde{\cal V}}^z=2 \tilde v,\quad \d_m{\tilde {\cal V}}^z=0
\eea
since the original CKV ${\cal V}$  does not depend of $z$. 
Integrability conditions for the last two equations imply that ${\tilde v}$ must be constant, so the CKV ${\cal V}$ must be homothetic. A simple example of a homothetic KV comes from rescaling of the flat space by a constant factor:
\bea
ds^2=\eta_{MN}dx^Mdx^N,\quad 
{\cal V}_M dx^M=\eta_{MN}x^N dx^M,\quad
v=1.
\eea

To summarize, for every homothetic CKV we find the complete set of transformations,
\be\label{SolCKVzDep}
{\tilde {\cal V}}^m={\cal V}^m,\quad {\tilde v}=v=\mbox{const},\quad 
{\tilde {\cal V}}^z=2zv+const
\ee
that produces a CKV after T duality. Non--homothetic conformal Killing Vectors are destroyed by T duality.

\subsection{Killing tensors in the NS sector}
\label{SecKTdual}

In this subsection we study the behavior of Killing tensors (KT) under $O(d,d)$ transformations, which include boosts, T dualities and rotations, and then extend the construction to the full NS sector by incorporating transformations involving S dualities. 

As discussed in section \ref{SecSeparKT} equation \eqref{KTeqn4} has reducible solution spanned by combinations of the metric and Killing vectors,
\bea\label{KtrivA}
K^{triv}_{MN}=e_0 g_{MN}+\sum_{ij} e_{ij}V^{(i)}_MV^{(j)}_N,
\eea
with constant coefficients $e_0$, $e_{ij}$. In section \ref{SecKVDuality} we showed that Killing vectors are preserved by the $O(d,d)$ transformations if conditions (\ref{BConstrCovar}) are satisfied. This implies that the expression (\ref{KtrivA}) for the ``trivial Killing tensor'' holds for the entire $O(d,d)$ orbit. Here we will focus on non--trivial Killing tensors, which can be either destroyed or modified by T duality, and we identify a subset of $O(d,d)$ transformations which do not lead to destruction of a nontrivial KT. The non--trivial Killing tensors can be found either by solving equation (\ref{KTeqn4}) or by separating the Hamilton--Jacobi equation \cite{Carter}, and the second approach is more convenient for the study of T duality. The relationship between Killing tensors and separation of the massive Hamilton--Jacobi equation has been reviewed in section \ref{SecKTandHJ}, and in this subsection these results will be extended to charged solutions. An alternative approach based on dimensional reduction of KT equation is discussed in Appendix \ref{AppDRKT4}. 

In subsection \ref{SecKTOdd} we focus on the $O(d,d)$ orbit which generates fundamental strings from pure metric, and in subsection \ref{SecKTBeyondOdd} these results are extended to general F1--NS5 solutions. As we will see, existence of KT imposes certain restrictions on the Kalb--Ramond field, and they are discussed in subsection \ref{SecCondsBfield}. Finally in subsection \ref{SecCondsBfieldDR} we use an alternative method (dimensional reduction) to derive the covariant form of the constraint on the $B$ field.

\subsubsection{Killing tensors and $O(d,d)$ transformations}
\label{KTOdd}
\label{SecKTOdd}

We begin with a pure metric that solves source--free Einstein equations in $D$ dimensions, admits a Killing tensor, and has $d$ cyclic directions $\phi^a$. Such geometry can be written in a reduced form:
\bea\label{TorusMetr}
ds^2=G_{ab}(d\phi^a+V^a_m dx^m)(d\phi^b+V^b_n dx^n)+h_{mn}dx^m dx^n.
\eea
This metric has an obvious $GL(d)$ symmetry that rotates cyclic directions into each other, but in supergravity this symmetry is enhanced to $O(d,d)$, which acts on the metric and on the Kalb--Ramond $B$ field \cite{Odd,GenMetric}. This symmetry is extended further to $O(D,D)$ via the Double Field Theory (DFT) formalism \cite{DFT}, which is reviewed in Appendix \ref{AppDFT}.

Specifically, a $2D\times 2D$ matrix written in $D\times D$ blocks
\bea
M=\left[
\begin{array}{cc}
G^{-1} & -G^{-1}B \\
BG^{-1} & G-BG^{-1}G
\end{array}
 \right]
\eea
is transformed under a global $O(D,D)$ as 
\bea\label{Mrot}
M\rightarrow \Omega M\Omega^T,
\eea
where
\bea
\Omega \eta\Omega^T=\eta, \qquad \eta=\left[
\begin{array}{cc}
0 & 1 \\
1 & 0
\end{array}
 \right].
\eea
Here $\eta$ is a metric for a group $O(D,D)$.

Since we are starting with a pure metric, the initial matrix $M$ is given 
by\footnote{Note that $g^{ab}$, $q^{am}$ and $h^{mn}$ are the components of $D\times D$ matrix $G^{-1}$.}
\bea\label{ModuliPureMetr}
M=\left[
\begin{array}{cc|cc}
g^{ab}&q^{am}&0&0\\
q^{ma}&h^{mn}&0&0\\
\hline
0&0&G_{ab}&G_{am}\\
0&0&G_{ma}&G_{mn}\\
\end{array}
\right].
\eea
Parameterizing the $O(d,d)$ rotations by $d\times d$ matrices $A,C,D,E$ as
\bea\label{OddTrans}
\Omega=\left[
\begin{array}{cc|cc}
A&0&E&0\\
0&I_{D-d}&0&0\\
\hline
C&0&D&0\\
0&0&0&I_{D-d}\\
\end{array}
\right],\quad 
\left[
\begin{array}{cc}
A^T&C^T\\
E^T&D^T
\end{array}\right]
\left[\begin{array}{cc}
0&I_d\\
I_d&0
\end{array}\right]
\left[
\begin{array}{cc}
A&E\\
C&D
\end{array}\right]=
\left[\begin{array}{cc}
0&I_d\\
I_d&0
\end{array}\right]
\eea
we find the transformed metric with upper indices
\bea\label{RotateMDFT}
\Omega M \Omega^T=
\left[
\begin{array}{c|c}
\begin{array}{cc}
AgA^T+EGE^T&A q\\
qA^T&h
\end{array}&
\bullet\\
\hline
\bullet&\bullet
\end{array}
\right]
\eea
Here and below $G$ denotes a $d\times d$ matrix with components $G_{ab}$.
The survival of the Killing tensor under transformation with arbitrary $A$ and $B$ implies that the following four quantities must separate:
\bea\label{SepCondPureMetr}
fg^{ab},\quad fq^{am},\quad fh^{mn},\quad fG_{ab}.
\eea
The first three conditions are satisfied before the $O(d,d)$ transformation since metric (\ref{TorusMetr}) had a Killing tensor. Separation in the dual frame requires  $fG_{ab}$ to separate with {\it the same} function $f$. 
Combining this with results of section \ref{SecKTandHJ} we arrive at the following conclusion:
\begin{enumerate}[(1)]
\item{Every KT is associated with a unique function $f$, which can be determined from the HJ equation or from eigenvalues, and with corresponding variables $(x,y)$.}
\item{T dualities and rotations in a sector spanned by cyclic coordinates $\phi_a$ do not spoil separation of variables for a given KT if and only if
\bea\label{DefTranslDir}
\d_x\d_y[f G_{ab}]=0.
\eea
}
\end{enumerate}
So far we have separated coordinates into cyclic and non--cyclic, but equation (\ref{DefTranslDir}) suggests a more refined distinction: among cyclic coordinates $\phi^a$ we identify the subsector where  (\ref{DefTranslDir}) holds and call the corresponding cyclic directions translational, and the remaining directions will be called rotational\footnote{Strictly speaking one should define coordinates are rotational and translational with respect to a particular Killing tensor: the same cyclic coordinate might by translational for one KT and rotational for another. Since we are dealing with one tensor at a time referring to a direction as simply translational should not cause confusions.}. A simple example demonstrates the origin of these names: in the metric 
\bea
ds^2=dr^2+r^2 d\theta^2+r^2\sin^2\theta (d\phi^1)^2+(d\phi^2)^2
\eea
coordinate $\phi^2$ would be called translational and coordinate $\phi^1$ would be called rotational since in this case $x=r$, $y=\theta$, and $f=r^2$. For many aspects of our discussion rotational coordinates appear on the same footing as non--cyclic ones. 

Once we have demonstrated that the Killing tensor is not destroyed by the $O(d,d)$ transformations as long as expressions (\ref{SepCondPureMetr}) separate, we can ask about transformation laws for this tensor. Recall that Killing vectors with upper components were unaffected by the $O(d,d)$ transformations, but Killing tensor has a more interesting behavior. The third expression in (\ref{SepCondPureMetr}) indicates that the separation function cannot be affected by the $O(d,d)$ transformations since $h^{mn}$ is invariant under them. This implies simple relations for the Killing tensors before and after T duality\footnote{For simplicity we are focusing on Killing tensor which separates two non--cyclic coordinates $x$ and $y$. Generalization to ore coordinates is straightforward, but the notation becomes cumbersome.
}:
\bea\label{KillTilde}
K^{MN}&=&X^{MN}-f_xg^{MN}, \quad {\tilde K}^{MN}={\tilde X}^{MN}-f_x{\tilde g}^{MN}.
\eea
We use tildes to denote the expressions after T duality. As discussed in section \ref{SecSeparKT}, separation in the original metric implies that
\bea
g^{MN}=\frac{1}{f}\left[X^{MN}+Y^{MN}\right],\nonumber
\eea
and the last condition in (\ref{SepCondPureMetr}) leads to an additional relation
\bea
G_{ab}=\frac{1}{f}\left[{\hat X}_{ab}+{\hat Y}_{ab}\right],
\eea
where $X,{\hat X}$ are functions of $x$ and $Y,{\hat Y}$ are functions of $y$. Then transformation (\ref{RotateMDFT}),
\bea
{\tilde g}^{ab}=\left[AgA^T+EGE^T\right]^{ab},\quad
{\tilde g}^{am}=A_{ab}g^{bm},\quad {\tilde g}^{mn}=g^{mn}\nonumber
\eea
gives
\bea\label{XforF1}
{\tilde X}^{ab}=\left[AXA^T+E{\hat X}E^T\right]^{ab},\quad 
{\tilde X}^{am}=A_{ab}X^{bm},\quad {\tilde X}^{mn}=X^{mn}.
\eea
Along with (\ref{KillTilde}) this completely determines the transformation of the Killing tensor under the action of $O(d,d)$.


To summarize, we have demonstrated that transformation (\ref{RotateMDFT}) preserve the Killing tensor as long as all directions $\phi^a$ in (\ref{TorusMetr}) are chosen to be translational, and all cyclic rotational directions are absorbed in $h_{mn}$. Notice, however, that some components on the Killing tensor are modified according to (\ref{KillTilde}), (\ref{XforF1}). Transformations (\ref{RotateMDFT}) allow 
one to generate a large class of charged solutions of supergravity starting from a simple neutral ``seed'', and this technique has been used to generate large classes of charged black holes in \cite{Sen,Cvetic}. 
One can also start with a ``seed'' which already contains a nontrivial Kalb--Ramond field, and the generalization of our analysis is straightforward. 

Suppose that metric (\ref{TorusMetr}) is supported by the $B$ field and the dilaton which are invariant under translations in $\phi$ directions:
\bea
\d_{\phi^a}e^{2\Phi}=0,\quad {\cal L}_{\phi^a} B=0.
\eea
Then application of the rotation \eqref{Mrot} with $\Omega$ given by \eqref{OddTrans} to the initial moduli matrix\footnote{Note that $Q$ is the full inverse metric, for example $Q^{aM}B_{Mb}=g^{as}B_{cb}+q^{as}B_{sb}$.}
\bea
M=\left[
\begin{array}{cc|cc}
g^{ab}&q^{am}&-Q^{aM}B_{Mb}&-Q^{aM}B_{Mm}\\
q^{nb}&h^{nm}&-Q^{nM}B_{Mb}&-Q^{nM}B_{Mm}\\
\hline
B_{aM}Q^{Mb}&B_{aM}Q^{Mm}&G_{ab}-B_{aM}Q^{MN}B_{Nb}&G_{am}-B_{aM}Q^{MN}B_{Nm}\\
B_{nM}Q^{Mb}&B_{nM}Q^{Mm}&G_{nb}-B_{nM}Q^{MN}B_{Nb}&G_{nm}-B_{nM}h^{MN}B_{Nm}\\
\end{array}
\right]
\eea
gives\footnote{Recall that indices of rotational matrices appearing in (\ref{OddTrans}) go only over specific subsets $A_{as},E_{am},C_{ma},D_{mn}$, so for example $(AQ)_a{}^M=A_{ab}Q^{bM}$.}
\bea
\Omega M \Omega^T=\left[
\begin{array}{c|c}
\begin{array}{cc}
AgA^T-AQBE^T + EBQA^T+E(G-BQB)E^T & Aq+EBQ\\
hA^T-QBE^T&h
\end{array}&
\bullet\\
\hline
\bullet&\bullet
\end{array}
\right].\nonumber
\eea
The new metric admits a Killing tensor if and only if the following combinations of the original quantities separate:
\bea\label{SepCondB}
\boxed{
fg^{ab},\quad fB_{aM}g^{Mb},\quad fB_{aM}g^{Mm},\quad  f(g_{ab}-B_{aM}g^{MN}B_{Nb}),
\quad fg^{am},\quad   fg^{mn}.}
\eea

In spite of the appearances, conditions (\ref{SepCondB}) are invariant under gauge transformations of the $B$ field. We will demonstrate this for the most interesting case where $B_{aM}$ has both legs in the cyclic directions (one of them translational and the other one is either translational or rotational). Indeed, separability of the second and third expressions in coordinates $(x,y)$ implies that
\bea\label{Feb7a}
\d_x\d_y(fg^{NM}B_{Mb})=0,
\eea
next recalling that that $\d_x\d_y(fg^{NM})=0$, the last condition can be rewritten in the gauge--invariant form:
\bea\label{Hcond1}
\d_y(f g^{NM})H_{xMb}+\d_x (f g^{NM}) H_{y Mb}+f g^{NM}\d_xH_{y Mb}=0.
\eea
Similarly, separability of the fourth expression in (\ref{SepCondB}) can be rewritten as
\bea\label{Hcond}
\d_x\d_y (fg_{ab})-fg^{MN}H_{y aM}H_{xNb}-fg^{MN}H_{xaM}H_{y Nb}=0.
\eea
By construction, constraints on the $B$ field for any point on an $O(d,d)$ trajectory passing through a pure metric are just separability conditions for the initial metric \eqref{SepCondPureMetr}.

\subsubsection{Conditions on the $B$ field from dimensional reduction}
\label{SecCondsBfieldDR}

So far we have been studying transformation of Killing tensors under $O(d,d)$ rotations using separation of HJ equation. Now we will use an alternative approach based on dimensional reduction to derive the unique covariant form of the constraint on the $B$ field, and the result is given by \eqref{ConstrBKT}.

Let us start with a standard Killing tensor equation
\bea\label{KTEq}
\nabla_M K_{NP}+\nabla_N K_{MP}+\nabla_P K_{MN}=0,
\eea
and perform dimensional reduction of the metric along $z$ direction:
\bea
ds^2=e^C[dz+A_i dx^i]^2+\hat{g}_{ij} dx^i dx^j.
\eea
The details of such reduction are given in Appendix \ref{AppDRKT4}, in particular $mnp$ components of the Killing tensor equation \eqref{KTEq}
\bea
\hat{\nabla}^m K^{np}+\hat{\nabla}^n K^{mp}+\hat{\nabla}^p K^{mn}=0
\eea
transform under T duality into 
\bea
\hat{\nabla}^m \tilde{K}^{np}+\hat{\nabla}^n \tilde{K}^{mp}+\hat{\nabla}^p \tilde{K}^{mn}=0.
\eea
We conclude that the KT equation is not modified by the $B$ field, in contrast to Killing-Yano tensor case, which will be discussed in section \ref{SectionModifiedKYT}. Next we look at the $mnz$ components
\bea\label{mnzKT}
{\hat g}^{ma}\left[\hat\nabla_a(e^{-C}K^n{}_{z})+F_{ba}K^{nb}\right]+(m\leftrightarrow n)=0.
\eea
Under T duality along $z$ direction $F_{mn}$ transforms into $H_{mnz}$ ($H=dB$), so we conclude that T dual counterpart of \eqref{mnzKT} should give an equation involving the $B$ field. As demonstrated in Appendix \ref{AppDRKT4}, the only covariant form of such equation is
\bea\label{ConstrBKT}
{\tilde H}_{AMP}\tilde K_N{}^A+{\tilde H}_{ANP}\tilde K_M{}^A=e^{C/2}{\tilde\nabla}_M[e^{-C/2}{\tilde W}_{NP}]+e^{C/2}{\tilde\nabla}_N[e^{-C/2}{\tilde W}_{MP}].
\eea
Recall that we had a similar expression as a constraint on the $B$ field for a Killing vector \eqref{BConstrCovar}.

Notice that the equation \eqref{mnzKT} has an interesting interpretation in terms of Lie derivatives. As
 shown in Appendix \ref{AppDRKT4} for the KT constructed from squaring a Killing vector as 
 $K^{mn}=V^m V^n$, equation \eqref{mnzKT} reduces to a combination of Lie derivatives of $A_m$ (recall that $F_{mn}=\d_m A_n-\d_n A_m$) along the Killing vector $V^m$ 
\bea
lhs=V^n \mathcal{L}_V A^m+V^m \mathcal{L}_V A^n.
\eea

To summarize we have used dimensional reduction to demonstrate that requirement of covariance of Killing tensor under T duality leads to the unique constraint on the $B$ field \eqref{ConstrBKT} similar to the equation on the $B$ field satisfied by Killing vectors. We will now discuss the behavior of Killing tensors under the U--duality group that extends $O(d,d)$ transformations, and demonstrate that covariance under such dualities leads to additional constraints on the Kalb--Ramond field.



\subsubsection{Extension beyond $O(d,d)$}
\label{SecKTBeyondOdd}

\begin{figure}
    \centering
    \includegraphics[width=1\textwidth]{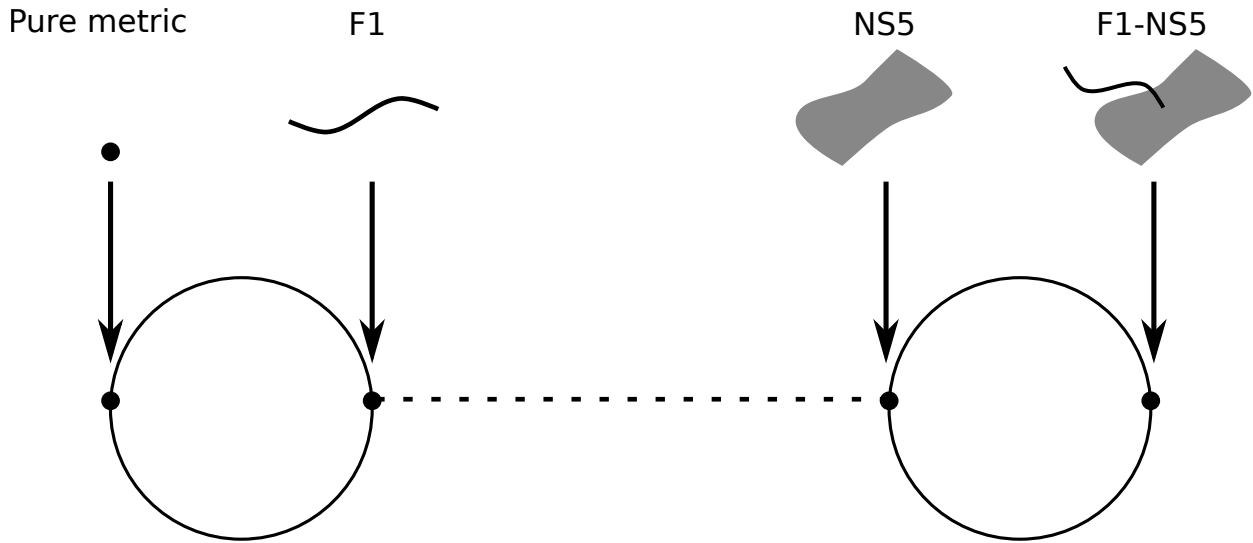}
    \caption{Pictorial representation of the duality chain (\ref{DualChainShort}).
    Applying 
    $O(d,d)$ transformations (the left solid circle) to a pure metric, one produces solutions 
    of the `F1 type', then the `bridge'  (dashed line) discussed in section 
    \ref{SecKTBeyondOdd} connects the F1 geometry with a pure NS5. Additional 
    $O(d,d)$ transformations, represented by the solid circle on the right, produce the general F1--NS5 solution.}
    \label{fig:F1NS5}
\end{figure}

In this article we are studying the symmetries of the NS sector of string theory\footnote{Solutions for the Ramond--Ramond fluxes are also interesting, but our construction of the modified Killing--Yano tensors discussed in section \ref{SectionModifiedKYT} needs further generalization to include such geometries.}, and so far we have only discussed the geometries related to pure metric by $O(d,d)$ transformations. Inclusion of S duality allows one to produce more general NS--NS backgrounds, and in this subsection our construction is extended to such geometries. 

In the context of black hole physics $O(d,d)$ transformation are often used to generate solutions with electric $B$ field\footnote{The most notable exceptions from this rule are gravity duals of non--commutative field theories \cite{nonCom}, beta--deformation of pure geometry \cite{Beta}, and generation of NS5--brane from KK monopole. From our perspective, all these operations give the solution of type `F1'.}, so we will call them `F1 geometries', even if they do not describe fundamental strings. To generate NS5 branes from black holes one has to use a specific combination of T and S dualities, and we will denote the resulting geometry by `NS5', even though it can contain more general fluxes.  This chain of dualities is shown in Figure \ref{fig:F1NS5}.

To generate the `NS5 geometry' we begin with a ten--dimensional metric reduced on $T^p\times T^4$:
\bea\label{Jan13a}
ds_P^2=H_{\alpha\beta}[dy^\alpha+Y^\alpha][dy^\beta+Y^\beta]
+G_{ab}(dz^a+A^a)(dz^b+A^b)+h_{mn}dx^mdx^n
\eea
To generate a magnetic NS flux, we perform the following dualities \cite{Multiwound,Paradox}\footnote{A detailed discussion of this duality map will be presented in the next section, where a more involved chain (\ref{DualChain}) will be used to add charges to various black holes.}: 
\bea\label{DualChainShort}
P\stackrel{T_{y}}{\longrightarrow}F\stackrel{S}{\longrightarrow}
D1\stackrel{T_z}{\longrightarrow}D5\stackrel{S}{\longrightarrow}NS5.
\eea
Notice that various labels just indicate the type of flux (i.e., F1 is an electric $B$--field, D5 is a magnetic $C^{(2)}$ and so on) rather than presence of branes. 

T dualities along $y$ directions produces F1 solution, and subsequent S duality gives
\bea
ds_{D1}^2&=&\sqrt{\mbox{det}\,H}\left[{\tilde H}_{\alpha\beta}dy_\alpha dy_\beta+G_{ab}(dz^a+A^a)(dz^b+A^b)+h_{mn}dx^mdx^n\right],\nonumber\\
e^{2\Phi}&=&{\mbox{det}\,H},\quad C^{(2)}=dy_\alpha\wedge Y^\alpha ,\quad {\tilde H}_{\alpha\beta}=[H^{-1}]_{\alpha\beta}.\nonumber
\eea
The outcome of four T dualities along $z$ directions depends on the presence of $z^a$ in $Y^\alpha$. If $Y^\alpha$ has no legs along $z$ directions, then T dualities produce a six--form, which can be dualized back to $C^{(2)}$. Any leg pointing in $z$ direction leads to $C^{(4)}$, and this RR flux can\rq{}t be removed by S duality. Thus to end up with NS system we require $Y$ to point only in the non--compact directions. Then T dualities along $z$ directions give
\bea
ds_{D5}^2&=&\sqrt{\mbox{det}\,H}\left[{\tilde H}_{\alpha\beta}dy_\alpha dy_\beta+h_{mn}dx^mdx^n\right]+\frac{1}{\sqrt{\mbox{det}\,H}}{\tilde G}_{ab}dz_adz_b,\nonumber\\
e^{2\Phi}&=&\frac{1}{{\mbox{det}}\,H {\mbox{det}}\,G},\quad C^{(2)}=dy_\alpha\wedge Y^\alpha\wedge\prod (dz^a+A^a),\quad
B=dz_a\wedge A^a,\nonumber
\eea
where ${\tilde G}$ is the inverse matrix of $G$. To avoid the RR fields after S duality, we must require $A^a=0$, this leads to the final result:
\bea\label{NS5metr}
ds_{NS5}^2&=&\sqrt{\mbox{det}\,G}\left[\mbox{det}\,H{\tilde H}_{\alpha\beta}dy_\alpha dy_\beta+{\mbox{det}\,H}h_{mn}dx^mdx^n+{\tilde G}_{ab}dz_adz_b\right],\nonumber\\
e^{2\Phi}&=&{{\mbox{det}}\,H {\mbox{det}}\,G},\quad 
B=dy_\alpha\wedge Y^\alpha\wedge\prod (dz^a+A^a).
\eea

Separation of the Hamilton--Jacobi equation in the geometry (\ref{Jan13a}) implies (among other things) the separation of
\bea
f h^{mn}\d_m S\d_n S,\quad f H^{\alpha\beta},
\eea
and for the geometry (\ref{NS5metr}) we need 
\bea
\frac{\tilde f}{{\mbox{det}\,H}\sqrt{{\mbox{det}~}G}}h^{mn}\d_m S\d_n S,
\quad
\frac{\tilde f}{{{\mbox{det}\,H}\sqrt{{\mbox{det}~}G}}}{\tilde H}^{\alpha\beta},\qquad 
\frac{\tilde f}{\sqrt{{\mbox{det}~}G}}{\tilde G}^{ab}
\eea
to separate for some function ${\tilde f}$. Setting
\bea\label{TildeNSf}
{\tilde f}={f}{{\mbox{det}\,H}}{\sqrt{{\mbox{det}~}G}}
\eea
we must require 
\bea\label{SeparCondNS5}
\d_x\d_y[{f}h^{mn}\d_m S\d_n S]=0,
\quad
\d_x\d_y[f{\tilde H}^{\alpha\beta}]=0,\ \d_x\d_y{\tilde f}=0,\ 
\d_x\d_y[f{{\mbox{det}\,H}}{\tilde G}^{ab}]=0.
\eea
The first condition is automatic, the second one is similar to the requirement for T duality (recall that 
${\tilde H}=H^{-1}$), and the last two relations are new. As before, the old and the new Killing tensors are expressed as (\ref{KillTilde})
\bea\label{KillTildeA}
K^{MN}=X^{MN}-f_xg^{MN}, \quad {\tilde K}^{MN}={\tilde X}^{MN}-{\tilde f}_x{\tilde g}^{MN},
\eea
although now tildes refer to the NS5 system. Repeating the steps which led to (\ref{XforF1}), we find
\bea\label{XforNS5}
{\tilde X}^{\alpha\beta}=\left[f H^{\alpha\beta}\right]_x,\
{\tilde X}^{ab}=\left[f{{\mbox{det}\,H}}{\tilde G}^{ab}\right]_x,\
{\tilde X}^{mn}=X^{mn},\quad {\tilde f}_x=[{f}{{\mbox{det}\,H}}{\sqrt{{\mbox{det}~}G}}]_x.
\eea
Equations (\ref{TildeNSf}), (\ref{KillTildeA}), (\ref{XforNS5}) give the Killing tensor ${\tilde K}$ in terms of the the original metric, in particular, we observe that the expression for ${\tilde K}$  in terms of $K$ is rather complicated. This reinforces the principle introduced in section \ref{SecKTandHJ}: to study the Killing tensors and their transformations under dualities, it is convenient to begin with finding the eigenvectors and eigenvalues of the tensors since the map (\ref{XforNS5}) between $X$ and ${\tilde X}$ is relatively simple. 
Several explicit examples of Killing tensors for F1--NS5 systems are presented in Appendix \ref{App5D6DandKT}.

\subsubsection{Conditions on the $B$ field from separation of variables}
\label{SecCondsBfield}

Equation \eqref{SeparCondNS5} gives the separability condition for the NS5 metric, and now we present the constraints on the $B$ field. In Section \ref{SecKTOdd} such restrictions were found by requiring separability of 
the metrics on any $O(d,d)$ orbit which starts from a pure metric, and now we impose the same requirement on the $O(d,d)$ orbit staring from an NS5 solution\footnote{The first orbit generates fundamental strings and momentum, and the second one generates F1--NS5--P system}. We will find that separability of F1--NS5--P geometries is guaranteed by (\ref{SeparCondNS5}) and constraints (\ref{Hcond2f}), (\ref{Hcond2fields}), (\ref{Hcond2sep}) on the Kalb--Ramond field of the original F1 system.

We start with constraints \eqref{Hcond1} and \eqref{Hcond} derived for the F1 orbit
\bea\label{Hconds}
&&\d_y(f g^{mM})H_{xMb}+\d_x (f g^{mM}) H_{y Mb}+f g^{mM}\d_xH_{y Mb}=0,\nn
&&\d_x\d_y (fg_{ab})-fg^{MN}H_{y aM}H_{xNb}-fg^{MN}H_{xaM}H_{y Nb}=0,
\eea
and require them to hold for NS5 solutions as well. Then using the relation between metrics for F1 and NS5 (\ref{NS5metr}),
\bea
g_{MN}^{NS5}=F g_{MN}^{F1}, \quad f^{NS5}=F f^{F1},\quad 
F\equiv \sqrt{\mbox{det}G}\,\mbox{det}H=e^{-2\Phi_{F1}}
\eea
and electric--magnetic duality transformation, we can rewrite (\ref{Hconds}) in terms of the metric and the $B$ field for F1. The detailed calculations presented in the Appendix \ref{AppCondBNS5} give
\bea\label{Hcond2fa}
&&\d_x\d_y[g_{ab}fF]+\frac{f}{F}g_{ab}\Big[\d_x\ln F \d_y\ln F+ \frac{1}{2}H_{xMN}H_y{}^{MN}\Big]=0
\eea
and
\bea\label{Hcond2f}
&&\d_y(f g^{mM})H_{xMb}+\d_x (f g^{mM}) H_{y Mb}+f g^{mM}\d_xH_{y Mb}=0,\\
&&\d_y(f g^{mM})\tilde{H}_{xMb}+\d_x (f g^{mM}) \tilde{H}_{y Mb}+\frac{1}{F}f g^{mM}\d_x(F\tilde{H}_{y Mb})=0,\nonumber
\eea
where $\tilde{H}=\star_6H^{(F1)}$ is the Hodge dual dual of $H^{(F1)}$ with respect to the metric $h_{mn}$.

Interestingly, in all examples we have considered, two terms in equation (\ref{Hcond2fa}) vanish separately, and perhaps such `coincidence' is guaranteed by equations of motion of supergravity for the NS5 brane, but we have not investigated this further. Vanishing of the first term in equation (\ref{Hcond2fa}) implies separation of a very interesting duality--invariant quantity
\bea
g^{(F1)}_{ab}f^{(F1)}e^{-2\Phi_{F1}}=g^{(NS5)}_{ab}f^{(NS5)}e^{-2\Phi_{NS5}}.
\eea
Then vanishing of the second term in \eqref{Hcond2fa} implies a relation in the F1 frame:
\bea\label{Hcond2fields}
\d_x\Phi \d_y\Phi+\frac{1}{8} H_{xMN}H_y{}^{MN}=0.
\eea

To summarize, the separability of the F1--NS5--P geometries obtained form the F1 system is guaranteed by equation (\ref{SeparCondNS5}), conditions (\ref{Hcond2f}), (\ref{Hcond2fields}) on the $B$ field of the original F1 system, and 
\bea\label{Hcond2sep}
\d_x\d_y[g_{ab}fe^{-2\Phi}]=0.
\eea

\subsection{T duality and the modified Killing--Yano equation}
\label{SectionModifiedKYT}

In this subsection we investigate the behavior of (conformal) Killing--Yano tensors under T dualities. We will show that generically T duality destroys Killing--Yano tensors, but there is a unique modification of the KYT equation which is invariant under T duality. For the geometries without Kalb--Ramond field, this modified Killing--Yano (mKY) equation reduces to the standard one (\ref{DefKYT}), but in general it also  contains contributions from the $B$ field. To motivate the mKYT equation, we apply T duality to  a pure metric. This leads to the unique modification of KYT equation in the dual frame, and we will demonstrate that such modification remains invariant under any combination of diffeomorphisms and T dualities.
 
Let us start with a standard equation for the Killing--Yano tensor (\ref{KYTdef})  
\bea\label{KYTdefAA}
\nabla_M Y_{NP}+\nabla_N Y_{MP}=0
\eea
and perform a dimensional reduction of the metric along $z$ direction:
\be\label{DimRedSetup2}
ds^2=e^C[dz+A_idx^i]^2+\hat g_{ij}dx^idx^j.
\ee
In the first step of our analysis we also assume that geometry (\ref{DimRedSetup2}) has a trivial Kalb--Ramond field.
The details of the reduction are given in Appendix \ref{AppDimRed1}, in particular, the $(mnp)$ component of the KY equation can be read off from (\ref{DimRedL}) by setting $L=Y$:
\bea\label{mnpComp}
\nabla^mY^{np}+\h F^{mp}Y^n{}_z+(m\leftrightarrow n)=0,
\eea
where $F=dA$ is the field strength associated with graviphoton. 
We will now look for the modification of the KYT equation in the dual frame that satisfies five requirements:
\begin{enumerate}[(1)]
\item The equation should be linear in the dual Killing-Yano tensor ${\tilde Y}$.
\item Its $(mnp)$ component must reproduce (\ref{mnpComp}) and other components must be consistent with dimensional reduction of (\ref{KYTdefAA}).
\item The equation must be invariant under gauge transformations of the $B$ field.
\item The new terms to be at most linear in  $B$ field since equations (\ref{mnpComp}) are linear in $F_{ab}$.  This implies that the modified KY equation should be linear in $H_{MNP}$.
\item The square of the modified KYT should give a Killing tensor in the dual frame. 
\end{enumerate}
As demonstrated in the in Appendix \ref{AppBeqKYT}, there exists a unique modification of equation (\ref{KYTdefAA}) which satisfies all these requirements, and it reads\footnote{The only alternative corresponds to changing the sign of $H$ in (\ref{KYTBfield}) and sign of $\tilde Y^n{}_z$ in (\ref{KYTtransYcomp}).}
\be\label{KYTBfield}
\boxed{
{\tilde\nabla}_M {\tilde Y}_{NP}+
{\tilde\nabla}_N {\tilde Y}_{MP}+
\frac{1}{2}\tilde H_{MPA}{\tilde g}^{AB}{\tilde Y}_{NB}+
\frac{1}{2}\tilde H_{NPA}{\tilde g}^{AB}
{\tilde Y}_{MB}=0.}
\ee
Moreover, the Kalb--Ramond field in the dual frame satisfies a constraint
\bea\label{RestBKYTGuess}
{{\tilde H}_{Q[MN}{\tilde Y}_{P]}^{\ \ Q}+\d_{[P} C\,{\tilde Y}_{MN]}=
-\d_{[P} {\tilde W}_{MN]}}
\eea
with some antisymmetric tensor ${\tilde W}_{MN}$.
Under the T duality the components of the mKYT transform as
\be\label{KYTtransYcomp}
\tilde Y^{mn}=Y^{mn}, \quad \tilde Y^n{}_z=e^{-C}Y^n{}_z.
\ee 
The counterpart of the constraint (\ref{RestBKYTGuess}) in the original metric (\ref{DimRedSetup2}) is 
\be\label{PureMetrConst}
dC\wedge dY=0.
\ee
Notice that (\ref{KYTBfield}) can be interpreted as a standard KYT equation with connection modified by torsion \cite{Strominger}
\bea
\Gamma_{MN}^P\rightarrow \Gamma_{MN}^P-\frac{1}{2}H^P{}_{MN}.
\eea
In Appendix \ref{AppComplexStructure} we discuss transformation of K{\"a}hler structure under T duality and demonstrate that a counterpart of the transformation 
(\ref{KYTtransYcomp}) maps the K{\"a}hler form into complex structure satisfying the Strominger's system for manifolds with torsion \cite{Strominger}.

Although equation (\ref{KYTBfield}) was derived by applying T duality to a pure metric, the result is invariant under any combination of T dualities and diffeomorphisms. In Appendix 
\ref{AppBeqKYT} we demonstrate that T duality maps any solution $Y_{MN}$ of (\ref{KYTBfield}) in an arbitrary geometry (\ref{DimRedSetup2})  supported by the $B$ field into a solution 
${\tilde Y}_{MN}$ of the same equation in the dual frame. The transformation (\ref{KYTtransYcomp}) between tensors can be viewed as an extension of Buscher's rules to Killing--Yano tensors.
The constraint (\ref{RestBKYTGuess}) is generalized as
\bea\label{ConstrBcompBdy}
&&
g^{ms}\d_sC {\hat Y}^n{}_{z}-g^{ns}\d_sC{\hat Y}^m{}_z+
g^{ns}G_{sr}Y^{rm}-g^{ms}G_{sr}Y^{rn}=0,\\
&&\quad G_{mn}\equiv e^{C/2}F_{mn}-e^{-C/2}{\tilde F}_{mn},\quad 
{\hat Y}_z{}^s\equiv e^{-C/2}Y_z{}^s\,,
\nonumber
\eea
where $F_{mn}$ and ${\tilde F}_{mn}$ are graviphotons in the original and dual frames. 
Notice that ${\hat Y}_z{}^s$ remains invariant under T duality, and $G_{mn}$ 
changes sign.

To summarize, we have demonstrated that the requirement of covariance under T duality leads to the unique equation (\ref{KYTBfield}) for the KYT, and the original equation (\ref{KYTdefAA}) is transformed into the system (\ref{KYTBfield})--(\ref{RestBKYTGuess}). In other words, unlike the KV and KT equations which are unaffected by the Kalb--Ramond field, the equation for the Killing--Yano tensor is modified, which is not very surprising since fermions interact with the $B$ field. In all three cases (KV, KT, mKYT) the Kalb--Ramond field satisfies additional constraints in the dual frame (see (\ref{BConstrCovar}), 
(\ref{ConstrBKT}), (\ref{RestBKYTGuess})). 

Although Ramond--Ramond fluxes appeared in the intermediate stages of the duality chain (\ref{DualChainShort}), neither the initial nor the final point contained such fields. Unfortunately an extension of our analysis to Ramond--Ramond backgrounds leads to certain complications, which we now discuss. Starting with a pure metric and performing a T duality, we find the new mKYT from (\ref{KYTtransYcomp}):
\be
\tilde Y^{mn}=Y^{mn}, \quad \tilde Y^n{}_z=e^{-C}Y^n{}_z.
\ee 
Since the mKYT equation is written in the string frame, S duality induces a conformal rescaling of such metric, so generically the modified Killing--Yano tensor is destroyed by such operation. To save it we have two option for the equation after the duality:
\begin{enumerate}[(a)]
\item Postulate that in the presence of the Ramond--Ramond fluxes, the covariant derivatives appearing in the mKYT should be computed using 
$g'_{MN}=e^{-\Phi}g_{MN}$ rather that $g_{MN}$, and $H_3$ should be replaced by $F_3$. While consistent with S duality, this prescription  does not reduce to the standard KYT in the NS--NS backgrounds with non--trivial dilaton, so it should be abandoned. 
\item Postulate that the modified KYT equation survives S duality only if the constraint 
\bea\label{ConstrDilat}
g^{AB}\d_B\Phi Y_{AM}=0 
\eea
is satisfied. Then the discussion presented in the Appendix \ref{AppCKYT} implies that the mKYT transforms according to (\ref{eqnA8})
\bea
Y'_{NP}=e^{-3\Phi/2}{\tilde Y}_{NP},
\eea
where ${\tilde Y}_{NP}$ satisfies equation (\ref{KYTBfield}) before S duality, and $\Phi$ is the dilaton for the NS system. 
\end{enumerate}
Although option (b) is not ruled out, the constraint (\ref{ConstrDilat}) is rather restrictive. Moreover, even assuming that this constraint is satisfied, and equation (\ref{KYTBfield}) does hold for the type IIB theory with replacement $H_3\rightarrow F_3$, an additional T duality to type IIA supergravity leads to rather unusual structures. By applying the dimensional reduction and T duality to Ramond--Ramond background, we found that the KY equation in the dual frame mixes tensors of different ranks. For example, starting with mKYT $Y_{MN}$ one produces an equation that mixes $Y^{(1)}_M$ and $Y^{(3)}_{MNP}$. This is not surprising since something similar happens for components of $F_3$, but KYT become rather complicated. While it would be very interesting to study the properties of such objects with mixed ranks and perhaps embed them in the democratic formalism \cite{Democratic}, this direction will not be pursued here.

\bigskip

Finally we comment on behavior of conformal Killing(--Yano) tensors. As demonstrated in section \ref{SubsectionCKV}, T duality introduces $z$--dependence in conformal Killing vectors, so such dependence should be allowed in  CKT as well.  Dimensional reduction for a relatively simple case $A_m=0$ is performed in Appendix \ref{AppCKT}, where we demonstrate that generically CKTs are destroyed by T duality. 
However, the CKT does survive the duality if two additional conditions (\ref{CKTConstr1}) and (\ref{CKTConstr2}) are satisfied. The same conclusion holds for a conformal mKYT: it survives T duality only in very special cases. 


\section{Examples of the modified KYT for F1--NS5 system}
\label{SecExamplesF1NS5}
\renewcommand{\theequation}{5.\arabic{equation}}
\setcounter{equation}{0}

In this section we present several examples of the modified Killing--Yano tensors introduced in section \ref{SectionModifiedKYT}. As we saw in section \ref{SecMyersPerry}, the ordinary Killing--Yano tensors exist for a large class of black holes described by the Myers--Perry solutions, and these geometries automatically solve our modified equation since they do not have a Kalb--Ramond field. However, string theory provides a very nice generating technique that allows one to start with a known solution of general relativity and construct black holes with various charges by applying string dualities \cite{Sen,Cvetic,Cvetic5D}. In this article we are focusing only on the NS--NS sector of string theory, so we will use the special cases of the general techniques introduced in \cite{Sen,Cvetic,Cvetic5D} to produce black holes with fundamental string and NS5--brane charges\footnote{The geometries containing D--branes are also interesting, but the full theory of modified Yano--Killing tensors for such solutions has not been developed yet. In particular, as we mentioned in section \ref{SectionModifiedKYT}, some D--brane backgrounds would contain Yano--Killing tensors of mixed ranks, and we hope to return to a detailed study of such objects in the future.}. For such special cases, it is convenient to specify the duality transformations more explicitly. 

We will start with a rotating black hole in $d<10$ dimensions and boost it in one of the $10-d$ direction. Then application of T duality along that direction produces a non--extremal fundamental string. To arrive 
at an NS5--brane (and more generally at a combination of strings and NS5--branes), one has to apply a more sophisticated procedure introduced in \cite{Multiwound,Paradox}:
\begin{enumerate}[1.]
\item Start with a rotating Myers--Perry black hole with mass $m$ in $d<6$ dimensions, perform a trivial embedding into the ten--dimensional type IIA supergravity, and identify a five--dimensional torus $T^4\times S^1$ orthogonal to the black hole. 
\item Perform a boost by $\alpha$ along $S^1$ direction\footnote{Following \cite{Multiwound,Paradox}, we will call the corresponding coordinate $y$ and parameterize the boost by $\alpha$, where 
$\tanh\alpha\equiv v/c$.} and T--dualize along $S^1$. This produces a black fundamental string wrapping one of the compact directions. 
\item Perform an S duality followed by four T dualities along $T^4$ and another S duality. The resulting metric describes a non--extremal rotating NS5 brane. 
\item Perform another boost by $\beta$ in the $S^1$ direction followed by T duality. This gives a non--extremal F1--NS5 system with mass $m$ and charges
\bea
Q_1=m\sinh^2\beta,\qquad Q_5=m\sinh^2\alpha.
\eea
\end{enumerate}
For future reference we summarize the duality chain using a simple diagram:
\bea\label{DualChain}
\mbox{BH}\ \rightarrow\
\mbox{P}_\alpha\ \rightarrow\
\mbox{F1}_\alpha\ \rightarrow\ \mbox{D1}_\alpha\ \rightarrow\ \mbox{D5}_\alpha
\ \rightarrow\ \mbox{NS5}_\alpha\ \rightarrow\ 
\left(\begin{array}{c}
\mbox{NS5}_\alpha\\ P_\beta
\end{array}\right)\ \rightarrow 
\left(\begin{array}{c}
\mbox{NS5}_\alpha\\ F1_\beta
\end{array}\right).
\eea
In this section we use $y$ to denote the $S^1$ direction. Notice that if we are adding only the F1 charge, the duality chain stops after the first two steps, and four--dimensional torus is not needed. Thus such charge can be added to the Myers--Perry black hole in $d<10$ dimensions\footnote{This construction also works for the embedding of the $d$--dimensional Myers--Perry black hole to the bosonic string as long as $d<26$.}, and we derive the explicit expression for the corresponding mKYT in section \ref{SecMPF1}. On the other hand, addition of the NS5 charge needs $T^4\times S^1$, so it only works for black holes with $d<6$. Since we are interested in asymptotically--flat geometries, the BTZ black hole \cite{BTZ} will not appear in the discussion, so $d$ can take only two values ($d=4,5$). These cases are discussed in sections \ref{SecEx5d} and \ref{SecEx6d}. Our results are summarized in table \ref{Table1}.

\begin{table}
\centering
    \begin{tabular}{ | l | c | c | c | c |}
    \hline
    {}		&  \multicolumn{2}{|c|}{4D} & \multicolumn{2}{|c|}{5D} \\ \hline
    {}		& extremal 	 & non--extremal& extremal & non-extremal  \\ \hline
    F1      & 	M		 & 			M	& M 	   & M 		       \\ \hline
    NS5     & 	M		 & 			--	& M 	   & M 		       \\ \hline
 F1--NS5($Q_1=Q_5$)	& 	-- & 		--	& C,M	   & M	           \\ \hline
 F1--NS5($Q_1\ne Q_5$)& 	--		 & 	--	& M 	   & M 		       \\ 
    \hline
    \end{tabular}
    \caption{Summary of the results for the F1--NS5 system constructed from four-- and five--dimensional black holes using the procedure (\ref{DualChain}). Here M denotes the modified KYT and C correspond to the conformal KYT.}
    \label{Table1}
\end{table}


\subsection{Charged Myers--Perry black hole}

\label{SecMPF1}

In our first example we add charges to the Myers--Perry black hole discussed in section 
\ref{SecMyersPerry} by applying the duality chain (\ref{DualChain}) and discuss the modified Killing--Yano tensor for the resulting solution. The transition from F1 to NS5 in 
(\ref{DualChain}) involves the electric--magnetic duality, which depends on the dimension of the black hole, so it is convenient to study individual black holes separately, and we will do that in sections \ref{SecEx5d}, \ref{SecEx6d}. In this section we will focus the first two 
{\it algebraic} steps in the duality chain (\ref{DualChain}) to generate a rotating black hole with F1 charge. 

As demonstrated in Appendix \ref{AppChargedMP}, the charged Myers--Perry black hole admits a family of modified Killing--Yano tensors, which generalizes 
(\ref{AllFramesMP})--(\ref{HInFrame}): the tensors are still given by 
(\ref{Kub2}), (\ref{HInFrame})\footnote{In this subsection we have to distinguish between $e^a=e^a_M dx^M$ and 
$e_a=e_a^M\d_M$, so the frame indices are written in the appropriate places. In the rest of the paper we abuse notation and write $e_a=e^a_M dx^M$ to simplify formulas.}
\bea\label{ChMPHYT}
Y^{(2n-2p)}=\star\left[\wedge h^p\right],\quad
h=re^r\wedge e^t+\sum_k\sqrt{-x_k}e^{x_k}\wedge e^k\,,
\eea
but the frames are modified 
\bea\label{ChMPframe}
e^r&=&\sqrt{\frac{FR}{R-mr}}dr,\quad e^{x_i}=\sqrt{-\frac{(r^2-x_i)d_i}{4x_iH_i}}dx_i,\nn
e^t&=&\frac{1}{h_1}
\sqrt{\frac{R-mr}{FR}}\left[\ch_\alpha dt+\sh_\alpha dy+\sum_k\frac{G_k}{a_kc_k^2}d\phi_k\right],\nn
e^y&=&\frac{1}{h_1}\left[\sh_\alpha dt+\ch_\alpha dy-
\frac{mr\sh_{\alpha}}{FR}dt
-\frac{mr\sh_{\alpha}\ch_\alpha}{FR}\sum_k\frac{d\phi_k}{a_kc_k^2}\right],\\
e^i&=&\frac{1}{h_1}\sqrt{\frac{H_i}{d_i(r^2-x_i)}}\left[\ch_\alpha dt+\sh_\alpha dy+
\sum_k\frac{G_k(r^2+a_k^2)}{c_k^2a_k(x_i+a_k^2)}\left\{1+
\frac{mr\sh^2_\alpha (r^2-x_i)}{FR(r^2+a_k^2)}
\right\}d\phi_i\right].\nonumber
\eea
The expressions for $c_i,d_i,H_i,G_i,(FR)$ are still given by (\ref{AllFramesMP}), and 
\bea\label{h1even}
h_1=1+\frac{m\sh_\alpha^2}{FR}.
\eea
Expressions for the inverse frames exhibit a clear separation between non--cyclic coordinates $(r,x_i)$:
\bea\label{ChMPframeDwn}
e_r&=&\sqrt{\frac{R-mr}{FR}}\d_r,\quad 
e_{x_i}=\sqrt{-\frac{4x_iH_i}{d_i(r^2-x_i)}}\d_{x_i},\quad
e_y=\ch_\alpha\d_y-\sh_\alpha\d_t,
\nn
e_t&=&-\sqrt{\frac{R^2}{FR(R-mr)}}\left[\ch_\alpha \d_t-\frac{\sh_\alpha}{R}(R-mr)\d_y
-\sum_k\frac{a_k}{r^2+a_k^2}\d_{\phi_k}\right],\\
e_i&=&-\sqrt{\frac{H_i}{d_i(r^2-x_i)}}\left[\ch_\alpha\d_t-\sh_\alpha \d_y-\sum_k\frac{a_k}{x_i+a_k^2}\d_{\phi_k}
\right].\nonumber
\eea
For the odd dimensions we find 
\bea\label{ChMPframeOdd}
e^r&=&\sqrt{\frac{FR}{R-mr^2}}dr,\quad e^{x_i}=\sqrt{\frac{d_i(r^2-x_i)}{4H_i}}dx_i,\nn
e^t&=&\frac{1}{h_1}
\sqrt{\frac{R-mr^2}{FR}}\left[\ch_\alpha dt+\sh_\alpha dy+\sum_k\frac{a_kG_k}{c_k^2}d\phi_k\right],\nn
e^y&=&\frac{1}{h_1}\left[\sh_\alpha dt+\ch_\alpha dy-
\frac{m\sh_{\alpha}}{FR}dt
-\frac{m\sh_{\alpha}\ch_\alpha}{FR}\sum_k\frac{a_kd\phi_k}{c_k^2}\right],\\
e^i&=&\frac{1}{h_1}\sqrt{-\frac{H_i}{x_id_i(r^2-x_i)}}\left[\ch_\alpha dt+\sh_\alpha dy+
\sum_k\frac{G_ka_k(r^2+a_k^2)}{c_k^2(x_i+a_k^2)}\left\{1+
\frac{m\sh^2_\alpha (r^2-x_i)}{FR(r^2+a_k^2)}
\right\}d\phi_k\right],\nonumber\\
e^\psi&=&\frac{1}{h_1}\sqrt{\frac{\prod a_i^2}{r^2\prod x_k}}\left[\ch_\alpha dt +
\sh_\alpha dy+\sum \frac{G_k(r^2+a_k^2)}{c^2_k a_k}\left\{1-
\frac{a^2_k m\sh_\alpha^2}{FR(r^2+a_k^2)} \right\}d\phi_k \right].\nonumber
\eea
and
\bea\label{ChMPframeDwnOdd}
e_r&=&\sqrt{\frac{R-mr^2}{FR}}\d_r,\quad 
e_{x_i}=\sqrt{\frac{4H_i}{d_i(r^2-x_i)}}\d_{x_i},\quad
e_y=\ch_\alpha\d_y-\sh_\alpha\d_t,
\nn
e_t&=&-\sqrt{\frac{R^2}{FR(R-mr^2)}}\left[\ch_\alpha \d_t-\frac{\sh_\alpha}{R}(R-mr^2)\d_y
-\sum_k\frac{a_k}{r^2+a_k^2}\d_{\phi_k}\right],\\
e_i&=&-\sqrt{-\frac{H_i}{x_id_i(r^2-x_i)}}\left[\ch_\alpha\d_t-\sh_\alpha \d_y-\sum_k\frac{a_k}{x_i+a_k^2}\d_{\phi_k}
\right],\nonumber\\
e_\psi&=&-\sqrt{\frac{\prod a_i^2}{r^2\prod x_k}}\left[\ch_\alpha\d_t-\sh_\alpha \d_y-\sum_k\frac{1}{a_k}\d_{\phi_k}
\right].\nonumber
\eea
The expressions for $c_i,d_i,H_i,G_i,(FR)$ are still given by (\ref{FROdd}), (\ref{OtherOdd}), and $h_1$ is given by 
(\ref{h1even}).

\subsection{F1--NS5 system from the Kerr black hole.}
\label{SubsectionExamples5D}
\label{SecEx5d}

Application of the duality chain (\ref{DualChain}) to the Kerr black hole (\ref{Kerr4D}) gives a rotating F1--NS5 system, and the complete solution is presented in Appendix \ref{App4DKerr} (see equation (\ref{NonExtr5D})). Explicit calculations show that the modified Killing--Yano equation (\ref{KYTBfield}) does not have nontrivial solutions\footnote{As shown in table \ref{Table1}, extremal F1--NS5 and non--extremal NS5 also don't admit  mKYT.}, so in this subsection we will focus on two special cases when the mKYT exists: the non--extremal fundamental string and the extremal NS5 brane. In the first case the existence of solution is guaranteed by the general construction presented in section \ref{SectionModifiedKYT} as long as condition (\ref{PureMetrConst}) is satisfied, and in the second case the mKYT comes from solving the Killing equations.

Application of the first two steps in the duality sequence (\ref{DualChain}) to Kerr geometry (\ref{Kerr4D}) leads to the system which we called F1$_\alpha$, and the corresponding geometry describes a non--extremal fundamental string with charge $Q_1=2m\sh_\alpha^2$:
\bea\label{KerrString}
ds^2&=&\frac{dy^2}{h_\alpha}+\frac{\rho^2}{\Delta}dr^2+\rho^2 d\theta^2-\left[\frac{\Delta}{\rho^2}+
\frac{4(mr \sh_\alpha)^2}{\rho^4h}\right](\ch_\alpha dt-as_\theta^2 d\phi)^2\nonumber\\&&
\qquad+\frac{s_\theta^2}{\rho^2}\left[(r^2+a^2)d\phi-
a\ch_\alpha dt\right]^2+(\sh_\alpha dt)^2\\
B_2&=&\frac{2mr \sh_\alpha}{\rho^2h}\left[\ch_\alpha dt-as_\theta^2 d\phi\right]\wedge dy,\qquad e^{2\Phi}=\frac{1}{h_\alpha}.
\nonumber
\eea
Here we defined
\bea
&&\rho^2=r^2+a^2c^2_\theta, \quad \Delta=r^2+a^2-2mr, \quad 
h_\alpha=1+\frac{2mr\sh_\alpha^2}{\rho^2}.
\nonumber
\eea
Transformation (\ref{KYTtransYcomp}) leads to the modified Killing--Yano tensor for (\ref{KerrString})
\bea\label{mKYTF15D}
Y&=&\frac{1}{h_\alpha}\left\{rs_\theta d\theta\wedge\left[(r^2+a^2)d\phi-a\ch_\alpha dt\right]+
ac_\theta dr\wedge \left[\ch_\alpha dt-as_\theta^2 d\phi\right]\right\}
\nonumber\\
&&+\frac{\sh_\alpha}{h_\alpha}(ac_\theta dr-ars_\theta d\theta)\wedge dy+
\frac{h_\alpha-1}{h_\alpha}r^3s_\theta d\theta\wedge d\phi.
\eea
To compare it with \eqref{Kerr4DKT}, we construct the Killing tensor $K_{MN}=-Y_{MA}{Y^A}_{N}$, define the frames as eigenvectors of this tensor, and rewrite the answer in terms of them: 
\bea\label{aa12}
ds^2&=&-e_t^2+e_y^2+e_r^2+e_\theta^2+e_\phi^2,\nn 
Y&=&r e_\theta\wedge e_\phi+a c_\theta e_r\wedge e_t,\quad
K=r^2[e_\theta^2+e_\phi^2]+(ac_\theta)^2[e_t^2-e_r^2]\nn
e_t&=&\frac{\sqrt{\Delta}}{\rho h_\alpha}\left(\sh_\alpha dy+\ch_\alpha dt-as^2_\theta d\phi\right),\quad e_r=\frac{\rho}{\sqrt{\Delta}}dr,\quad 
e_\theta=\rho d\theta,\\
e_y&=&\frac{1}{h_\alpha}\left[\ch_\alpha dy+\sh_\alpha
\left(1-\frac{2mr}{\rho^2}\right)dt+
\frac{amr s^2_\theta \sh_{2\alpha}}{\rho^2}d\phi \right],\nn
e_\phi&=&\frac{s_\theta}{\rho h_\alpha}\left(-a\sh_\alpha dy-a\ch_\alpha dt+(r^2+a^2+2mr\sh^2_\alpha)d\phi \right). \nonumber
\eea
Notice that eigenvalues of the Killing tensor and mKYT do not depend on the boost parameter $\alpha$.

The duality sequence (\ref{DualChain}) involves D-branes supported by Ramond--Ramond flux, and the analysis presented in section \ref{SectionModifiedKYT} does not apply to T duality performed in such systems. It would be interesting to generalize our discussion of mKYT to the geometries with Ramond--Ramond fields, but such analysis goes beyond the scope of this article. Instead we applied the duality chain (\ref{DualChain}) to the Kerr black hole and solved the mKYT equations for the resulting F1--NS5 geometry. We found that the mKYT does not exist in the system involving NS5 branes unless one takes an extremal limit and sets the F1 charge to zero:
\bea
m\rightarrow 0,\ Q_1\rightarrow 0,\quad \mbox{fixed}\quad Q_5=2m\sinh^2\alpha.
\eea
The resulting geometry,
\bea\label{KerrNS54d}
ds^2&=&-dt^2+h\left[\frac{\rho^2}{r^2+a^2}dr^2+\rho^2 d\theta^2+(r^2+a^2) s_\theta^2d\phi^2+dy^2\right],\nonumber\\
B_2&=&\frac{Q_5(r^2+a^2) c_\theta}{\rho^2} d\phi \wedge dy,\quad e^{2\Phi}=h, \quad h=1+\frac{Q_5 r}{r^2+a^2},
\eea
admits the unique mKYT
\bea\label{mKYTNS55D}
Y&=&hdy\wedge(rs_\theta d\theta-c_\theta dr)+hd\phi\wedge [rs_\theta^2 dr+(r^2+a^2)s_\theta c_\theta d\theta]\nonumber\\
&=&hd\left[rc_\theta dy-\frac{1}{2}(r^2+a^2) s_\theta^2 d\phi\right]
\eea
which was found by the direct calculation. Introducing convenient frames, we can rewrite this KYT and its square as
\bea\label{KYT4DNS5Extr}
Y&=&e_r\wedge e_y-e_\theta \wedge e_\phi, \qquad K\equiv -Y_{MA}Y^A{}_N dx^Mdx^N=e_r^2+e_y^2+e^2_\theta +e^2_\phi,\nn
e_t&=&dt,\quad 
e_r=\sqrt{\frac{\rho^2+Qr}{r^2+a^2}}dr,\quad e_\theta=\sqrt{\rho^2+Qr},\\
e_{y}&=&\frac{1}{\rho^2}\sqrt{(r^2+a^2)(\rho^2+Qr)}\left[\cos\theta dy+r\sin^2\theta d\phi \right],\nn
e_{\phi}&=&
\frac{\sin\theta}{\rho^2}\sqrt{(\rho^2+Qr)}\left[r dy-(r^2+a^2)\cos\theta d\phi \right].\nonumber
\eea
Notice that square of the KYT gives the spacial part of the metric, which can be viewed as a linear combination of two `trivial' Killing tensors: one coming form the metric and one built from the square of the Killing vector $\d_t$.

An additional T duality along $y$ direction in (\ref{KerrNS54d}) produces a metric of the extremal KK monopole, and application of (\ref{KYTtransYcomp}) to (\ref{mKYTNS55D}) gives the standard KYT for the monopole:
\be
Y=dr\wedge[(Q+r\sin^2\theta)  d\phi +\cos\theta dy]+d\theta\wedge[\cos\theta\sin\theta(a^2+r^2) d\phi-r\sin\theta dy].
\ee
In the frames we find
\bea
Y&=&e_r\wedge e_y-e_\theta \wedge e_\phi,\qquad K=e_r^2+e_y^2+e^2_\theta +e^2_\phi,\nn
e_t&=&dt,\quad e_r=\sqrt{\frac{\rho^2+Qr}{r^2+a^2}}dr,\quad e_\theta=\sqrt{\rho^2+Qr},\nn
e_{y}&=&\sqrt{\frac{r^2+a^2}{\rho^2+Qr}}\left[\cos\theta dy+(Q+r\sin^2\theta)d\phi\right],\\
e_{\phi}&=&\frac{\sin\theta}{\sqrt{\rho^2+Qr}}\left[r dy -\cos\theta(r^2+a^2)d\phi \right].\nonumber
\eea
Once again, the KYT squares to a `trivial'' Killing tensor.


\subsection{F1--NS5 system from the five--dimensional black hole.}
\label{SubsectionExamples6D}
\label{SecEx6d}

Application of the duality chain (\ref{DualChain}) to the five--dimensional black hole gives another example of the rotating F1--NS5 system, the complete geometry was found in 
\cite{Cvetic5D,GuiMathSax}, and it is given by equation \eqref{NonExtr6D}. This subsection  discusses the modified Killing--Yano tensor for this solution.

Recalling that even the neutral five--dimensional black hole had the KYT of rank three rather than two (see section \ref{SecKYTranks}), we should look at the obvious extension of  \eqref{KYTBfield} to such objects\footnote{The discussion presented in Appendix \ref{AppBeqKYT} trivially extends to KY tensors of arbitrary rank.}:
\bea\label{KYTBfieldrank3}
{\nabla}_M { Y}_{NPQ}+
{\nabla}_N {Y}_{MPQ}+
\frac{1}{2} H_{MPA}{ g}^{AB}{Y}_{NB Q}+
\frac{1}{2} H_{MQA}{ g}^{AB}{ Y}_{NPB}\nonumber\\
+\frac{1}{2} H_{NPA}{ g}^{AB}{Y}_{MB Q}+
\frac{1}{2} H_{NQA}{ g}^{AB}{Y}_{MPB}=0.
\eea
The general construction of section \ref{SectionModifiedKYT} guarantees existence of the mKYT for $\alpha=0$ (as long as constraint (\ref{PureMetrConst}) is satisfied), but the generation of the NS5 branes goes through Ramond--Ramond fluxes, which can potentially destroy the modified KYT. Remarkably, the tensor survives, and solution of \eqref{KYTBfieldrank3} for the geometry \eqref{NonExtr6D} is 
\bea\label{mKYT6DNonExtr}
Z^{-1}Y&=&-ad[r^2\cos^2\theta] dt d\psi -a\mu_A\mu_B d[(r^2+a^2-M)\sin^2\theta]d\phi dy
\\ 
&&+
a\mu_A d[(r^2+a^2-M)\sin^2\theta]  dt  d\phi-a\mu_Bd[r^2\cos^2\theta] dy d\psi 
+\sigma d[\sin^2\theta] d\phi  d\psi \nonumber
\eea
with
\bea
Z&=&\frac{r^2+A^2+a^2c_\theta^2}{r^2+B^2+a^2c_\theta^2},\quad
\sigma=\frac{(a^2 - M)A^2 B^2 - [a^2 +  A^2 + B^2 ] M r^2 - M r^4}{\sqrt{A^2+M}\sqrt{B^2+M}},
\nn
\mu_A&=&\frac{A}{\sqrt{M+A^2}},\quad \mu_B=\frac{B}{\sqrt{M+B^2}},\quad
s_\theta=\sin\theta,\quad c_\theta=\cos\theta,\\
A&=&\sqrt{M}\sinh\alpha,\qquad B=\sqrt{M}\sinh\beta.\nonumber
\eea
Although expression (\ref{mKYT6DNonExtr}) is already relatively simple, we also rewrite it in frames to connect to the general analysis presented in section \ref{SecKYTranks}. Constructing the Killing tensor $K_{MN}=-Y_{MA}{Y^A}_{N}$ and defining the frames as its eigenvectors, we find 
\bea\label{5DmKYT}
ds^2&=&-e_t^2+e_y^2+e_r^2+e_\theta^2+e_\phi^2+e_\psi^2,\nn
 Y&=&\left(a\sqrt{2 A^2 + 2Mc^2_{\theta}}\,e_r\wedge e_t+\sqrt{2M(A^2+r^2)-2a^2A^2}e_\theta\wedge e_\phi\right)\wedge e_\psi,\nn
 K&=&-\frac{1}{2}Y_{MA}{Y^A}_{N}=[M(A^2+r^2)-a^2A^2][e^2_\theta+e^2_\phi]+
 a^2 (A^2 + M c_\theta^2)[e^2_t-e^2_r]\nn
 &&+
[a^2(A^2+M c_\theta^2)-(M(A^2+r^2)-a^2A^2)]e_\psi^2,
 \eea
 where the frames are given by 
 \bea
e_t&=&\frac{r}{\rho^2H_1} \sqrt{\frac{M\Delta\rho^2H_5}{M(A^2+r^2)-a^2A^2}}
\left[\ch_\beta dt+\sh_\beta dy-a(\sh_\alpha c_\theta^2d\psi-\ch_\alpha s_\theta^2 d\phi)\right],\nn
e_y&=&\frac{1}{2\rho^2 H_1}\Bigg[
2\sh_\beta\left(\rho^2-M\right)dt+2\rho^2\ch_\beta dy+aM\sh_{2\beta}(
\sh_\alpha c_\theta^2d\psi-\ch_\alpha s_\theta^2 d\phi) \Bigg], \nn
e_r&=&\sqrt{\frac{A^2+\rho^2}{\Delta}}dr,\quad e_\theta=\sqrt{A^2+\rho^2}d\theta,\\
e_\phi&=&\frac{s_\theta c_\theta}{\rho H_1}\sqrt{\frac{MH_5}{A^2+Mc^2_{\theta}}}\Bigg[a\ch_\beta dt+a\sh_\beta dy+
 (B^2+r^2)(\sh_\alpha d\psi + \ch_\alpha d\phi)+a^2\ch_\alpha d\phi \Bigg],\nonumber\\
e_\psi&=&\frac{1}{\rho^2H_1\sqrt{A^2+Mc^2_{\theta}}\sqrt{M(A^2+r^2)-a^2A^2}}
\Bigg[
\nn
&&[r^2+(a^2-M)c_\theta^2]\left(\frac{1}{2}aM\sh_{2\alpha}
(\ch_\beta dt+\sh_\beta dy)+
Ma^2\sh_\alpha s_\theta^2 d\phi
\right)
\nn
&&+\left[M(r^2+A^2)(r^2+B^2) + MA^2 c_\theta^2 r^2-A^2 B^2 a^2 \right]
(\sh_\alpha s_\theta^2 d\phi-\ch_\alpha c_\theta^2 d\psi)\Bigg].\nonumber
\eea
For $\alpha=0$ we find
\bea
e_t&=&\frac{1}{\rho^2H_1} \sqrt{{\Delta\rho^2}}
\left[\ch_\beta dt+\sh_\beta dy+as_\theta^2 d\phi\right],\nn
e_y&=&\frac{1}{2\rho^2 H_1}\Bigg[
2\sh_\beta\left(\rho^2-M\right)dt+2\rho^2\ch_\beta dy-aM\sh_{2\beta} s_\theta^2 d\phi \Bigg], \nn
e_r&=&\sqrt{\frac{\rho^2}{\Delta}}dr,\quad e_\theta=\sqrt{\rho^2}d\theta,\quad
e_\psi=r\cos\theta d\psi,\\
e_\phi&=&\frac{s_\theta }{\rho H_1}\Bigg(a\ch_\beta dt+a\sh_\beta dy+
(a^2+M\sh^2_\beta+r^2)d\phi \Bigg).\nonumber
\eea
This is the special case of (\ref{ChMPframeOdd}) for $n=1$ and one rotation parameter. 
Finally we give the expression for the mKYT (\ref{5DmKYT}) in the extremal limit ($M=0$ with fixed $A,B$):
\bea
Z^{-1}Y
&=&-\frac{1}{2}d\left[(dt+dy)\left\{(r^2c_\theta^2)d\psi-
(r^2+a^2)s_\theta^2 d\phi\right\}+ABa\cos^2\theta d\phi d\psi\right].
\eea


\subsection{Conformal Killing--Yano tensors}

We conclude this section with discussing the CKYT for rotating F1--NS5 systems. 
Explicit calculations show that the geometry obtained by application of (\ref{DualChain}) to the Kerr solution (\ref{Kerr4D}) does not have CKYT. On the other hand 
F1--NS5 system constructed from the five--dimensional black hole (\ref{5DKerrNtrFrm}) does admit a CKYT if and only if $Q_1=Q_5$. In this case the metric has the form
\bea
ds^2&=&-e_t^2+e_\phi^2+e_y^2+e_\psi^2+e_r^2+e_\theta^2,\nonumber\\
e_{t}&=&\frac{\sqrt{\Delta}}{\rho}(dt-as_\theta^2 d\phi),\quad 
e_{\phi}=\frac{s_\theta}{\rho}\left[(r^2+a^2+Q)d\phi-adt\right],\\
e_{r}&=&\frac{\rho}{\sqrt{\Delta}}dr,\quad e_{\bf\theta}=\rho d\theta,\quad 
e_\psi=\frac{c_\theta}{\rho}[(Q+r^2)d\psi-a dy],\quad e_y=\frac{r}{\rho}[dy+ac_\theta^2 d\psi],
\nonumber\\
\Delta&=&r^2+a^2-M,\quad \rho^2=r^2+a^2c_\theta^2+Q,\nonumber
\eea
and the corresponding CKYT and CKT are given by
\bea\label{CKYT_D1D5_NonExtr}
{\cal Y}&=&\rho (e_r\wedge e_t\wedge e_y+e_\theta \wedge e_\phi \wedge e_\psi),\quad Z=\frac{1}{\rho^2}(ac_\theta e_\psi-re_y)\wedge (\sqrt{\Delta}e_t+as_\theta e_\phi),\nn
{\cal K}&=&\rho^2[e_t^2-e_r^2-e_y^2+e_\theta^2+e_\psi^2+e_\phi^2], \quad W=-d[r^2-a^2c_\theta^2].
\eea
Since $W$ is a total derivative, the general prescription (\ref{KTfromCKT}) can be used to construct a standard Killing tensor
\bea
K=-[2(ac_\theta)^2+Q][-e_t^2+e_r^2+e_y^2]+[2r^2+Q][e_\theta^2+e_\psi^2+e_\phi^2].
\eea
Conformal Killing tensors for four-- and five--dimensional black holes discussed in this section were constructed in \cite{CveticLarsen} via separation of variables.

\section{Discussion}\label{SectionDiscussion}
\renewcommand{\theequation}{5.\arabic{equation}}
\setcounter{equation}{0}

In this article we analyzed hidden symmetries of stringy geometries and their behavior under string dualities. In particular, we demonstrated that in the presence of the Kalb--Ramond field the equation for the Killing--Yano tensor is modified as (\ref{KYTBfield}), and this is the unique modification consistent with string dualities. The transformations laws for the Killing vectors, tensors, and Killing--Yano tensors are given by \eqref{KVtransf}, (\ref{KillTilde})--(\ref{XforF1}), (\ref{KYTtransYcomp}). We have also demonstrated that nontrivial Killing tensors in arbitrary number of dimensions are always associated with ellipsoidal coordinates, and we used this observation to construct the (modified) Killing(--Yano) tensors for the Myers--Perry black hole (\eqref{AllFramesMP}, \eqref{Kub2}, \eqref{HInFrame}), its charged version \eqref{ChMPHYT}--\eqref{ChMPframe}, and for several examples of F1--NS5 geometries (\eqref{KYT4DNS5Extr}, \eqref{mKYT6DNonExtr}--\eqref{5DmKYT}). 

This work has several implications. First and foremost, the modified equation for the Killing--Yano tensor  (\ref{KYTBfield}) provides a new powerful tool for studying symmetries of stringy geometries, which can extend the successful applications of the standard Killing--Yano tensors to physics of black holes \cite{CveticLarsen, KYTbh}. Also, the understanding of hidden symmetries developed in this article can be used to extend the `no-go theorems' for integrability \cite{ChL} to backgrounds without supersymmetry. Finally, the explicit Killing--Yano tensors for the Myers--Perry black hole and its charged version constructed in sections \ref{SecMyersPerry} and \ref{SecMPF1} generalize most of the previously known examples and provide the largest known class of KYT.

\section*{Acknowledgments}

We thank Finn Larsen for useful discussions. This work is supported in part by NSF grant PHY-1316184.
\appendix


\section{Conformal transformations of Killing tensors}
\label{AppConfResc}
\renewcommand{\theequation}{A.\arabic{equation}}
\setcounter{equation}{0}

In this appendix we analyze the behavior of Killing vectors and tensors under conformal rescaling of the metric. In the context of string theory such rescalings appear when one goes from the string to the Einstein frame or when one compares the string frames before and after S duality. In this appendix we will find the restrictions on the dilaton which guarantee that Killing vectors and tensors survive after S duality. We study general conformal Killing vectors and tensors, and reduction to the standard objects is obtained by setting the conformal factors to zero.

\subsection{Killing vectors}\label{AppConfRescKV}

We begin with considering an equation for the conformal Killing vector (CKV):
\bea\label{AppConfRescCKV}
\nabla_M V_{N}+\nabla_N V_{M}=2g_{M N}v
\eea
and writing its counterpart in the rescaled metric:
\bea\label{AppConfRescConfResc}
g'_{M N}=e^C g_{M N}:\qquad \nabla'_M V'_{N}+\nabla'_N V'_{M}=2g'_{MN}v'.
\eea
Recalling the transformation of the connections,
\bea\label{AppConfRescConnResc}
(\Gamma^M_{NP})'=\Gamma^M_{NP}+\frac{1}{2}\left[\delta^M_P\d_N C+\delta^M_N\d_P C-
g_{N P}g^{M A}\d_A C\right].
\eea
we can rewrite the equation for $V'$ in terms of the original covariant derivatives:
\bea
\nabla_M(e^{-C} V'_{N})+\nabla_N (e^{-C}V'_{M})=
2g_{MN}(v'+\frac{1}{2}V'^{A}\d_A e^{-C}).
\eea
Comparing this to (\ref{AppConfRescCKV}), we find the transformation law for the CKV:
\bea\label{AppSdualKVa}
V'_M&=&e^C V_M\quad \Rightarrow\quad V'^M=g'^{MN}V'_N=V^M,\nn
v'&=&v-\frac{1}{2}V^A \d_A e^{-C}.
\eea
This implies that CKV always survives the conformal rescaling, but the KV (which must have $v=0$) disappears unless
\bea
V^A \d_A e^{-C}=0.
\eea
In the context of S duality and transition between string and Einstein frames, the last condition implies that Lie derivatives of the dilaton along the Killing vector must vanish, which is a very natural requirement.

\subsection{Killing(--Yano) tensors}\label{AppCKYT}

Next we look at transformation properties of the conformal Killing--Yano tensor, which satisfies equation
\bea\label{AppConfRescCKYT}
\nabla_M Y_{NP}+\nabla_N Y_{MP}=
2g_{MN}W_P-g_{MP}W_N-g_{NP}W_M.
\eea
Using \eqref{AppConfRescConnResc} we can rewrite the left hand side of (\ref{AppConfRescCKYT}) in the rescaled frame as
\bea
&&\nabla'_M Y'_{N P}+\nabla'_N Y'_{M P}=
\nabla_M Y'_{N P}-\frac{1}{2}\left[
\d_M C Y'_{N P}+\d_N CY'_{M P}-g_{M N}g^{AB}\d_B CY'_{AP}
\right]\nonumber\\
&&\qquad-\frac{1}{2}\left[
\d_M C Y'_{NP}+\d_P CY'_{NM}-g_{M P}g^{AB}\d_B CY'_{NA}
\right]+(M \leftrightarrow N)\nonumber\\
&&=\nabla_M Y'_{NP}-\frac{3}{2}\d_MCY'_{NP}+\frac{1}{2}g_{MN}g^{AB}\d_B C Y'_{AP}+\frac{1}{2}g_{MP}g^{AB}\d_B C Y'_{NA}+(M \leftrightarrow N)\nonumber
\eea
and the full equation becomes
\bea
&&e^{3C/2}\nabla_M (e^{-3C/2}Y'_{NP})+
e^{3C/2}\nabla_N (e^{-3/2C}Y'_{MP})\nonumber\\
&&\qquad=2g_{MN}(W'_P e^C-\frac{1}{2}g^{AB}\d_B C Y'_{AP})-
\left[g_{MP}(W'_N e^C-\frac{1}{2}g^{AB}\d_B C Y'_{AN})+(M\leftrightarrow N)\right]\nonumber.
\eea
To recover the original equation \eqref{AppConfRescCKYT}, we must set
\bea\label{eqnA8}
Y'_{NP}=e^{3C/2}Y_{NP}, \quad W'_M=e^{C/2}(W_M+\frac{1}{2}g^{AB}\d_B CY_{AM}).
\eea
The conformal Killing--Yano tensors of higher rank can be analyzed in a similar fashion, and for the rank $k$ tensor we find
\bea\label{May15c}
Y'_{M_1\dots M_k}&=&e^{(k+1)C/2}Y_{M_1\dots M_k}	,\\
W'_{M_2\dots M_k}&=&e^{(k-1)C/2}W_{M_2\dots M_k}+\frac{e^{(3k-5)C/2}}{2}g^{AB}\d_B CY_{A{M_2\dots M_k}}\nonumber
\eea
The same calculations show that for Killing tensors we have
\be\label{May15d}
K'_{MN}=e^{2C}K_{MN}, \quad W'_M=e^C(W_M+g^{AB}\d_B C K_{AM}).
\ee
Equations (\ref{May15c}) and (\ref{May15d}) summarize the behavior of Killing(--Yano) tensors under conformal rescalings.


\section{Killing tensors and ellipsoidal coordinates}
\renewcommand{\theequation}{B.\arabic{equation}}
\setcounter{equation}{0}

In this appendix we will justify the procedure for extracting separation of variables from a nontrivial Killing tensor and review an example of ellipsoidal coordinates and their degeneration.
 
\subsection{Ellipsoidal coordinates from Killing tensors}
\label{AppEllips}

As discussed in section \ref{SecSeparKT}, existence of a non--trivial Killing tensor leads to separation of variables, and in this appendix we will provide some details of the procedure for extracting the relevant coordinates and the separation function. 

We will focus on studying the reduced metric (\ref{KillLaMain}), and  to simplify the notation we will drop the subscript $red$. Assuming that non--cyclic coordinates separate, we find
\bea\label{KTApr22}
ds^2=\sum g_k dx_k^2,\quad 
K=\sum \Lambda_k g_k dx_k^2
\eea
where $g_k$ and $\Lambda_k$ are functions of all coordinates. 
Equations for the Killing tensor give
\bea\label{Apr22a}
\d_i\Lambda_i=0,\quad \d_j\ln g_i=\d_j\ln(\Lambda_i-\Lambda_j),\quad 
j\ne i
\eea
and there are no summations in these relations. We will now make an additional assumption of separability:
\bea\label{g1ElptSepar}
\d_j\d_k\ln g_m=0\quad\mbox{for different}\quad (i,j,k).
\eea
and determine the form of $g_k$ and $\Lambda_k$. The procedure involves several steps:
\begin{enumerate}[1.]
\item Equation (\ref{g1ElptSepar}) leads to factorization of $g_1$ 
\bea\label{g1Elpt}
g_1=\prod f_{1j}(x_1,x_j),\quad \d_j \ln f_{1j}(x_1,x_j)=\d_j \ln(\Lambda_1-\Lambda_j)
\eea
which implies factorization of
\bea\label{temp12}
\Lambda_1-\Lambda_2=f_{12}(x_1,x_2)g_{12}(x_1,x_3\dots).
\eea
The same expression can also be obtained by starting with $g_2$, but this leads to a different factorization:
\bea\label{temp12a}
\Lambda_1-\Lambda_2=f_{21}(x_2,x_1)g_{21}(x_2,x_3\dots).
\eea
Applying $\d_1\d_3$ to the logs of (\ref{temp12}), (\ref{temp12a}), we conclude that $x_1$ dependence factorizes in $g_{12}$. Absorbing the $x_1$--dependent factor in $f_{12}(x_1,x_2)$, we find
\bea
&&\Lambda_1-\Lambda_2=f_{12}(x_1,x_2)g_{12}(x_3\dots).\nonumber
\eea
The left--hand side of the last relation is killed by $\d_1\d_2$ (recall the first relation in (\ref{Apr22a})), so
\bea
f_{12}(x_1,x_2)=f_{12}^{(1)}(x_1)-f_{12}^{(2)}(x_2).
\eea
Repeating the same steps for $x_3,\dots, x_n$, we conclude that 
\bea\label{eqn23}
&&g_1=h_1(x_1)\prod_j [f_{1j}^{(1)}(x_1)-f_{1j}^{(j)}(x_j)],\nonumber\\
&&\Lambda_1-\Lambda_j=[f_{1j}^{(1)}(x_1)-f_{1j}^{(j)}(x_j)]g_{1j}(x_3\dots x_n),\quad
\d_j g_{1j}=0.
\eea
Since coordinate $x_1$ is not special, the last equation can be generalized:
\bea\label{eqn23a}
\Lambda_k-\Lambda_j=[f_{kj}^{(k)}(x_k)-f_{kj}^{(j)}(x_j)]g_{kj}(x_1\dots x_n),\quad
\d_j g_{kj}=\d_k g_{kj}=0.
\eea

\item
Assuming that $f_{1j}^{(j)}(x_j)$ are nontrivial functions of their arguments\footnote{This assumption of generality eventually leads to ellipsoidal coordinates for curved spaces. Relaxing this assumption, one arrives at degenerate cases, and some examples are presented in Appendix \ref{AllFlatEllips}. We conjecture that any degenerate case can be obtained by a singular limit of ellipsoidal coordinates, but we will not prove this statement. The proof for flat three dimensional space is implicitly contained in \cite{MorseFesh}.}, we can define new coordinates by setting
\bea
{\tilde x}_j\equiv f_{1j}^{(j)}(x_j),\quad j>1.
\eea
and dropping the tildes. We still have the freedom of making a linear transformation of $x_k$, which will be fixed later. Taking a second derivative of (\ref{eqn23}) with respect 
to $x_j$,
\bea
\d^2_j\Lambda_1=\d^2_j\Lambda_j+\d^2_j\Big([f_{1j}^{(1)}(x_1)-x_j]g_{1j}(x_2\dots x_n)\Big)=0.\nonumber
\eea
we conclude that $\Lambda_1$ is a linear polynomial in every coordinate  
$(x_2,\dots x_n)$.
Furthermore, since $\d_2\Lambda_2=0$ we find
\bea
\Lambda_2=\Lambda_1-[f_{12}^{(1)}(x_1)-x_2]g_{12}(x_3,\dots, x_n)=
\Lambda_1+[f_{12}^{(1)}(x_1)-x_2]\d_2 \Lambda_1\nonumber
\eea
and similarly
\bea
\Lambda_j=\Lambda_1+[f_{1j}^{(1)}(x_1)-x_j]\d_j \Lambda_1.
\eea
\item
Next we look at 
\bea
&&\Lambda_2-\Lambda_3=(f_{12}^{(1)}(x_1)-x_2)\d_2 \Lambda_1-
(f_{13}^{(1)}(x_1)-x_3)\d_3 \Lambda_1\nn
&&\quad=[f_{12}^{(1)}\d_2\Lambda_1-f_{13}^{(1)}\d_3\Lambda_1]_0+
x_3[f_{12}^{(1)}\d_2\d_3\Lambda_1+\d_3\Lambda_1]_0-
x_2[f_{13}^{(1)}\d_2\d_3\Lambda_1+\d_2\Lambda_1]_0\nonumber
\eea
Expressions in the square brackets are evaluated at $x_2=x_3=0$.  
Equation (\ref{eqn23a}) implies that $(x_2,x_3)$ dependence in the last equation must factorize, and this is possible only if
\bea\label{eqn23c}
f_{13}^{(1)}(x_1)=c_{32}f_{12}^{(1)}(x_1)+d_{32},\quad 
[f_{12}^{(1)}\d_2\Lambda_1-f_{13}^{(1)}\d_3\Lambda_1]_0=e_{32}
\eea
with {\it constant} $(c_{32},d_{32},e_{32})$. Similar arguments demonstrate that all 
$f_{1j}(x_1)$ are linear polynomials in $f_{12}^{(1)}(x_1)$, so by re-defining this coordinate,
\bea
x_1\rightarrow f_{12}^{(1)}(x_1),\nonumber
\eea
we conclude that all $f_{1j}(x_1)$ are linear functions of their arguments. For example,
\bea
f_{13}^{(1)}(x_1)-x_3=c_{32}x_1+d_{32}-x_3,\nonumber
\eea
so by making a linear transformation of $x_3$, we can simplify the last expression:
\bea
f_{13}^{(1)}(x_1)-x_3\rightarrow c_{32}(x_1-x_3).\nonumber
\eea
Repeating this for $(x_4\dots x_n)$, we find
\bea\label{eqn23b}
g_1=h_1(x_1)\prod_j [x_1-x_j],\qquad
\Lambda_j=\Lambda_1+[x_1-x_j]\d_j \Lambda_1.
\eea
\item
We will now demonstrate that polynomial $\Lambda_1(x_2,\dots,x_n)$ must be symmetric under interchange of any pair of its arguments. Without the loss of generality, we focus on $x_2$ and $x_3$ and write $\Lambda_1$ as
\bea
\Lambda_1=P_{1}x_2x_3+P_{2}x_2+P_{3}x_3+P_{4},
\eea
where $P_k$ are polynomials in $(x_4\dots x_n)$. The second equation in 
(\ref{eqn23c}) gives
\bea\label{eqn23e} 
e_{32}=x_1[\d_2\Lambda_1-\d_3\Lambda_1]_0=x_1[P_2-P_3].
\eea
Consistency of this relation requires $P_2=P_3$, i.e., symmetry of $\Lambda_1$ under the interchange of $x_2$ and $x_3$.
\item
Once we established that $\Lambda_1(x_2\dots x_n)$ is symmetric, it is convenient to introduce a ``generating" linear polynomial $\Lambda(x_1\dots x_n)$ symmetric in its arguments and define
\bea
\Lambda_1=\d_1 \Lambda.
\eea
Then the second relation in (\ref{eqn23b}) implies
\bea
\Lambda_j=\d_1\Lambda+(x_1-x_j)\d_1\d_j\Lambda=
\d_1\Lambda|_{x_j=0}+x_1\d_1\d_j\Lambda=
\d_j\Lambda|_{x_1=0}+x_1\d_1\d_j\Lambda=\d_j\Lambda.\nonumber
\eea
\end{enumerate}

To summarize, we have demonstrated that in the generic case existence of the Killing tensor in the non--cyclic part of the metric (\ref{KTApr22}) implies that 
\bea
g_k=h_1(x_k)\prod_{j\ne k} [x_k-x_j],\qquad
\Lambda_j=\d_j\Lambda,
\eea
where $\Lambda(x_1\dots x_n)$ is a linear polynomial in every $(x_1\dots x_n)$ symmetric under interchange of every pair of arguments. This completes the justification of (\ref{KillLaMain})--(\ref{SeparHJmain}), which summarize the extraction of the separable coordinates from a Killing tensor.

\subsection{Ellipsoidal coordinates in flat space}
\label{AllFlatEllips}

In section \ref{SecSeparKT} we demonstrated that separation of non--cyclic coordinates generically leads to ellipsoidal coordinates. Our derivation was based on the assumption of generality: we postulated that metric components have non--trivial dependence on all non--cyclic coordinates. If this assumption is dropped, one recovers degenerate cases of ellipsoidal coordinates, and in this appendix we will illustrate this using a well-known example of flat three--dimensional space. Degeneration in higher dimensions is very similar, but its detailed discussion is beyond the scope of this article.

Consider a flat three--dimensional space with a metric
\bea\label{EllAppFlat}
ds^2=dr_1^2+dr_2^2+dr_3^2.
\eea
The ellipsoidal coordinates $(x_0,x_1,x_2)$ are defined as three solutions of a cubic equation for $x$ \cite{Jacobi}:
\bea
\frac{r_1^2}{x-a}+\frac{r_2^2}{x-b}+
\frac{r_3^2}{x-c}=1,
\eea
Without the loss of generality we assume that non--degenerate coordinates have $a> b> c$ and the roots are arranged in the following order:
\bea\label{ElpsdRng}
x_0>a>x_1>b>x_2>c.
\eea
Cartesian coordinates $(r_1,r_2,r_3)$ can be expressed in terms of $(x_0,x_1,x_2)$ as
\bea\label{EllipsoidalCoords}
r_1=\left[\frac{(x_0-a)(x_1-a)(x_2-a)}{
(a-b)(a-c)}\right]^{1/2},\quad
r_2=\left[\frac{(x_0-b)(x_1-b)(x_2-b)}{
(b-a)(b-c)}\right]^{1/2},\quad
\eea
\bea
r_3=\left[\frac{(x_0-c)(x_1-c)(x_2-c)}{
(c-a)(c-b)}\right]^{1/2}.\nonumber
\eea
This transformation turns the metric (\ref{EllAppFlat}) into
\bea\label{EllAppElps}
ds^2&=&\frac{(x_0-x_1)(x_0-x_2)dx_0^2}{4(x_0-a)(x_0-b)(x_0-c)}+
\frac{(x_1-x_0)(x_1-x_2)dx_1^2}{4(x_1-a)(x_1-b)(x_1-c)}\nn
&&+\frac{(x_2-x_0)(x_2-x_1)dx_2^2}{4(x_2-a)(x_2-b)(x_2-c)}.
\eea
Shifting six quantities $(x_i,a,b,c)$ by $c$, one usually sets $c=0$, and we will follow this convention\footnote{In section \ref{SecMyersPerry} we use a different convention: $a=0$, $b=-a_1^2$, $c=-a_2^2$.}. 

The degenerate cases of the ellipsoidal coordinates are discussed in great detail in \cite{MorseFesh}\footnote{There are ten of them:
rectangular, oblate/prolate spheroidal, circular/elliptic/parabolic cylinder, spherical, conical, paraboloidal, and parabolic.}, and we will focus only on oblate spheroidal and spherical coordinates. Oblate spheroidal coordinates are obtained from \eqref{EllipsoidalCoords} by writing
\bea
x_0=a+\xi_0,\quad x_1=a-a\xi_1,\quad x_2=b \xi_2
\eea
and sending $b$ to zero. Then metric (\ref{EllAppElps}) becomes
\bea
ds^2&=&\frac{(\xi_0+a\xi_1)d\xi_0^2}{4\xi_0(\xi_0+a)}+
\frac{(\xi_0+a\xi_1)d\xi_1^2}{4\xi_1(1-\xi_1)}
+(\xi_0+a)(1-\xi_1)\frac{d\xi_2^2}{4\xi_2(1-\xi_2)}.
\eea
This expression has a very simple interpretation: $\xi_2$ gives rise to a new cyclic coordinate $\zeta$ ($\xi_2=\cos^2\zeta$), while $(\xi_0,\xi_1)$ form two--dimensional elliptic coordinates. This is in a perfect agreement with general analysis of non--cyclic directions presented in section \ref{SecSeparKT}.   

As a next example we consider spherical coordinates, which can be obtained by writing
\bea
b=a-\epsilon,\quad x_0=\xi_0,\quad x_1=a-\epsilon \xi_1,\quad x_2=a\xi_2,
\eea
sending $\epsilon$ to zero, and setting $a=0$ in the resulting expression. This gives
\bea
ds^2&=&\frac{d\xi_0^2}{4\xi_0}+
\frac{\xi_0(1-\xi_2)d\xi_1^2}{4\xi_1(1-\xi_1)}
+\frac{\xi_0d\xi_2^2}{4(1-\xi_2)\xi_2}=dr^2+r^2\sin^2\theta d\phi^2+d\theta^2.
\eea
We see that although $\xi_2$ (which is related to the polar angle $\theta$) remains a non--cyclic coordinate, it does not appear in $g_{11}$, so spherical coordinates violate one of the assumptions made in section \ref{SecSeparKT}. Nevertheless such parameterization can be obtained as a degenerate case of ellipsoidal coordinates, and we conjecture that any separable frame in the non--cyclic coordinates can be obtained as a similar singular limit from the systems derived in section \ref{SecSeparKTsub}. The proof of this conjecture is beyond the scope of this paper.



\section{Principal CKYT for the Myers--Perry black hole}
\label{AppPCKYT}
\renewcommand{\theequation}{C.\arabic{equation}}
\setcounter{equation}{0}

In section \ref{SecMyersPerry} we found a family of the Killing--Yano tensors (\ref{Kub2}) for the Myers--Perry black hole, and the construction was based on three statements:
\begin{enumerate}
\item The anti--symmetric tensor $h$ defined by (\ref{Kub1}) is a Conformal Killing--Yano tensor and the form (\ref{Kub1}) is closed. Such tensors are called Principal Conformal Killing--Yano tensors (PCKYT) \cite{Kub1}.
\item A wedge product of two PCKY tensors is again a PCKYT\footnote{Note that this is not true for KY tensors.}, so the expression $\wedge h^n$ is a PCKYT for any value of $n$.
\item If ${\cal Y}$ is a PCKYT then $Y=\star {\cal Y}$ is a Killing--Yano tensor.
\end{enumerate}
The proofs of these statements are scattered throughout the literature 
\cite{Carig,Kub1,KubThes}, and the goal of this appendix is to present a simpler derivation of properties 1-3. We will begin with properties 2 and 3 since they are not specific to the Myers--Perry black hole.

\bigskip

We begin with writing the condition $d{\cal Y}=0$ for a Principal Conformal Killing--Yano tensor ${\cal Y}$ of rank $p$:
\bea\label{Jun10}
\nabla_a {\cal Y}_{bcd\dots}-\nabla_b {\cal Y}_{acd\dots}-\nabla_c {\cal Y}_{bad\dots}+\dots=0.
\eea
There are $p$ terms in this equation. Using the defining relation (\ref{CKTdef}) for the CKYT,
\bea\label{Jun10a}
\nabla_b {\cal Y}_{acd\dots}=-\nabla_a {\cal Y}_{bcd\dots}+2 g_{ab}Z_{cd\dots}-
[g_{ca}Z_{bd\dots}+g_{cb}Z_{ad\dots}]+\dots,
\eea
equation (\ref{Jun10}) can be rewritten as
\bea\label{Jun10b}
\nabla_a {\cal Y}_{bcd\dots}=g_{ab}Z_{cd\dots}-g_{ac}Z_{bd\dots}+\dots=
p g_{a[b}Z_{cd\dots]}.
\eea
The PCKYT is defined as an object satisfying relations (\ref{Jun10}), (\ref{Jun10a}), but one can use the equivalent set of defining relation (\ref{Jun10}) and (\ref{Jun10b}) instead. In particular, we observe that any Killing--Yano tensor which is also closed must be covariantly constant. Such objects are closely related to complex structures on K\"ahler manifolds, which are discussed in the Appendix \ref{AppComplexStructure}. 

To prove property 2, we observe that a product of two PCKYT, ${\cal Y}^{(p)}\wedge {\cal Y}^{(q)}$ is closed, and it satisfies equation (\ref{Jun10b}) with 
\bea
Z^{(p+q)}=\frac{1}{p+q}\left[p Z^{(p)}\wedge {\cal Y}^{(q)}+(-1)^{p+q}q Z^{(q)}\wedge {\cal Y}^{(p)}\right].
\eea
To prove property 3, we consider
\bea
\nabla_m\left[\eps_{a_1\dots a_q}{}^{b_1\dots b_p}{\cal Y}_{b_1\dots b_p}\right]=
\eps_{a_1\dots a_q}{}^{b_1\dots b_p}p g_{m[b_1}Z_{b_2\dots b_p]}=
p \eps_{a_1\dots a_q m}{}^{b_2\dots b_p}Z_{b_2\dots b_p}.
\eea
Symmetrization over $(m,a_1)$ gives zero, so 
$Y_{a_1\dots a_q}\equiv \eps_{a_1\dots a_q}{}^{b_1\dots b_p}{\cal Y}_{b_1\dots b_p}$ is a Killing--Yano tensor. This completes the proof of properties 2 and 3 which hold for all spaces admitting PCKYT.

\bigskip

Next we focus on the Myers--Perry black hole and demonstrate that the closed form
\bea
h&=&\h \sum a_id\mu_i^2\wedge\left[a_i dt+(r^2+a_i^2)d\phi_i\right]+
\h dr^2\wedge\left[dt+a_i\mu_i^2 d\phi_i\right]\nonumber\\
&=&\h d[r^2+\sum a_i^2\mu_i^2]\wedge dt+\h\sum  d[a_i(r^2+a_i^2)\mu_i^2]\wedge d\phi_i
\eea
is a Conformal Killing--Yano tensor. The proof will go in two steps: first we will verify the CKYT equation for $m=0$, and then we will show that $m$ dependence does not affect the result.

For $m=0$ the geometry (\ref{MPodd}) is flat, and it is convenient to rewrite it in the Cartesian coordinates. In odd dimensions such coordinates are defined by
\bea
X_k+iY_k=\sqrt{r^2+a_k^2}\mu_k e^{i\phi_k},\quad ds^2=-dt^2+\sum [(dX_k)^2+(dY_k)^2],
\eea
and the two--form $h$ becomes
\bea
h=\h d[\sum (X_k^2+Y_k^2)]\wedge dt+\sum a_k dX_k\wedge dY_k\,.
\eea
This gives interesting relations for the derivatives of $h_{MN}$,
\bea
&&\nabla_M h_{NP}+\nabla_N h_{MP}=0,\quad\mbox{if}\quad (MNP)\ne t,\nn
&&\nabla_M h_{Nt}+\nabla_N h_{Mt}=2[\delta_{MN}-\delta_{Mt}\delta_{Nt}],\\
&&\nabla_M h_{tP}+\nabla_t h_{MP}=-[\delta_{MP}-\delta_{Mt}\delta_{Pt}],\nonumber
\eea
which can be summarized as an equation for the CKYT (\ref{CKTdef}):
\bea\label{PrincKYT}
\nabla_M h_{NP}+\nabla_N h_{MP}=2g_{MN}Z_P-g_{MP}Z_N-g_{NP}Z_M,\quad 
Z^M\d_M=\d_t.
\eea
The argument for even dimensions works in a similar way. 
This concludes the first part of the proof ($h$ is a CKYT for the flat space), and now we will demonstrate that (\ref{PrincKYT}) holds for $m\ne 0$ as well.

While it is possible to verify (\ref{PrincKYT}) using the explicit form of the Christoffel's symbols\footnote{Such `brute force' calculation is performed in the Appendix B.3 of \cite{KubThes}.}, this calculation is tedious and not very instructive since it does not take advantage of the high degree of symmetry of the Myers--Perry solution. We will use an alternative method based on spin connections, which gives the answer in an easier and more transparent way. First we rewrite (\ref{PrincKYT}) in terms of frame indices:
\bea\label{AddPCKYTframes}
&&\nabla_a h_{bc}+\nabla_b h_{ac}=2\eta_{ab}Z_c-\eta_{ac}Z_b-\eta_{bc}Z_a\\
&&h=re^{\hat r}\wedge e^{\hat t}+\sum_i\sqrt{-x_i}e^{{\hat x^i}}\wedge e^{\hat i},\quad Z_a=e_{at}
\nonumber
\eea
To derive the desired result we should analyze the $m$--dependence of
\bea\label{DefTcky}
T_{abc}\equiv\nabla_a h_{bc}+\nabla_b h_{ac}
\eea
Covariant derivatives of the objects with frame indices are evaluated using the standard relations
\bea
\nabla_a V^b=e_a^M\d_M V^b+\omega_{a,}{}^b{}_c V^c,\quad
\nabla_a W_b=e_a^M\d_M W_b-\omega_{a,}{}^c{}_b W_c,
\eea
and the spin connection $\omega_{a,}{}^b_{e}$ is related to the anholonomy coefficients $\Gamma_{a,be}$ by
\bea\label{SpinConn}
\omega_{c,ab}=\frac{1}{2}\left[\Gamma_{c,ab}+\Gamma_{b,ac}-\Gamma_{a,bc}\right],\quad
de^a=\frac{1}{2}\Gamma^a{}_{,bc}e^b\wedge e^c
\eea
In particular, the explicit expressions for $e^{\hat t}$ in (\ref{AllFramesMP}) and (\ref{AllFramesMPOdd}) imply that 
\bea
\Gamma^{\hat{t}}{}_{,\beta \hat{r}}=0.
\eea
Here and below the greek letters denote the frame indices excluding $(\hat r,\hat t)$. Although it is not obvious from $e^{\hat i}$ and $e^{\hat \phi_i}$, the anholonomy coefficients $\Gamma_{\alpha,\hat t\gamma}$ vanish as well. To see this, we use an 
alternative expression for $\Gamma$:
\bea
\Gamma_{a,bc}&=&(de_a)_{\mu\nu}e^\mu_be^\nu_c=(\d_\mu e_{a\nu}-\d_\nu e_{a\mu})e^\mu_be^\nu_c=
-e^\mu_be_{a\nu}\d_\mu e^\nu_c+e^\mu_ce_{a\nu}\d_\mu e^\nu_b,
\eea
which gives for (\ref{AllFramesMP}) and (\ref{AllFramesMPOdd}):
\bea
\Gamma_{\alpha,\hat t\gamma}&=&e^\mu_\gamma e_{\alpha\nu}\d_\mu e^\nu_{\hat{t}}=-\frac{1}{2}
e^\mu_\gamma e_{\alpha\nu} e^\nu_{\hat{t}} \d_\mu\ln F=-\frac{1}{2}
e^\mu_\gamma \eta_{\alpha\hat{t}}\, \d_\mu\ln F=0.
\eea
Next we use the frames (\ref{AllFramesMP}) and (\ref{AllFramesMPOdd}) to compute the anholonomy $\Gamma_{a,bc}$ coefficients and spin connections $\omega_{a,bc}$ in terms of their counterparts ${\tilde\Gamma}_{a,bc}$ and ${\tilde\omega}_{a,bc}$ for $m=0$. The simplicity of the $m$ dependence in the frames combined with relations $\Gamma_{\alpha,\hat t\gamma}=\Gamma^{\hat{t}}{}_{,\beta \hat{r}}=0$ allows us to write the answers without doing complicated calculations which are normally associated with evaluation of the spin connection. Introducing convenient notation
\bea
S=\left\{
\begin{array}{ll}
\sqrt{\frac{R-mr}{R}},&\mbox{even\ }d \\
\sqrt{\frac{R-mr^2}{R}},&\mbox{odd\ }d
\end{array}
\right.\,,
\eea
we can summarize the anholonomy coefficients as
\bea\label{SpConGam}
&&\Gamma^\alpha{}_{,\beta\gamma}={\tilde\Gamma}^\alpha{}_{,\beta\gamma},\quad
\Gamma^\alpha{}_{,\beta \hat r}=S{\tilde\Gamma}^\alpha{}_{,\beta \hat r},\quad
\Gamma^{\hat{r}}{}_{,\beta \hat{r}}={\tilde\Gamma}^{\hat{r}}{}_{,\beta \hat{r}},\nonumber\\
&&\Gamma^\alpha{}_{,\beta \hat{t}}=0,\quad
\Gamma^{\hat t}{}_{,\beta \hat t}={\tilde\Gamma}^{\hat t}{}_{,\beta t},\quad
\Gamma^{\hat{t}}{}_{,\beta \gamma}=S{\tilde\Gamma}^{\hat{t}}{}_{,\beta \gamma}\\
&&\Gamma^{\hat{t}}{}_{,\beta \hat{r}}=0,
\quad\Gamma^\alpha{}_{,\hat{t}\hat{r}}={\tilde\Gamma}^\alpha{}_{,\hat{t}\hat{r}},\quad
\Gamma^{\hat{t}}{}_{,\hat{t} \hat{r}}=S{\tilde\Gamma}^{\hat{t}}{}_{,\hat{t} \hat{r}}-\frac{1}{F}\d_r S,\quad \Gamma^{\hat{r}}_{\hat{r}\hat{t}}=0,
\nonumber
\eea
and the spin connections as
\bea\label{SpConOm}
&&\omega_{\alpha,\beta\gamma}={\tilde\omega}_{\alpha,\beta\gamma},\quad
\omega_{\hat{r},\alpha\beta}=S {\tilde\omega}_{\hat{r},\alpha\beta},\quad
\omega_{\alpha,\hat{r}\beta}=S{\tilde\omega}_{\alpha,\hat{r}\beta},\quad 
\omega_{\hat{r},\hat{r}\beta}={\tilde\omega}_{\hat{r},\hat{r}\beta},\nn
&&
\omega_{\hat{t},\hat{t}\alpha}={\tilde\omega}_{\hat{t},\hat{t}\alpha},\quad 
\omega_{\hat{t},\alpha\beta}=S{\tilde\omega}_{\hat{t},\alpha\beta},
\quad \omega_{\alpha,\beta \hat{t}}=S{\tilde \omega}_{\alpha,\beta \hat{t}},\\
&&\omega_{\alpha,\hat{r}\hat{t}}=
{\tilde\omega}_{\alpha,\hat{r}\hat{t}},\quad \omega_{\hat{r},\alpha \hat{t}}=
{\tilde\omega}_{\hat{r},\alpha \hat{t}},\quad \omega_{\hat{t},\alpha \hat{r}}=
{\tilde\omega}_{\hat{t},\alpha \hat{r}},\quad \omega_{\hat{r},\hat{r}\hat{t}}=0,\quad \omega_{\hat{t},\hat{t}\hat{r}}=\Gamma_{\hat{t},\hat{t}\hat{r}}.\nonumber
\eea
Substituting the expressions (\ref{SpinConn}) into (\ref{DefTcky}) and introducing ${\hat\d}_a\equiv e_a^M\d_M$, we find
\bea\label{AddJun8}
T_{abc}&=&{\hat\d}_a h_{bc}+{\hat\d}_b h_{ac}+\left[\omega_{a,be}+\omega_{b,ae}\right]h^e{}_c+
\omega_{a,ce}h_b{}^e+\omega_{b,ce}h_a{}^e\nn
&=&{\hat\d}_a h_{bc}+{\hat\d}_b h_{ac}+(\Gamma_{a,be}+\Gamma_{b,ae})h^e{}_c+
\omega_{a,ce}h_b{}^e+\omega_{b,ce}h_a{}^e
\eea
and the explicit expressions (\ref{SpConGam}), (\ref{SpConOm}) give
\bea
T_{\alpha\beta\gamma}={\tilde T}_{\alpha\beta\gamma},\quad
T_{\alpha {\hat m}{\hat n}}={\tilde T}_{\alpha {\hat m}{\hat n}},\quad 
T_{{\hat m}{\hat n}\alpha}={\tilde T}_{{\hat m}{\hat n}\alpha},\quad
T_{{\hat m}\beta\gamma}=S{\tilde T}_{{\hat m}\beta\gamma},\quad
T_{\alpha\beta {\hat m}}=S{\tilde T}_{\alpha\beta {\hat m}},
\eea
where $({\hat m},{\hat n})$ take values ${\hat t}$ or ${\hat r}$. The remaining components are
\bea
T_{\hat{r}\hat{r}\hat{r}}&=&0,\quad T_{\hat{t}\hat{t}\hat{t}}=0,\quad
T_{\hat{r}\hat{t}\hat{t}}=0,\quad T_{\hat{t}\hat{r}\hat{t}}=0,\nn
T_{\hat{r}\hat{r}\hat{t}}
&=&2{\hat\d}_{\hat{r}} h_{\hat{r}\hat{t}}+2\Gamma_{\hat{r},\hat{r}\hat{r}}h^{\hat{r}}{}_{\hat{t}}+
2\omega_{\hat{r},\hat{t}\hat{t}}h_{\hat{r}}{}^{\hat{t}}=2{\hat\d}_{\hat{r}} h_{\hat{r}\hat{t}}=S{\tilde T}_{\hat{r}\hat{r}\hat{t}},\\
T_{\hat{t}\hat{r}\hat{r}}
&=&{\hat\d}_{\hat{r}} h_{\hat{t}\hat{r}}+\Gamma_{\hat{t},\hat{r}\hat{t}}h^{\hat{t}}{}_{\hat{r}}+
2\omega_{\hat{t},\hat{r}\hat{t}}h_{\hat{r}}{}^{\hat{t}}={\hat\d}_{\hat{r}} h_{\hat{t}\hat{r}}=S{\tilde T}_{\hat{t}\hat{r}\hat{r}}\,.\nonumber
\eea
Recalling that 
\bea
Z_\alpha=e_{\alpha t}={\tilde Z}_\alpha,\quad Z_{\hat t}=e_{\hat t t}=S{\tilde Z}_{\hat t},\quad 
Z_{\hat r}=0,
\eea
we conclude that equation (\ref{AddPCKYTframes}),
\bea
T_{abc}=2\eta_{ab}Z_c-\eta_{ac}Z_b-\eta_{bc}Z_a,
\eea
is equivalent to 
\bea
{\tilde T}_{abc}=2\eta_{ab}{\tilde Z}_c-\eta_{ac}{\tilde Z}_b-\eta_{bc}{\tilde Z}_a,
\eea
which has been verified earlier. This completes the proof of the relation (\ref{PrincKYT}) for the Myers--Perry black hole and verification of statements 1-3 made in the beginning of this appendix.



\section{Dimensional reduction and T duality}\label{AppDimRed}
\renewcommand{\theequation}{D.\arabic{equation}}
\setcounter{equation}{0}

This appendix discusses dimensional reduction of equations for Killing vectors, Killing--(Yano) tensors and their conformal counterparts. Section \ref{AppDRconv} sets up the conventions, section \ref{AppDimRed1} discusses dimensional reduction of arbitrary tensors, and these results are applied to Killing vectors in section \ref{AppDRKV}, to symmetric Killing tensors in section \ref{AppDRKT4}, and to Killing--Yano tensors in section \ref{AppBeqKYT}. Conformal Killing tensors are discussed in section \ref{AppCKT}, conformal Killing vectors are analyzed in section \ref{SubsectionCKV} and some comments about conformal Killing--Yano tensors are made in the end of section \ref{SectionModifiedKYT}.

We demonstrate that equations for the KV and KT are consistent with T duality, but equation for the KYT should be modified, and we find the unique modification. Also we find that consistency between continuous symmetries and T duality leads to constraints on the Kalb--Ramond field if one is present, and such constraints suggest an interesting generalization of a standard Lie derivative along vector field to the derivative along Killing tensors. This construction is discussed in section \ref{AppLieDer}.

\subsection{Conventions}
\label{AppDRconv}

We begin with setting up the conventions. Consider a geometry which admits a Killing vector $\d_z$ and 
write the metric and the Kalb--Ramond field in the form
\bea\label{AppDimRedBdy}
\label{DimRedSetup}
ds^2&=&e^C[dz+A_mdx^m]^2+\hat g_{mn}dx^mdx^n,\nn
B&=&{\tilde A}_n dx^n\wedge [dz+\frac{1}{2}A_m dx^m]+
\frac{1}{2}{\hat B}_{mn}dx^m\wedge dx^m.
\eea
Here $(m,n)$ run over all coordinates excluding $z$, and an unusual notation for $B$ field will be justified below. Ramond--Ramond fields may also be present, but they will not affect our discussion. For future reference we also write the metric and its inverse in matrix form:
\bea\label{SetupBefore}
g_{MN}=\begin{pmatrix}
  e^C & e^C A_i\\
              e^CA_j & e^CA_iA_j+\hat{g}_{ij}\end{pmatrix}, \qquad
g^{MN}=\left(\begin{array}{cc}
	 e^{-C}+A_iA^i & -A^i\\
              -A^j & \hat{g}^{ij}\end{array}\right). 
\eea
Since $z$ is a cyclic coordinate in (\ref{DimRedSetup}), it is possible to perform T duality along this direction using the Buscher's rules
\bea\label{Buscher}\label{TdualityMap}
\tilde{g}_{zz}&=&\frac{1}{g_{zz}},\quad e^{2\tilde{\Phi}}=\frac{e^{2\Phi}}{g_{zz}},\quad
\tilde{g}_{m z}=\frac{B_{m z}}{g_{zz}},\qquad \tilde{B}_{m z}=\frac{g_{m z}}{g_{zz}},
\\
\tilde{g}_{mn}&=&g_{mn}-\frac{G_{m z}G_{n z}-B_{m z}B_{n z}}{g_{zz}},\quad
\tilde{B}_{mn}=B_{mn}-\frac{B_{m z}g_{n z}-g_{m z}B_{n z}}{g_{zz}}.\nonumber
\eea
Application of this procedure to  (\ref{DimRedSetup}) gives
\be\label{SetupAfter}
d\tilde s^2=e^{-C}(dz+{\tilde A}_mdx^m)^2+\hat g_{mn}dx^mdx^n, \quad 
\tilde{B}={A}_n dx^n\wedge [dz+\frac{1}{2}\tilde{A}_m dx^m]+
\frac{1}{2}{\hat B}_{mn}dx^m\wedge dx^m. 
\ee
Notice that $A_m$ and ${\tilde A}_m$ are interchanged by T duality making the notation (\ref{DimRedSetup}) very natural. 

\bigskip
\noindent
In this paper we use the following conventions:
\begin{itemize} 
\item capital letters run through all the coordinates, $\{M,N, ... \}= \{ 1,...,d\}$;
\item lower case letters run through all the coordinates except $z$, $\{ m, n, ... \}= \{ 1,...,d-1\}$;
\item objects after T duality are marked with tilde, e.g. $\tilde V_i$, $\tilde K_{mn}$;
\item objects not affected by T duality are marked by hat, e.g. $\hat g_{ij}$, $\hat\nabla_m$.
\end{itemize}


\subsection{Dimensional reduction and covariant derivatives}

\label{AppDimRed1}

In this appendix we will express covariant derivatives in the geometry (\ref{DimRedSetup}) 
in terms of derivatives on the base $d{\hat s}^2$ assuming that all objects are 
$z$--independent. 

We begin with analyzing covariant derivatives of a vector:
\bea
W_{MN}=\nabla_M V_N.
\eea
The connections corresponding to the metric  (\ref{DimRedSetup}) are:
\bea
&&\Gamma^z_{zz}=\frac{1}{2}A^a\d_a e^C,\quad
\Gamma^m_{zz}=-\frac{1}{2}{\hat g}^{ma}\d_a e^C,\quad
\Gamma^z_{mz}=\frac{1}{2}\left[\d_m C-A^a e^C F_{ma}-2A^aA_{[a}\d_{m]}e^C\right],
\nonumber\\
&&\Gamma^m_{nz}=\frac{1}{2}{\hat g}^{ma}(e^C F_{na}-A_{n}\d_{a}e^C),\quad
\Gamma^z_{mn}=-A_a {\hat\Gamma}^a_{mn}+
\frac{1}{2}e^{-C}(\d_m g_{nz}+\d_n g_{mz})\\
&&\Gamma^s_{mn}={\hat\Gamma}^s_{mn}-\frac{1}{2}A^s(\d_m g_{nz}+\d_n g_{mz}).\nonumber
\eea
Indices of the gauge field $A_i$ are raised using ${\hat g}^{ij}$, and ${\hat\Gamma}^s_{mn}$ denotes Christoffel symbols on the base.

Explicit calculations give various components of $W$:
\bea\label{DimRedVectApp}
W_{zz}&=&\frac{1}{2}V^a\d_a e^C,
\nonumber\\
{W_z}^n&=&\frac{1}{2}{\hat g}^{nb}e^C F_{ab}V^a-
\frac{1}{2}{\hat g}^{na}V_z\d_a C,\\
{W^n}_z&=&\hat{g}^{na}\d_a V_z-\frac{1}{2}\hat{g}^{na}V_z \d_a C-\frac{1}{2}\hat{g}^{nb}e^C F_{ba}V^a,\nonumber\\
W^{mn}&=&{\hat \nabla}^m V^n
+\frac{1}{2}\hat{g}^{ma}\hat{g}^{nb} F_{ab}V_z.\nonumber
\eea
All components of $W_{MN}$ can be obtained by taking linear combinations of the expressions written above, for example,
\bea\label{WzmDown}
W_{zm}&=&g_{ma}{W_z}^a+g_{mz}{W_z}^z=g_{ma}{W_z}^a+\frac{g_{mz}}{g_{zz}}\left[W_{zz}-g_{az}{W_z}^a\right]=
A_m W_{zz}+{\hat g}_{ma}{W_z}^a\nonumber\\
&=&-\frac{1}{2}V_z\d_m C-\frac{1}{2}e^C F_{ma}V^a+\frac{1}{2}A_m V^a\d_a e^C.
\eea
The relation (\ref{DimRedVectApp}), (\ref{WzmDown}) are used in section \ref{SecKVDuality}. While discussing conformal Killing vectors in section \ref{SubsectionCKV} we also need generalization of (\ref{DimRedVectApp}) to derivatives of a $z$--dependent vector:
\bea\label{AppzdepKVred}
W_{zz}&=& \d_z V_z+\h  V^a\d_a e^C,\nn
W^m{}_{z}+W_{z}{}^m&=& \hat{g}^{ma}\left[\d_aV_z-\d_a CV_z-e^C F_{ab}V^b+\hat{g}_{ab}\d_z V^b-A_a\d_z V_z\right],\\
W^{mn}+W^{nm}&=& \hat\nabla^m V^n+\hat\nabla^n V^m-A^m\d_z V^n-A^n\d_z V^m.\nonumber
\eea

Once the action of covariant derivatives on various types of indices is specified, their application to a tensor of rank 2 becomes straightforward:
\bea\label{DimRedL}
\nabla_z L_{zz}&=&\frac{1}{2}[{L^a}_z+{L_z}^a]\d_a e^C,\nonumber\\
\nabla_z {L_z}^n&=&\frac{1}{2}{L^{an}}\d_a e^C+\frac{1}{2}{\hat g}^{nb}e^C F_{ab}{L_z}^a-
\frac{1}{2}{\hat g}^{na}L_{zz}\d_a C,\nonumber\\
\nabla_z L^{mn}&=&\frac{1}{2}[{\hat g}^{mb}L^{an}+{\hat g}^{nb}L^{ma}]e^C F_{ab}-
\frac{1}{2}[{\hat g}^{ma}{L_z}^n+{\hat g}^{na}{L^m}_z]\d_a C,\nonumber\\
\nabla^n {L_{zz}}&=&\hat{g}^{na}\d_a L_{zz}-\hat{g}^{na}L_{zz} \d_a C-\frac{1}{2}\hat g^{nb}e^C F_{ba}[{L^a}_z+{L_z}^a],
\\
\nabla^m {L_z}^n&=&{\hat\nabla}^m {L_z}^n-\frac{1}{2}\hat{g}^{ma}{L_z}^n \d_a C-\frac{1}{2}\hat{g}^{mb}e^C F_{ba}L^{an}
+\frac{1}{2}\hat{g}^{ma}\hat{g}^{nb} F_{ab}L_{zz},\nonumber\\
{\nabla}^m L^{np}&=&{\hat \nabla}^m L^{np}
+\frac{1}{2}\hat{g}^{ma}\hat{g}^{nb} F_{ab}{L_z}^p
+\frac{1}{2}\hat{g}^{ma}\hat{g}^{pb} F_{ab}{L^n}_z.\nonumber
\eea
These formulas are used in section \ref{SecKillingsDualities} to study the reduction of Killing--(Yano) tensors.

\subsection{Dimensional reduction for Killing vectors}
\label{AppDRKV}

In this subsection we will consider the behavior of Killing vectors under T duality. We will start with an object which satisfies the Killing equation
\be\label{AppKVeq}
\nabla_M V_N+\nabla_N V_M=0,
\ee
in the geometry (\ref{AppDimRedBdy}) supported by the NS--NS fields.  
T duality along $z$ direction gives the geometry (\ref{SetupAfter}) which has the same form with replacements
\bea\label{TdualityMapBdy}
C\rightarrow -C,\quad A\leftrightarrow {\tilde A},\quad e^{2\phi}\rightarrow e^{2\phi-C},\quad
\mbox{fixed}\quad {\hat g}_{mn},\ {\hat B}_{mn},
\eea 
If present, Ramond--Ramond fields would also transform under such duality, but such fields will not affect our analysis. 

Let us assume that before T duality geometry (\ref{AppDimRedBdy}) admitted a Killing vector that satisfied equation
\bea\label{KVeqnBdy}
Z_{MN}=0,\qquad Z_{MN}\equiv \nabla_M V_N+\nabla_N V_M.
\eea
As demonstrated in section \ref{AppDimRed1}, equation (\ref{KVeqnBdy}) can be written as a system\footnote{This follows from (\ref{DimRedVectApp}) by noticing that $Z_{MN}=W_{MN}+W_{NM}$.}
\bea\label{Oct3}
Z_{zz}&=&V^a\d_a e^C=0,\nonumber\\
Z^{mn}&=&{\hat \nabla}^m V^n+{\hat \nabla}^n V^m=0,\\
{Z_z}^m&=&{\hat g}^{ma}\d_a(e^{-C}V_z)-
{\hat g}^{mb} F_{ba}V^a=0.\nonumber
\eea
T duality (\ref{TdualityMapBdy}) leaves the first two equations invariant as long as we make identification
\bea
{\tilde V}^a=V^a, 
\eea
and it maps the last equation (\ref{Oct3}) into a restriction on the $B$ field:
\bea
{\hat g}^{ma}\d_a {\tilde W}_z+{\hat g}^{mb} 
{\tilde H}_{baz}V^a=0,
\qquad {\tilde W}_z\equiv -e^{-C}V_z.
\eea
Similarly, before the T duality we must have
\bea
{\hat g}^{ma}\d_a {W}_z+{\hat g}^{mb} 
{H}_{baz}V^a=0,
\qquad {W}_z\equiv -e^{C}{\tilde V}_z.
\eea
The last equation is a $(mz)$ component of a covariant relation:
\bea\label{BfieldKV}
H_{MNS}V^S=\nabla_M W_N-\nabla_N W_M,
\eea
as now we will discuss its origin and implications coming from the remaining components. 

To give a geometrical interpretation of (\ref{BfieldKV}) we look at a Lie derivative of the $B$ field along the Killing vector $V$:
\bea
{\cal L}_V B_{MN}&=&V^A\nabla_A B_{MN}+B_{AN}\nabla_M V^A+B_{MA}\nabla_N V^A\nonumber\\
&=&
V^A H_{MNA}-\nabla_M(V^AB_{AN})+\nabla_N(V^AB_{AM})\nonumber
\eea
and recall that if $V^A$ is a Killing vector, then this derivative must be a pure gauge, i.e.,
\bea\label{KillDerBfield}
{\cal L}_V B_{MN}=\nabla_M W'_N-\nabla _N W'_M
\eea
for some vector $W'_M$. Combining the last two relations, we find 
\bea
V^A H_{MNA}=\nabla_M (W'_N+V^AB_{AN})-\nabla _N (W'_M+V^AB_{AM}),\nonumber
\eea
which coincides with (\ref{BfieldKV}) if we define
\bea\label{WgaugeKV}
W_N=W'_N+V^AB_{AN}.
\eea
At this point we have demonstrated that condition (\ref{BfieldKV}) comes from requiring that the Lie derivative of the $B$ field is a pure gauge, and we found the T duality map for various components of $V$ and $W$:
\bea
{\tilde V}^a=V^a,\quad {\tilde W}_z= -e^{-C}V_z,\quad 
{\tilde V}_z= -e^{-C}W_z.
\eea
To complete the proof that the system
\bea\label{KVArray}
\left\{
\begin{array}{l}
\nabla_M V_N+\nabla_N V_M=0\\
H_{MNS}V^S=\nabla_M W_N-\nabla_N W_M
\end{array}
\right.
\eea
remains invariant, we have to analyze the $(mn)$ components of the last equation and find the map between $W_m$ and ${\tilde W}_m$. 

Let us start with a $B$ field that satisfies the constraint (\ref{BfieldKV}) in the original frame. In particular this implies 
\bea
\nabla_m W_n-\nabla_n W_m=H_{mna}V^a+H_{mnz}V^z=
[d{\hat B}+\frac{1}{2}d({\tilde A}\wedge A)]_{mna}V^a+
{\tilde F}_{mn}V^z.
\eea
Assuming that the counterpart of this relation after T duality is also satisfied, we can subtract it from the last relation to find
\bea
&&\nabla_m (W_n-{\tilde W}_n)-
\nabla_n (W_m-{\tilde W}_m)=[d({\tilde A}\wedge A)]_{mna}V^a+
{\tilde F}_{mn}V^z-F_{mn}{\tilde V}^z\nonumber\\
&&\qquad={\tilde F}_{mn}[e^{-C}V_z-A_aV^a]-
F_{mn}[e^C{\tilde V}_z-{\tilde A}_a V^a]+[d({\tilde A}\wedge A)]_{mna}V^a\nonumber\\
&&\qquad={\tilde F}_{mn}e^{-C}V_z-F_{mn}e^C{\tilde V}_z-
[{\tilde F}_{ma}A_n-F_{ma}{\tilde A}_n-(m\leftrightarrow n)]V^a.
\eea
Using the last equation in (\ref{Oct3}) and its counterpart after T duality, we can simplify the last bracket:
\bea
&&\nabla_m (W_n-{\tilde W}_n)-
\nabla_n (W_m-{\tilde W}_m)\nonumber\\
&&\qquad={\tilde F}_{mn}e^{-C}V_z-F_{mn}e^C{\tilde V}_z-
[\d_m(e^{C}{\tilde V}_z)A_n-\d_m(e^{-C}V_z){\tilde A}_n-(m\leftrightarrow n)]
\nonumber\\
&&\qquad=\d_m[{\tilde A}_n e^{-C}V_z-A_n e^C{\tilde V}_z]-
\d_n[{\tilde A}_m e^{-C}V_z-A_m e^C{\tilde V}_z].
\eea
We conclude that the system (\ref{KVArray}) remains invariant under T duality if the standard rules (\ref{TdualityMapBdy}) are supplemented by 
\bea
&&{\tilde V}^a=V^a,\quad {\tilde W}_z= -e^{-C}V_z,\quad 
{\tilde V}_z= -e^{-C}W_z,\nonumber\\
&&{\tilde W}_n=W_n-{\tilde A}_n e^{-C}V_z-A_n W_z+\d_n f,
\eea
where $f$ is an arbitrary function. The last line can also be written as
\bea
{\tilde W}^n=W^n+{\hat g}^{na}\d_a f,
\eea
and the transformation law can be made symmetric between $V$ and $W$ by setting $f=0$.

\subsection{Dimensional reduction of the Killing tensor equation}
\label{AppDRKT4}

Next we look at the equation for the Killing tensor:
\bea\label{AppKTEq}
M_{MNP}=0,\quad M_{MNP}\equiv\nabla_M K_{NP}+\nabla_N K_{MP}+\nabla_P K_{MN}, 
\quad K_{MN}=K_{NM}.
\eea
Assuming that geometry (\ref{DimRedSetup}) does not have a $B$ field and that all components of $K_{MN}$ are $z$--independent, we can use (\ref{DimRedL}) to perform dimensional reduction along $z$ direction:
\bea\label{DimRedKeqn}
M_{zzz}&=&{K^a}_z\d_a e^C,\nonumber\\
{M_{zz}}^p&=&{K^{an}}\d_a e^C+2{\hat g}^{pb}e^C F_{ab}{K_z}^a-
2{\hat g}^{pa}K_{zz}\d_a C+{\hat g}^{pa}\d_a K_{zz},\\
M^{mn}{}_z&=&{\hat\nabla}^m {K_z}^n+{\hat\nabla}^n {K_z}^m
+[{\hat g}^{mb}K^{an}+{\hat g}^{nb}K^{ma}]e^C F_{ab}-
[{\hat g}^{ma}{K_z}^n+{\hat g}^{na}{K_z}^m]\d_a C,\nonumber\\
M^{mnp}&=&{\hat \nabla}^m K^{np}+{\hat \nabla}^n K^{mp}+{\hat \nabla}^p K^{mn}.\nonumber
\eea
and match equations for the Killing tensor before and after the duality:
\bea\label{AppAllKTEqs}
\begin{array}{|r|l|l|}
\hline
zzz&K_z^{\ t}\d_t e^C=0&\tilde K_z{}^t\d_t e^{-C}=0\Big.\\
zzp&2{\hat g}^{pa}F_{ba}K_z^{\ b}e^{-C}= \d_a e^{-C} K^{ap}- 
{\hat g}^{pa}\d_a (e^{-2C}K_{zz})&
\d_a e^C \tilde K^{ap}-{\hat g}^{pa}\d_a(e^{2C}\tilde K_{zz})=0\Big.\\
mnz&{\hat g}^{ma}\left[\hat\nabla_a(e^{-C}K^n{}_{z})+F_{ba}K^{nb}\right]+(m\leftrightarrow n)=0&
\hat\nabla^m(e^{C}\tilde K^n{}_{z})+(m\leftrightarrow n)=0\Big.\\
mnp&{\hat\nabla}^mK^{np}+{\hat\nabla}^nK^{mp}+{\hat\nabla}^p K^{mn}=0&
{\hat\nabla}^m\tilde K^{np}+{\hat\nabla}^n\tilde K^{mp}+{\hat\nabla}^p\tilde K^{mn}=0
\Big.
\\
\hline
\end{array}\nonumber
\eea
From $mnp$ components we obtain 
\bea
\tilde K^{mn}=K^{mn}.
\eea
Next we rewrite the $(mnz)$ components before T duality using the relation 
${\tilde H}_{mnz}=F_{mn}$:
\bea
g^{ma}\left[{\tilde H}_{abz}K^{nb}-{\hat\nabla}_a (e^{-C}K^n_{\ z})\right]+
(m\leftrightarrow n)=0.
\eea
Using the general reduction (\ref{DimRedL}) {\it after} duality, we find
\bea
{\tilde\nabla}^m {L_z}^n&=&{\hat\nabla}^m {L_z}^n+\frac{1}{2}g^{ma}{L_z}^n \d_a C
\quad\Rightarrow\quad
{\hat\nabla}^m {L_z}^n=e^{C/2}{\tilde\nabla}^m [e^{-C/2}{L_z}^n],\nonumber
\eea
and applying this relation to ${L_z}^n=e^{-C/2}K^n_{\ z}$, we find a constraint on the Kalb--Ramond field after duality.\footnote{Recall that ${\tilde K}^{mn}={K}^{mn}$, 
so we can write the left hand side of (\ref{addGuee}) in terms of dual variables.}
\bea\label{addGuee}
{\tilde g}^{ma}{\tilde H}_{abz}{\tilde K}^{nb}+
{\tilde g}^{na}{\tilde H}_{abz}{\tilde K}^{mb}=e^{C/2}{\tilde\nabla}^m [e^{-C/2}K^n_{\ z}]+
e^{C/2}{\tilde\nabla}^n [e^{-C/2}K^m_{\ \ z}]
\eea
The only covariant extension of this equation for the $B$--field is\footnote{As a consistency check, we note that the trivial 
Killing tensor $\tilde K_{MN}=g_{MN}$ does not give any restriction on the $B$ field.}
\bea\label{AppBEqGuessFinal}
{\tilde H}_{AMP}\tilde K_N{}^A+{\tilde H}_{ANP}\tilde K_M{}^A=e^{C/2}{\tilde\nabla}_M[e^{-C/2}{\tilde W}_{NP}]+e^{C/2}{\tilde\nabla}_N[e^{-C/2}{\tilde W}_{MP}].
\eea
Equation (\ref{addGuee}) recovers the $(mnz)$ component of this constraint, but other components require additional analysis. Here we just mention that the constraint (\ref{AppBEqGuessFinal}) admits a special solution
\be\label{SpecSolnConstr}
\begin{aligned}
&\tilde K^n{}_z=0, \quad W^n{}_z=-e^{-C}K^n{}_z, \quad W_{zz}=0,\quad W_{mn}=0,\\
&F_{pa}K_n{}^a-F_{na}K_p{}^a+2F_{np}\tilde K_z{}^z=\tilde \nabla_n (-e^{-C}K_{zp})-\tilde \nabla_p (-e^{-C}K_{zn}),\\
&\d_a e^C g_{pb}K^{ab}-\d_p(e^{2C}\tilde K_{zz})=0.
\end{aligned}
\ee
To summarize, we found that T duality maps equations for KT to a combination of the same equation and a constraint on the $B$ field:
\bea\label{AppClaimKT}
\nabla_{(M} K_{NP)}=0 
\Longleftrightarrow\left\{
\begin{array}{l}
 \nabla_{(M} \tilde K_{NP)}=0 ,\\
H_{A P (M}\tilde K_{N)}{}^A+e^{C/2}\nabla_{(M}[e^{-C/2}W_{N)P}]=0.
\end{array}\right.
\eea

\subsubsection{Lie derivative along KT}
\label{AppLieDer}

Note that the third equation in (\ref{DimRedKeqn}) has an interesting interpretation in terms of Lie derivatives. To see this, we rewrite the $M^{mn}{}_{z}$ as
\bea
0&=&{\hat g}^{ma}\left[{\hat\nabla}_a(e^{-C}K^n{}_{z})+
({\hat\nabla}_b A_a-{\hat\nabla}_a A_b)K^{nb}\right]+(m\leftrightarrow n)\\
&=&
g^{ma}\left[{\tilde\nabla}_a 
(e^{-C}K^n_{\ z}-A_b K^{nb})+{\hat\nabla}_b A_a K^{nb}+A_b{\hat\nabla}_a K^{nb}\right]+
(m\leftrightarrow n)\nonumber\\
&=&
\left[{{\hat\nabla}}^m
(e^{-C}K^n_{\ z}-A_b K^{nb})+
(m\leftrightarrow n)\right]+{\hat\nabla}_b A^m K^{nb}+
{\hat\nabla}_b A^n K^{mb}-A_b{\hat\nabla}^b K^{mn}.\nonumber
\eea
At the final step we used the equation for the Killing tensor. The last equation implies 
an interesting relation for the Killing tensor
\bea\label{AppLieTenA}
{
\nabla_a A^m K^{na}+
\nabla_a A^n K^{ma}-A_a\nabla^a K^{mn}=\nabla^m W^n+\nabla^n W^m,}
\eea
which generalizes the expression (\ref{KillDerBfield}) involving the Lie derivative of the $B$ field along a Killing vector. Specifically, rewriting (\ref{AppLieTenA}) as
\bea\label{AppLieTenB}
A_a\nabla^a K_{mn}-
K_m^{\ a}\nabla_a A_n-K_n^{\ a}\nabla_a A_m=-\nabla^m W^n-\nabla^n W^m
\eea
we are tempted to interpret the left--hand side of the last equation as a ``Lie derivative of $A_m$ along a Killing tensor". Although the analogy with the usual Lie derivative has limitations (for example, the rank of the lhs is higher than the rank of $A_m$), equation (\ref{AppLieTenB}) does reduce to the combination of Lie derivative if Killing tensor has a form $K^{mn}=\la^m\la^n$:
\bea
lhs&=&\la^n\la^a\nabla_a A^m +
\nabla_a A^n \la^m\la^a-A_r\nabla^a (\la^m\la^n)
\nonumber\\
&=&\la^n\left[\la^a\nabla_a A^m-A_a\nabla^a\la^m\right]+
\la^m\left[\la^a\nabla_a A^n-A_a\nabla^a\la_n\right]
\nonumber\\
&=&
\la^n\left[\la^a\nabla_a A^m+A_a\nabla^m\la^a\right]+
\la^m\left[\la^a\nabla_a A^n+A_a\nabla_n\la^a\right]\\
&=&\la^n{\cal L}_\la A^m+\la^m{\cal L}_\la A^n.\nonumber
\eea
It would be interesting to  investigate the relation between (\ref{AppLieTenB}) and Lie derivatives further.


\subsection{Extension to CKT}
\label{AppCKT}

In this appendix our results are extended to the conformal Killing tensor assuming that the original geometry has vanishing $B$ field and that there is no mixture between $z$ and other coordinates.  Starting with equation for the CKT,
\be
3\nabla_{(M} {\cal K}_{NP)}=g_{MN}W_P+g_{MP}W_N+g_{NP}W_M,
\ee
and performing reduction with $A_m=0$, we find 
\bea
zzz:&&\d_z {\cal K}_{zz}+{{\cal K}_z}^m\d_m e^C=e^C W_z\nonumber\\
mnz:&&\left[{\tilde\nabla}^m{{\cal K}^n}_z-{{\cal K}^n}_z\nabla^n C\right]+
(m\leftrightarrow n)+\d_z {\cal K}^{mn}=W_z g^{mn}
\nonumber\\
zzp:&&{\cal K}^{ap}\d_a e^C-2{\cal K}_{zz}\nabla^p C+\nabla^p {\cal K}_{zz}+
2\d_z {{\cal K}_z}^p=e^C W^p\\
mnp:&&\nabla^m{\cal K}^{np}+\nabla^n{\cal K}^{mp}+
\nabla^n{\cal K}^{mp}=W^m g^{np}+W^n g^{mp}+W^p g^{nm}
\nonumber
\eea
Motivated by the discussion of the CKV in subsection \ref{SubsectionCKV} we allowed the components of CKT to depend on the $z$ coordinate. We will assume that $\d_z=0$ before T duality, but the $z$--dependence appears afterward. 

To satisfy the $(mnp)$ equations before and after duality, we require
\bea
{\tilde W}^p=W^p,\qquad {\tilde {\cal K}}^{mn}={\cal K}^{mn}. 
\eea
Comparing $(mnz)$ equations before and after duality, and taking into account that 
$\d_z {\cal K}^{mn}=0$, we  set
\bea
{\tilde W}_z=e^{-2C}W_z+2ve^{-C},\quad 
{\tilde {\cal K}^n}_{\ \ z}=e^{-2C}{{\cal K}^n}_z+e^{-C}{\cal V}^n,
\eea
where ${\cal V}^n$ is a CKV with conformal factor $v$. 
Then $(zzz)$ equation after T duality gives
\bea
\d_z {\tilde {\cal K}}_{zz}&=&
2e^{-3C} W_z+e^{-2C}({\cal V}^a\d_a C+2v),\nonumber\\
{\tilde {\cal K}}_{zz}&=&
z\left[2e^{-3C} W_z+e^{-2C}({\cal V}^a\d_a C+2v)\right]+N_{zz},
\eea
where $N_{zz}$ is $z$--independent ``integration constant".

Comparing the $(zzp)$ equations before and after duality,
\bea\label{Jul25}
&&e^{-C}\nabla^p 
(e^{2C}\tilde {\cal K}_{zz})+e^C\nabla^p (e^{-2C}{\cal K}_{zz})+
2e^{C}\d_z {\tilde{\cal K}_z}^{\ p}=2W^p,\nonumber\\
&&e^{-C}\nabla^p 
(e^{2C}\tilde {\cal K}_{zz})-e^C\nabla^p (e^{-2C}{\cal K}_{zz})+
2e^{C}\d_z {\tilde {\cal K}_z}^{\ p}=2{\cal K}^{ap}\d_a C,
\eea
and assuming that $\d_z {\cal V}^n=0$ (and 
thus $\d_z {\tilde {\cal K}_z}^{\ p}=0$), we conclude that $z$--dependence disappears from the last two equations if
\bea
&&\d_p\left[2e^{-C} W_z+({\cal V}^a\d_a C+2v)\right]=0,
\nonumber\\
\label{CKTConstr1}
&&\d_p\left[2{\tilde W}^z+({\cal V}^a\d_a C-2v)\right]=0.
\eea
The last equation is a counterpart of the homothety condition for the CKV. The remaining equations are (\ref{Jul25}):
\bea\label{CKTConstr2}
&&e^{-C}\nabla^p 
(e^{2C}N_{zz})+e^C\nabla^p (e^{-2C}{\cal K}_{zz})=2W^p,\nonumber\\
&&e^{-C}\nabla^p 
(e^{2C}N_{zz})-e^C\nabla^p (e^{-2C}{\cal K}_{zz})=2{\cal K}^{ap}\d_a C.
\eea
To summarize, we have to satisfy two constraints (\ref{CKTConstr1}) and (\ref{CKTConstr2}) on 
constraints on $W^p$ and ${\cal K}^{tp}\d_t C$, then all equations can be solved.


\subsection{mKTY equation and the constraint on the $B$ field}
\label{AppBeqKYT}

This subsection is dedicated to the derivation of our main result: invariance of the modified Killing--Yano (mKYT) equation (\ref{KYTBfield}),
\bea\label{mKYTapp}
\nabla_M Y_{NP}+\h H_{MPA}g^{AB}Y_{NB}+(M \leftrightarrow N)=0,
\eea 
under the  T--duality transformations
. Starting with a geometry (\ref{AppDimRedBdy}) that admits a modified 
Killing--Yano tensor (mKYT) satisfying 
(\ref{mKYTapp}), we will show that the system (\ref{SetupAfter}) related to (\ref{AppDimRedBdy}) by T duality 
admits a mKYT ${\tilde Y}_{MN}$ with components
\bea\label{TransYaa}
{\tilde Y}^{mn}=Y^{mn},\qquad  {\tilde Y}_z{}^s= e^{-C}Y_z{}^s.
\eea

To demonstrate the invariance of the mKYT equation, we perform a dimensional reduction of
\bea
T_{MNP}\equiv \nabla_M Y_{NP}+\h H_{MPA}g^{AB}Y_{NB}+(M \leftrightarrow N)\,.
\eea
As discussed in section \ref{AppDimRed1}, it is sufficient to look only at components with covariant indices $z$ and contravariant indices $(m,n\dots)$, and since tensor $T_{MNP}$ is symmetric in the first two indices, we have to analyze five types of  components\footnote{Notice that $T_{zzz}=0$.}:
\bea\label{TensZnmp}
{T_{zz}}^p,\quad {T^m}_{zz},\quad {T^{mn}}_z,\quad {T_z}^{mp},\quad T^{mnp}.
\eea
and demonstrate that they are invariant under the T duality (\ref{TdualityMap}).

\bigskip

\noindent
{\bf 1. $(zzp)$ component.}

The first component in (\ref{TensZnmp}) is
\bea
{T_{zz}}^p&=&2\nabla_z Y_{z}{}^p+H_z{}^p{}_AY_z{}^{A}=
\d_ae^C Y^{ap}+g^{pa}e^C F_{ba}Y_z{}^b+H_{zsa}g^{sp}Y_z{}^a{}\\
&&=\d_ae^C Y^{ap}+g^{pa}e^C F_{ba}Y_z{}^b+\tilde F_{ab}g^{ap}Y_z{}^b.
\nonumber
\eea
Here we used expression (\ref{DimRedL}) for the covariant derivative $\nabla_z L_{z}{}^p$ of 
an arbitrary rank--2 tensor. Rewriting the last equation  as
\bea\label{TzzpOne}
e^{-C}{T_{zz}}^p=\d_aC Y^{ap}+g^{pa}F_{ba}Y_z{}^b-g^{ap}e^{-C}\tilde F_{ba}Y_z{}^b,
\eea
we observe that is it invariant under the T duality transformation (\ref{TdualityMap}) if we require that 
\bea
Y^{mn}\rightarrow Y^{mn},\qquad Y_z{}^m\rightarrow e^{C}Y_z{}^m.
\eea
To keep track of the last rescaling in the remaining equations, we introduce
\bea\label{hatY}
{\hat Y}_z{}^m\equiv e^{-C/2}Y_z{}^m
\eea
that remains invariant under T duality. Then equation (\ref{TzzpOne}) becomes more symmetric:
\bea\label{TzzpTwo}
{T_{zz}}^p=\d_aC Y^{ap}+g^{pa}F_{ba}e^{C/2}{\hat Y}_z{}^b-g^{ap}e^{-C/2}\tilde F_{ba}{\hat Y}_z{}^b
\eea
and invariance of equation ${T_{zz}}^p=0$ under T duality becomes explicit.
\bigskip 

\noindent
{\bf 2. $(mzz)$ component.}

The second component in (\ref{TensZnmp}),
\bea\label{Tmzz}
{T^m}_{zz}=\nabla_z {Y^m}_z+\frac{1}{2}{H^m}_{zA}{Y^A}_z=-
\frac{1}{2}{T_{zz}}^m,
\eea
is also invariant under T duality.

\bigskip 

\noindent
{\bf 3. $(mnz)$ component.}

The third component of (\ref{TensZnmp}) is
\bea
{T^{mn}}_z&=&\nabla^m {Y^n}_z+\h {H^m}_{zA}Y^{An}+(m \leftrightarrow n)
\nonumber\\
&=&\hat\nabla^m Y^n{}_z-\h g^{ma}\d_a C Y^n{}_z-\h g^{ma} e^C F_{ar} Y^{nr}+\h g^{ma}H_{azb}Y^{nb}+ (m\leftrightarrow n)\nonumber
\eea
Here we used (\ref{DimRedL}) to express $\nabla^m Y^n{}_z$ in terms of the covariant derivative $\hat\nabla^m Y^n{}_z$ in the reduced metric ${\hat g}_{mn}$. 
Rewriting the last equation in terms of the field strengths $(F_{ij},{\tilde F}_{ij})$,
\bea\label{TmnzOne}
{T^{mn}}_z=
{\hat\nabla}^m Y^n{}_z-\h g^{ma}\d_a C Y^n{}_z-\h g^{ma} \left[e^C F_{ar}+
\tilde F_{ab}\right]Y^{nb}+ (m\leftrightarrow n),
\eea
and expressing the result in terms of ${\hat Y}$ defined by (\ref{hatY}), we find
\bea\label{TmnzTwo}
{T^{mn}}_z=
{\hat\nabla}^m [{\hat Y}^n{}_z]-\h g^{ma} \left[e^{C/2} F_{ab}+
e^{-C/2}\tilde F_{ab}\right]Y^{nb}+ (m\leftrightarrow n).
\eea
Clearly this expression is invariant under T duality. 

\bigskip

\noindent
{\bf 4. $(zmp)$ component.}

To simplify the fourth component of (\ref{TensZnmp}) we again use (\ref{DimRedL}):
\bea
{T_z}^{mp}&=&\nabla^m Y_{z}{}^p+\nabla_z Y^{mp}+\h H^{mpA}Y_{Az}+
\h {H_z}^{pA}{Y^m}_A\nonumber\\
 &=&{\hat\nabla}^m Y_{z}{}^p-\h g^{ma}\d_aC Y_{z}{}^p-\h g^{ma}e^CF_{ab}Y^{bp}
\nonumber\\
&&-\h g^{ma}\d_aCY_z{}^p+\h g^{ma}e^CF_{ba} Y^{bp}-\h g^{pa}\d_aCY^m{}_z+\h g^{pa}e^CF_{ba}Y^{mb}\nonumber\\
&&+\h {H^{mp}}_a{Y^a}_{z}+\h {H^{mp}}_z{Y^z}_{z}+\h {{H_z}^{p}}_{a}{Y}^{ma}
\nonumber
\eea
Using expressions
\bea
{H^{mp}}_s{Y^s}_{z}&=&g^{ma}g^{pb}\left[H_{abs}-A_a H_{zbs}-A_b H_{azs}\right]{Y^s}_{z}=
g^{ma}g^{pb}\left[H_{abs}-A_a {\tilde F}_{bs}-A_b {\tilde F}_{sa}\right]{Y^s}_{z}\,,
\nonumber\\
{H^{mp}}_z{Y^z}_{z}&=&g^{ma}g^{pb}{\tilde F}_{ab}~\frac{1}{g_{zz}}\left[Y_{zz}-g_{zs}{Y^s}_z\right]=-g^{ma}g^{pb}A_s{\tilde F}_{ab}{Y^s}_z\,,
\nonumber\\
{{H_z}^{p}}_{s}{Y}^{ms}&=&g^{pa}{\tilde F}_{as}{Y}^{ms}\,,\nonumber
\eea
we find
\bea\label{TzmpOne}
{T_z}^{mp}&=&{\hat\nabla}^m Y_{z}{}^p-g^{ms}\d_sC Y_{z}{}^p+g^{ms}e^CF_{sr}Y^{pr}
-\h g^{ps}\d_sCY^m{}_z+\h g^{ps}e^CF_{rs}Y^{mr}\nonumber\\
&&+\frac{1}{2}g^{ma}g^{pb}\left[H-A\wedge {\tilde F}\right]_{abs}{Y^s}_{z}+
\frac{1}{2}g^{pa}{\tilde F}_{as}{Y}^{ms}
\eea
Recalling the expression for $H$ in terms of duality--invariant ${\hat B}$ (see 
(\ref{DimRedSetup}), (\ref{SetupAfter})),
\bea
H=d{\hat B}+\frac{1}{2}({\tilde A}\wedge A),
\eea
we observe that 
\bea\label{HhatDef}
{\hat H}\equiv H-A\wedge {\tilde F}=d{\hat B}-\frac{1}{2}[A\wedge \tilde F+{\tilde A}\wedge F]
\eea
is invariant under T duality. To demonstrate the invariance of (\ref{TzmpOne}), we rewrite that expression as
\bea\label{TzmpTwo}
{T_z}^{mp}&=&{\hat\nabla}^m {\hat Y}_{z}{}^p
+\frac{1}{2}g^{ms}(e^{C/2}F_{sr}+e^{-C/2}{\tilde F}_{sr})Y^{pr}+\frac{1}{2}g^{ma}g^{pb}{\hat H}_{abs}{{\hat Y}^s}{}_{z}\nonumber\\
&&+\frac{1}{2}\left[g^{ms}\d_sC {\hat Y}^p{}_{z}-g^{ps}\d_sC{\hat Y}^m{}_z+
g^{ps}G_{sr}Y^{rm}-g^{ms}G_{sr}Y^{rp}\right]\\
G_{sr}&\equiv& e^{C/2}F_{sr}-e^{-C/2}{\tilde F}_{sr}.\nonumber
\eea
The first line of this equation is invariant under T--duality, while the second line changes sign. Thus to make ${T_z}^{mp}$ invariant, we must impose a constraint on $F_{mn}$ and 
${\tilde F}_{mn}$:
\bea\label{ConstrBcomp}
{\tilde S}_{}^{mp}&\equiv& 
\left[g^{ms}\d_sC {\hat Y}^p{}_{z}-g^{ps}\d_sC{\hat Y}^m{}_z+
g^{ps}G_{sr}Y^{rm}-g^{ms}G_{sr}Y^{rp}\right]=0,\\
G_{sr}&\equiv& e^{C/2}F_{sr}-e^{-C/2}{\tilde F}_{sr}\nonumber
\eea
The physical meaning of this constraint is discussed in section \ref{SectionModifiedKYT}. 

\bigskip

\noindent
{\bf 5. $(mnp)$ component.}

The final component of (\ref{TensZnmp}) gives\footnote{We used (\ref{DimRedL}) to express 
$\nabla^m Y^{np}$ in terms of $\hat\nabla^m Y^{np}$.}
\bea\label{Sept29}
T^{mnp}=\hat\nabla^m Y^{np} + \h F^{mp} Y^n{}_z + \h H^{mp}{}_{A}Y^{nA}+ (m \leftrightarrow n).
\eea
Simplifying the term that involves flux
\bea
&&H^{mp}{}_{D}Y^{nD}=g^{mA}g^{pB} H_{AB D}Y^{nD}\nonumber\\
&&=g^{mz}g^{pb}H_{zbc}Y^{nc}+g^{ma}g^{pz} H_{azc}Y^{nc}+g^{ma}g^{pb} H_{abz}Y^{nz}+g^{ma}g^{pb} H_{abc}Y^{nc}\nonumber\\
&&=-A^mg^{pb}\tilde F_{bc}Y^{nc}+g^{ma}A^{p} \tilde F_{ac}Y^{nc}+g^{ma}g^{pb} \tilde F_{ab}(e^{-C}Y^n{}_z-A_cY^{nc})+g^{ma}g^{pb} H_{abc}Y^{nc}\nonumber\\
&&=g^{ma}g^{pb}e^{-C} \tilde F_{ab}Y^n{}_z+
g^{ma}g^{pb}Y^{nc}(H_{abc}-A_a\tilde F_{bc}+A_b \tilde F_{ac}-A_c \tilde F_{ab})
\nonumber\\
&&=g^{ma}g^{pb}e^{-C} \tilde F_{ab}Y^n{}_z+
g^{ma}g^{pb}Y^{nc}(H-A\wedge \tilde F)_{abc}
\eea
and recalling expression (\ref{HhatDef}) for the duality--invariant $\hat H$, we find
\bea
H^{mp}{}_{A}Y^{nA}=g^{ma}g^{pb}e^{-C} \tilde F_{ab}Y^n{}_z+
g^{ma}g^{pb}Y^{nc}{\hat H}_{abc}
\eea
Then equation (\ref{Sept29}) becomes
\bea\label{TmnpOne}
T^{mnp}=\tilde\nabla^m Y^{np} +\h g^{ma}g^{pb}\Big[[F_{ab}+e^{-C} \tilde F_{ab}] Y^n{}_z +Y^{nc}{\hat H}_{abc}\Big]
+ 
(m \leftrightarrow n),
\eea
and rewriting it in terms of ${\hat Y}$
\bea\label{TmnpTwo}
T^{mnp}=\tilde\nabla^m Y^{np} +\h g^{ma}g^{pb}\Big[[
e^{C/2}F_{ab}+e^{-C/2} \tilde F_{ab}] {\hat Y}^n{}_z +Y^{nc}{\hat H}_{abc}\Big]
+ 
(m \leftrightarrow n)
\eea
make the invariance under T duality explicit. 

The constraint (\ref{ConstrBcomp}) treat $z$ direction in a special way, but it would be nice to write it in a covariant form. This can be accomplished in an important special case when $F_{mn}=0$, which implies the T--dual configuration has no $B$ field. Then (\ref{ConstrBcomp}) reduces to
\be\label{AppmKYTMismatch}
H_{naz}g^{ab}Y_{bp}+H_{zpa}g^{ab}Y_{nb}+[\d_n C\check{Y}_{zp}-\d_p C\check{Y}_{zn}]=0.
\ee
The unique covariant form of this relation is
\be\label{AppRestBKYTGuess}
\begin{aligned}
H_{AMP}\tilde Y_N{}^A-H_{ANP}\tilde Y_M{}^A-H_{AMN}\tilde Y_P{}^A-(\d_M C \tilde Y_{NP}-\d_N C \tilde Y_{MP}-\d_P C \tilde Y_{N M})\\=\d_M W_{NP}-\d_N W_{MP}-\d_P W_{N M},
\end{aligned}
\ee
where $W$ is auxiliary field introduced to satisfy the $mnp$ components of the last equation, which would be too restrictive otherwise.

\bigskip

To summarize, we have demonstrated that all independent components of $T_{MNP}$ given by 
(\ref{TensZnmp}) can be written in a way that makes invariance under T duality (\ref{TdualityMap}) very explicit (see (\ref{TzzpTwo}), (\ref{Tmzz}), (\ref{TmnzTwo}), (\ref{TzmpTwo}), (\ref{TmnpTwo})), as long as constraint (\ref{ConstrBcomp}) is satisfied. 

\subsubsection{KT from mKYT}

Finally we show that the modified Killing-Yano equation reduces to a standard Killing tensor equation. To do so we begin with the modified equation for KYT 
\bea
\nabla_M Y_{NP}+\nabla_N Y_{MP}+
\frac{1}{2}H_{MPA}{Y_N}^A+
\frac{1}{2}H_{NPA}{Y_M}^A=0
\eea
and construct various combinations:
\bea
&&{Y_B}^P\left[\nabla_M Y_{NP}+\nabla_N Y_{MP}+
\frac{1}{2}H_{MPA}{Y_N}^A+
\frac{1}{2}H_{NPA}{Y_M}^A\right]=0,\nonumber\\
&&{Y_N}^P\left[\nabla_M Y_{BP}+\nabla_B Y_{MP}+
\frac{1}{2}H_{MPA}{Y_B}^A+
\frac{1}{2}H_{BPA}{Y_M}^A\right]=0,\nonumber\\
&&{Y_M}^P\left[\nabla_B Y_{NP}+\nabla_N Y_{BP}+
\frac{1}{2}H_{BPA}{Y_N}^A+
\frac{1}{2}H_{NPA}{Y_B}^A\right]=0.\nonumber
\eea
Adding these equations, we find the standard Killing tensor equation
\bea
&&\nabla_M K_{BN}+\nabla_N K_{MB}+
\nabla_B K_{MN}+
\frac{1}{2}\left[H_{MPA}({Y_N}^A{Y_B}^P+{Y_N}^P{Y_B}^A)+
perm\right]=0,\nonumber\\
&&\nabla_M K_{BN}+\nabla_N K_{MB}+
\nabla_B K_{MN}=0.
\eea
Here 
\bea
K_{MN}\equiv {Y_M}^A Y_{NA}.
\eea
To summarize, we demonstrated that the standard relation ``KT=KYT$^2$'' persists for the modified Killing--Yano tensors as well.  




\section{The restrictions on the $B$ field from the F1 $\to$ NS5 duality chain}
\label{AppCondBNS5}
\renewcommand{\theequation}{E.\arabic{equation}}
\setcounter{equation}{0}

In Section \ref{KTOdd} we derived the restrictions on the metric and the B-field \eqref{SepCondB}
by requiring separability of the Hamilton--Jacobi equation along all $O(d,d)$ orbits which start with a pure metric. 
In this section we will extend those results to $O(d,d)$ orbits starting with NS5 solutions (thus generating the 
entire F1--NS5--P family) and show that separability leads to additional constraints (\ref{Hcond2f}), (\ref{Hcond2fields}), (\ref{Hcond2sep}) on the $B$ field. 

We start with conditions on the $B$ field \eqref{Hcond1} and \eqref{Hcond}
\bea\label{AppHconds}
&&\d_x\d_y (fg_{ab})-fg^{MN}H_{y aM}H_{xNb}-fg^{MN}H_{xaM}H_{y Nb}=0,\\
&&\d_y(f g^{mM})H_{xMb}+\d_x (f g^{mM}) H_{y Mb}+f g^{mM}\d_xH_{y Mb}=0.\nonumber
\eea
Next we consider the first equation and require this constraint to hold on the entire $O(d,d)$ orbit containing NS5 brane. Comparing  
\eqref{AppHconds} for F1 orbit with its counterpart for NS5, we find
\bea\label{AppHcond21}
&&\d_x\d_y (fg_{ab})-fH_{y aM} H_{x}{}^M{}_b-fH_{xaM} H_{y}{}^M{}_b=0,\\
&&\d_y \d_x [F^2fg_{ab}]-fF (H^{(NS5)}_{y aM} H^{(NS5)}_{x}{}^M{}_{b}+H^{(NS5)}_{xaM} H^{(NS5)}_{y}{}^M{}_{b})=0.\nonumber
\eea
Here we used the transformation law for the metric and defined a convenient function $F$
\bea\label{AppF1NS5trans}
g_{MN}^{NS5}=F g_{MN}^{F1}, \quad f^{NS5}=F f^{F1},\quad F\equiv \sqrt{\mbox{det}G}\,\mbox{det}H.
\eea
Expressions without superscript in (\ref{AppHcond21}) refer to the fundamental string. The field strengths of the Kalb--Ramond fields for NS5 and F1 systems are related by the electric--magnetic duality
\bea
H^{(NS5)}_{y aM}=\frac{1}{7!} e^{2\Phi_{NS5}}e_{y a M}{}^{x NP z_1z_2z_3z_4} G^{(NS5)}_{x NP z_1z_2z_3z_4}=
\frac{1}{7!} e^{2\Phi_{NS5}}e_{y a M}{}^{xNP z_1z_2z_3z_4} H^{(F1)}_{x NP}.
\eea
In particular, the product of the field strengths is
\bea\label{AppEqH2}
H^{(NS5)}_{yaM}H^{(NS5)}_{x}{}^M{}_b&=&
\frac{e^{4\Phi_{NS5}}}{(3!)^2}e_{ya M}{}^{x NP} e_{x}{}^M{}_b{}^{y}{}_A{}^{B} H_{x NP} H_{y}{}^A {}_{B}=\frac{e^{4\Phi_{NS5}}}{(2!)^2}e_{a M}{}^{NP} e^M{}_{bA}{}^{B} H_{x NP} H_{y}{}^A {}_{B} \nonumber\\
&=&e^{4\Phi_{NS5}}\left[-H_{xbM}H_y{}^M{}_a-\frac{1}{2}H_{xMN}H_y{}^{MN}g_{ab}^{(NS5)}\right]_{g^{(NS5)}}.
\eea
In the last line all indices are contracted with $g_{MN}^{(NS5)}$. In terms of the F1 metric we find
\bea
H^{(NS5)}_{y aM} H^{(NS5)}_{x}{}^M{}_{b}=
F\left[-H_{xbM}H_y{}^M{}_a-\frac{1}{2}H_{xMN}H_y{}^{MN}
g_{ab}\right].
\eea
We can now rewrite the conditions \eqref{AppHcond21} in terms of the F1 fields:
\bea\label{AppHcond22}
&&\d_x\d_y(f g_{ab})-fH_{y aM} H_{x}{}^M{}_b-fH_{xaM} H_{y}{}^M{}_b=0,\\
&&\d_x\d_y\Big[f g_{ab}F^2 \Big]+fF^2(H_{y aM}H_x{}^M{}_b+(x\to y)+H_{xMN}H_y{}^{MN}g_{ab})=0.\nonumber
\eea
Subtracting the first equation from the second one we get the relation
\bea
\d_x\d_y\Big[f g_{ab}F^2 \Big]+F^2(\d_x\d_y[f g_{ab}]+fH_{xMN}H_y{}^{MN}g_{ab})=0,\nonumber
\eea 
which can be rewritten as
\bea\label{AppHcond2f}
\d_x\d_y[g_{ab}fF]+\frac{f}{F}g_{ab}\Big[\d_x\ln F \d_y\ln F+ \frac{1}{2}H_{xMN}H_y{}^{MN}\Big]=0.
\eea
Remarkably in all our examples the two terms entering this expression vanish separately, so we conjecture that this will always happen for the systems obtained from fundamental stings via the  duality chain, although we will not attempt to prove this fact. Recalling that  $F=e^{-2\Phi_{F1}}$,  we conclude that vanishing of the first term in (\ref{AppHcond2f}) implies separation of the duality--invariant expression
\bea
g^{(F1)}_{ab}f^{(F1)}e^{-2\Phi_{F1}}=g^{(NS5)}_{ab}f^{(NS5)}e^{-2\Phi_{NS5}}.
\eea
In other words vanishing of the first term in \eqref{AppHcond2f} can be written as 
\bea\label{AppHcond2sep}
\d_x\d_y\left[g_{ab}fe^{-2\Phi}\right]=0
\eea
in {\it every frame} containing only NS--NS fields. Vanishing of the second term in \eqref{AppHcond2f} gives the relation in the F1 frame
\bea\label{AppHcond2fields}
\d_x\Phi \d_y\Phi+\frac{1}{8} H_{xMN}H_y{}^{MN}=0.
\eea

Now we consider the the second condition in \eqref{AppHconds}
\bea
\d_y(f g^{mM})H_{xMb}+\d_x (f g^{mM}) H_{y Mb}+f g^{mM}\d_xH_{y Mb}=0.
\eea
Writing it for F1 and for NS5, and using \eqref{AppF1NS5trans} we get
\bea\label{AppHcond1f}
&&\d_y(f g^{mM})H_{xMb}+\d_x (f g^{mM}) H_{y Mb}+f g^{mM}\d_xH_{y Mb}=0,\\
&&\d_y(f g^{mM})\tilde{H}_{xMb}+\d_x (f g^{mM}) \tilde{H}_{y Mb}+\frac{f}{F} g^{mM}\d_x(F\tilde{H}_{y Mb})=0.\nonumber
\eea
Here $\tilde{H}=\star_6H^{(F1)}$ is six--dimensional Hodge dual of the field strength for F1. Note that the first equation (and its dual counterpart) can be written in two different ways (using $\d_xH_{y Mb}=\d_y H_{xMb}$). The difference gives equation of motion for the $B$ field
\bea
g^{mM}\Big[\d_x(e^{2\Phi_{NS5}}\tilde{H}_{y Mb}) -\d_y( e^{2\Phi_{NS5}}\tilde{H}_{x Mb}) \Big]=0,\quad e^{2\Phi_{NS5}}=\mbox{det}H\sqrt{\mbox{det}G}.
\eea

To summarize we have found two additional constrains \eqref{AppHcond2f}, \eqref{AppHcond1f} on the $B$ field that guarantee separability of F1--NS5. Remarkably in the studied examples the first condition decouples into two very simple equations - separation condition \eqref{AppHcond2sep} and the field equation \eqref{AppHcond2fields}.




\section{Modified KY tensor for the charged Myers--Perry black hole}
\label{AppChargedMP}
\renewcommand{\theequation}{F.\arabic{equation}}
\setcounter{equation}{0}

In section \ref{SecMPF1} we presented the modified Killing--Yano tensor for the charged counterpart of the Myers--Perry black hole. In this appendix we will outline the derivation of (\ref{ChMPHYT})--(\ref{ChMPframe}).

We begin with the original Myers--Perry metric and its Killing--Yano tensor written in terms of frames (\ref{AllFramesMP}) and apply the first two steps in the duality chain 
(\ref{DualChain}). The boost leads to replacements
\bea
\begin{array}{l}
dt\rightarrow \ch_\alpha dt+\sh_\alpha dy\\
dy\rightarrow  \ch_\alpha dy+\sh_\alpha dt\\
\end{array},\qquad
\begin{array}{l}
\d_t\rightarrow \ch_\alpha\d_t-\sh_\alpha\d_y\\
\d_y\rightarrow \ch_\alpha\d_y-\sh_\alpha\d_t
\end{array}
\eea
in the frames (\ref{AllFramesMP}), but it does not modify the expressions 
(\ref{Kub2}), (\ref{HInFrame}). T duality along $y$ direction leaves the contravariant components $g^{mn}={\hat g}^{mn}$ and $Y^{mn}$ invariant, so it is reasonable to assume that neither expressions (\ref{Kub2}), (\ref{HInFrame})
nor components of $e_A$ which don't involve $y$ are modified.  In other words, we will assume that after T duality the frames have the form
\bea\label{ChMPframePrm}
e_r&=&\sqrt{\frac{R-mr}{FR}}\d_r,\quad 
e_{x_i}=\sqrt{-\frac{4x_iH_i}{d_i(r^2-x_i)}}\d_{x_i},\quad
e_y=C_y\ch_\alpha\d_y-\sh_\alpha\d_t,
\nn
e_t&=&\sqrt{\frac{R^2}{FR(R-mr)}}\left[\ch_\alpha \d_t-C_t{\sh_\alpha}\d_y
-\sum_k\frac{a_k}{r^2+a_k^2}\d_{\phi_k}\right],\\
e_i&=&\sqrt{\frac{H_i}{d_i(r^2-x_i)}}\left[\ch_\alpha\d_t-C_i\sh_\alpha \d_y-\sum_k\frac{a_k}{x_i+a_k^2}\d_{\phi_k}
\right]\nonumber
\eea
with some functions $(C_y,C_t,C_i)$. This assumption will be justified by the explicit calculation that recovers transformation rules (\ref{DimRedSetup}), (\ref{SetupAfter}) and (\ref{TransYaa}) and determines the functions $(C_y,C_t,C_i)$.

We begin with recovering the relation ${\tilde g}^{ym}=0$, which must hold after T duality. Equations (\ref{ChMPframePrm}) give
\bea\label{TempGyM}
{\tilde g}^{ym}\d_m&=&
-C_y\ch_\alpha\sh_\alpha\d_t+
{\frac{R^2}{FR(R-mr)}}C_t{\sh_\alpha}\left[\ch_\alpha \d_t-
\sum_k\frac{a_k}{r^2+a_k^2}\d_{\phi_k}\right]\nn
&&-
\sum_i\left[{\frac{(-x_i)H_i}{d_i(r^2-x_i)}}C_i\sh_\alpha\left[\ch_\alpha\d_t-\sum_k\frac{a_k}{x_i+a_k^2}\d_{\phi_k}
\right]\right]=0.
\eea
Coefficients in front of $\d_t$ and all $\d_{\phi_k}$ must vanish, so we find $n$ equations for $(n+1)$ variables $(C_y,C_t,C_i)$, which are completely determined up to one overall factor. Thus it is sufficient to guess the solution and check the result. 
To determine the coefficients $(C_y,C_t,C_i)$ we set $m=0$ in the boosted frames before T duality, which can be extracted from (\ref{ChMPframePrm}) by setting $C_y=C_t=C_i=1$. This gives the off--diagonal components before T duality
\bea\label{TempGyM1}
{g}^{yp}\d_p|_{m=0}&=&-\ch_\alpha\sh_\alpha\d_t+
{\frac{R}{F}}{\sh_\alpha}\left[\ch_\alpha \d_t-
\sum_k\frac{a_k}{r^2+a_k^2}\d_{\phi_k}\right]\nn
&&-
\sum_i\left[{\frac{(-x_i)H_i}{d_i(r^2-x_i)}}\sh_\alpha\left[\ch_\alpha\d_t-\sum_k\frac{a_k}{x_i+a_k^2}\d_{\phi_k}
\right]\right]=0.
\eea
The last expression must vanish since for $m=0$ time and $y$ coordinate enter the Myers--Perry metric (\ref{MPeven}) only through the boost--invariant combination $-dt^2+dy^2$. Comparison of (\ref{TempGyM}) with (\ref{TempGyM1}) gives the unique expressions for the unknown functions in terms of $C_y$:
\bea
C_i=C_y, \quad C_t=\frac{R-mr}{R}C_y.
\eea
To determine the last remaining coefficient we compute ${\tilde g}^{yy}$:
\bea
{\tilde g}^{yy}&=&C_y^2\left[\ch^2\alpha-\frac{(R-mr)}{FR}\sh^2\alpha+
\sum_i\left[{\frac{(-x_i)H_i}{d_i(r^2-x_i)}}\sh^2\alpha\right]\right]\nonumber\\
&=&C_y^2\left[1+\frac{mr}{FR}\sh^2\alpha\right].
\eea
To simplify this expression we again used the trick of setting $m$ to zero. For the boosted version of (\ref{MPeven}) we find
\bea
g_{yy}=1+\frac{mr}{FR}\sh^2\alpha.
\eea
Matching this with ${\tilde g}^{yy}$, we conclude that $C_y=1$.

To summarize, we have demonstrated that the frames (\ref{ChMPframePrm}) with 
\bea\label{CframeDual}
C_i=C_y=1, \quad C_t=\frac{R-mr}{R}
\eea
reproduce the metric after T duality and expression (\ref{ChMPHYT}) recovers the correct components $Y^{mn}$, it only remains to check that the correct transformation of $Y_z{}^s$ is also recovered. 

According to our conjecture (\ref{ChMPHYT}), the mKYT in the original and T dual frames are given by
\bea
Y^{(p)}=\sum A_{a_1,\dots a_p}e^{a_1}\wedge \dots\wedge e^{a_p},\quad
{\tilde Y}^{(p)}=\sum A_{a_1,\dots a_p}{\tilde e}^{a_1}\wedge \dots\wedge {\tilde e}^{a_p}
\eea
with the {\it same} coefficients $A_{a_1,\dots a_p}$. The original frames $e^{a}$ are given by  (\ref{ChMPframePrm}) with $C_i=C_y=C_t=1$, and the dual frames $\tilde e^{a}$ have different values of coefficients  (\ref{CframeDual}). Observing that 
\bea
\tilde e^{a}_y=\frac{1}{h_1}e^{a}_y,\quad \tilde e_{a}^m=e_{a}^m,
\eea
we find the perfect agreement with transformation (\ref{KYTtransYcomp}),
\bea
{\tilde Y}^{m_1\dots m_p}= Y^{m_1\dots m_p},\qquad 
{\tilde Y}_z{}^{m_2\dots m_p}= e^{-C}{Y}_z{}^{m_2\dots m_p},
\eea
since
\bea
e^C\equiv g_{yy}=1+\frac{mr}{FR}\sh^2\alpha=h_1.
\eea
This concludes the derivation of the Killing--Yano tensors (\ref{ChMPHYT}), (\ref{ChMPframe}), (\ref{ChMPframeDwn}) for the charged Myers--Perry black holes in even dimensions. The arguments for the odd dimensions are identical, and the answer is given by (\ref{ChMPframeOdd}), (\ref{ChMPframeDwnOdd}).


\section{Killing tensors for the F1--NS5 system}
\label{App5D6DandKT}
\renewcommand{\theequation}{G.\arabic{equation}}
\setcounter{equation}{0}

In this appendix we will present some technical details of calculations leading to the Killing tensors for the examples discussed in section \ref{SecExamplesF1NS5}. 

\subsection{F1--NS5 from the four--dimensional Kerr metric}
\label{App4DKerr}

Starting with Kerr metric (\ref{Kerr4D}) and using the duality chain (\ref{DualChain}), we generate the F1--NS5 solution
\bea\label{NonExtr5D}
ds^2&=&\frac{1}{h_{\beta}}dy^2+\frac{4ma\sh_\beta \sh_\alpha \cos\theta}{\rho^2 h_\beta}dzdy-
\left(1-\frac{2mr \ch^2_\beta}{\rho^2 h_{\beta}}\right)dt^2
-\frac{4mra \ch_\alpha \ch_\beta \sin^2\theta}{\rho^2 h_{\beta}}dtd\phi
\nn
&&+\Bigg[
(r^2+a^2)h_\alpha+\frac{2mra^2\sin^2\theta}{\rho^2}-
\frac{(2mar \ch_\alpha \sh_\beta \sin\theta)^2}{\rho^4 h_{\beta}}
 \Bigg]\sin^2\theta d\phi^2\\
&&+\frac{h_\alpha \rho^2}{\Delta}dr^2+h_\alpha \rho^2 d\theta^2+\Big[1+
\frac{2m\sh^2_\alpha(2m \sh^2_\beta+r)}{\rho^2 h_\beta}\Big]dz^2,\nn
B_2&=& \frac{mr \sh_{2\beta}}{h_\beta \rho^2} dy\wedge dt - 
\frac{2amr \ch_\alpha \sh_\beta\sin^2\theta }{h_\beta\rho^2} dy\wedge d\phi 
+ \frac{2am\cos\theta \ch_\beta \sh_\alpha}{h_\beta \rho^2} dt\wedge d z\nn
&& - \frac{m\cos\theta 
\sh_{2\alpha}(a^2+2mr \sh^2_\beta+r^2)}{h_\beta \rho^2} d\phi\wedge dz,\nn
e^{2\Phi}&=&\frac{h_\alpha}{h_\beta},\nn
\Delta&=&r^2+a^2-2mr, \quad \rho^2=r^2+a^2\cos^2\theta, \quad 
h_\alpha=1+\frac{2mr \sh^2_\alpha}{\rho^2}, \quad h_\beta=1+\frac{2mr \sh^2_\beta}{\rho^2}.\nonumber
\eea
The charges associated with NS5 branes and fundamental strings are defined by 
\be\label{ChargesDef}
Q_5=2A^2=2m\sinh^2\alpha, \quad Q_1=2B^2=2{m}\sinh^2\beta
\ee
The nontrivial Killing tensor for (\ref{NonExtr5D}) can be extracted either from solving a system of differential equations (\ref{KTeqnDef}) or by separating variables in the massive Hamilton--Jacobi equation. The second approach is easier and more instructive, so we begin with equation
\be\label{HJMass}
g^{MN}\frac{\d S}{\d x^M}\frac{\d S}{\d x^N}+\mu^2=0,
\ee
multiply it by $\rho^2 h_\alpha$, and rewrite the result as a system of two differential equations
\bea\label{5Dlambda}
\Lambda&=&(2A^2+r)(2B^2+r)(\d_y S)^2-\Big[\frac{(r^2+2A^2r+a^2)(r^2+2B^2r+a^2)}{\Delta}-\frac{a^2}{2}\Big](\d_t S)^2\nn
&&-\frac{4ar\sqrt{(A^2+m)(B^2+m)}}{\Delta}\d_tS\d_\phi S -\frac{a^2}{\Delta}(\d_\phi S)^2+\Delta (\d_r S)^2 +r^2(\d_z S)^2\nn
&&+\mu^2 (2B^2r+r^2),\\
\Lambda&=&-a^2c^2_\theta(\d_y S)^2+4aABc_\theta \d_zS \d_y S+ \frac{a^2c_{2\theta}}{2}(\d_t S)^2- \frac{1}{s^2_\theta}(\d_\phi S)^2- (\d_\theta S)^2\nn
&&-a^2c^2_\theta(\d_z S)^2-\mu^2a^2c^2_\theta.\nonumber
\eea 
In general $\Lambda$ can depend on all coordinates, but for separable solutions,
\bea
S=-Et+J\phi+p_z z+p_y y+S_r(r)+S_\theta(\theta)
\eea
this function must be constant. This constant gives rise to a Killing tensor
\bea\label{KT5dApp}
K^{MN}\d_M\d_N=-a^2c^2_\theta\d^2_y+4aABc_\theta \d_z\d_y+
\frac{a^2c_{2\theta}}{2} \d^2_t-\frac{1}{s^2_\theta}\d^2_\phi-\d^2_\theta-a^2c^2_\theta\d^2_z
+a^2c^2_\theta g^{MN}\d_M \d_N.
\nn
\eea
Here we removed $\mu^2$ from (\ref{5Dlambda}) using the relation
\bea
g^{MN}\d_M S\d_N S+\mu^2=0.\nonumber
\eea
This Killing tensor (\ref{KT5dApp}) is used in section \ref{SecEx5d}. Note that even though we found KT, the square root of \eqref{KT5dApp} does not solve either standard or modified KYT equation for arbitrary charges. The special cases for which modified KYT exists are discussed in subsection \ref{SubsectionExamples5D}.

\subsection{F1--NS5 from the five--dimensional black hole}

The chain of dualities (\ref{DualChain}) can also be applied to a five--dimensional black hole, but fortunately this procedure has been performed in \cite{GuiMathSax}\footnote{The metric has been constructed earlier in \cite{Cvetic5D} using different methods, and in the full solution (\ref{NonExtr6D}) for the extremal case was found in \cite{Multiwound}.}. Here we will focus on solution with one rotation which can be obtained by setting  $\delta_p=0, a_1=0, a_2=a$ in equation  (3.6) of \cite{GuiMathSax} and performing an S duality. The result reads
\bea\label{NonExtr6D}
ds^2 &=& - \left( 1- \frac{M}{f}
\right)\frac{dt^2}{H_{1}}  + \frac{ dy^2}{H_{1}} +   fH_{5}\left(\frac{ dr^2
}{r^2+a^2-M} +  d\theta^2\right) \nonumber \\
  &+& \left[ r^2H_{5} + \frac{a^2
K_{1}K_{5} \cos^2\theta}{H_{1}} \right] \cos^2\theta
d\psi^2 + \left[ (r^2+a^2) H_{5} -
\frac{a^2  K_{1}K_{5} \sin^2\theta}{H_{1}}
\right] \sin^2\theta d\phi^2 \nonumber \\
&+& \frac{M  }{fH_{1}}a^2\sin^4\theta d\phi^{2} + \frac{2\cos^2\theta}{fH_{1} } aAB dyd\psi +  \frac{2\sin^2\theta}{fH_{1} }
a\sqrt{A^2+M}\sqrt{B^2+M} dt d\phi +\sum_{i=1}^{4} dz_{i}^2 \nonumber \\
B_{2} &=& \frac{\cos^2\theta}{fH_{1}} aA \sqrt{B^2+M} dt \wedge d\psi + \frac{\sin^2\theta }{f H_{1}} aB \sqrt{A^2+M}dy\wedge  d\phi \nonumber \\
  &-& \frac{B\sqrt{B^2+M} }{fH_{1}}  dt\wedge dy - \frac{A\sqrt{A^2+M}}{f H_{1}}\left(r^2+a^2+B^2\right)\cos^2\theta d\psi\wedge d\phi,\nonumber\\
e^{2\Phi}&=&\frac{H_5}{H_1},\nn
f&=&r^2+a^2\cos^2\theta,\quad K_1=\frac{B^2}{f}, \quad K_5=\frac{A^2}{f},\quad H_i\equiv 1+K_i, ~ i=1,5.
\eea
Multiplying the Hamilton--Jacobi equation (\ref{HJMass}) for the metric (\ref{NonExtr6D}) by $f H_5$ and separating variables, we find
\bea\label{AppSep6D}
&&-\left(A^2+B^2+M+r^2+\frac{(A^2+M)(B^2+M)}{a^2-M+r^2}\right)(\d_t S)^2+\frac{2a\sqrt{A^2+M}\sqrt{B^2+M}}{a^2-M+r^2}\d_tS\d_\phi S\nonumber\\
&&+\frac{(A^2+r^2)(B^2+r^2)}{r^2}(\d_y S)^2-\frac{2aAB}{r^2}\d_y S\d_\psi S+(a^2-M+r^2)(\d_r S)^2\\
&&+\frac{a^2}{r^2}(\d_\psi S)^2-\frac{a^2}{a^2-M+r^2}(\d_\phi S)^2+(A^2+r^2)\mu^2=\nonumber\\
&&a^2\cos^2\theta(\d_t S)^2-a^2\cos^2\theta(\d_y S)^2
-(\d_\theta S)^2-\frac{1}{\cos^2\theta}(\d_\psi S)^2
-\frac{1}{\sin^2\theta}(\d_\phi S)^2-a^2\cos^2\theta\mu^2.\nonumber
\eea
This equation clearly separates in $\theta,r$ and gives rise to the Killing tensor
\bea\label{KT6dApp}
K^{MN}\d_M\d_N&=&a^2\cos^2\theta\d^2_t-a^2\cos^2\theta\d^2_y-\d^2_\theta-\frac{1}{\cos^2\theta}\d^2_\psi-\frac{1}{\sin^2\theta}\d^2_\phi\nn
&&+a^2\cos^2\theta g^{MN}\d_M S\d_N S.
\eea
In contrast to the F1--NS5 system constructed from the four--dimensional Kerr solution (there was no mKYT) the square root of \eqref{KT6dApp} give rises to a rank-3 modified Killing-Yano tensor discussed in subsection \ref{SubsectionExamples6D}.

\subsection{F1--NS5 from the Plebanski--Demianski solutions}

Our final example is F1--NS5 constructed from the Plebanski--Demianski metric \cite{PlebDem}:
\bea\label{Plebanski}
ds^2&=&\frac{p^2+q^2}{X}dp^2+\frac{p^2+q^2}{Y}dq^2+\frac{X}{p^2+q^2}(d\tau+q^2 d\sigma)^2-\frac{Y}{p^2+q^2}(d\tau-p^2 d\sigma)^2,\nn
X&=&\gamma-g^2-\epsilon p^2-\lambda p^4+2lp, \quad Y=\gamma+e^2+\epsilon q^2-\lambda q^4-2mq.
\eea
Here $\lambda$ is a cosmological constant, $e$ and $g$ are electric and magnetic charges (we will set these quantities to zero). The remaining constants $(\gamma, m , l , \epsilon)$ effectively comprise 3 real continuous parameters and one discrete parameter, since one can always rescale coordinates to set $\eps$ to one of three values $(+1,-1,0)$. The remaining  continuous parameters $(\gamma, m , l)$ are related to the angular momentum, mass, and the NUT charge. The Kerr solution (\ref{Kerr4D}) is recovered by setting 
\bea
\gamma=a^2,~ \epsilon=1-\lambda a^2,~ p=a \cos\theta,~ q=r,~ \tau=t-\frac{a}{1+\lambda a^2}\phi,~ \sigma=-\frac{1}{a(1+\lambda a^2)}\phi.\nonumber
\eea
In string theory applications one usually sets $e=g=0$, and since asymptotic flatness is a crucial part of our solution generating technique, we set $\la=0$ as well. Applying the chain of dualities (\ref{DualChain}) to such truncated version of \eqref{Plebanski} we get an 
F1--NS5 solution
\bea\label{F1NS5Pleb}
ds^2&=&\frac{f_\alpha}{X}dp^2+\frac{f_\alpha}{Y}dq^2+\frac{X-Y}{f_\beta}d\tau^2+2\frac{q^2X+p^2Y}{f_\beta}\ch_\alpha\ch_\beta d\tau d\sigma\nn
&-&\Big[ \frac{p^4Y-q^4X}{p^2+q^2}\ch_\alpha^2+XY\sh_\alpha^2+\frac{(q^2X+p^2Y)^2\ch_\alpha^2 \sh_\beta^2}{f_\beta(p^2+q^2)}\Big]d\sigma^2\nn
&+&\frac{p^2+q^2}{f_\beta}dy^2+\frac{4(mp-lq)\sh_\alpha\sh_\beta}{f_\beta}dy dz+\Big[\frac{f_\alpha}{p^2+q^2}+\frac{4(mp-lq)^2\sh_\alpha^2\sh_\beta^2}{f_\beta(p^2+q^2)} \Big]dz^2,\nn
B&=&\Bigg[ \frac{p^2+q^2+X-Y}{f_\beta}\ch_\beta\sh_\beta d\tau+\frac{q^2X+p^2Y}{f_\beta}\ch_\alpha\sh_\beta d\sigma \Bigg]\wedge dy\nn
&+&\Bigg[ \frac{2(lq-mp)}{f_\beta}\ch_\beta\sh_\alpha d\tau \\
&&- \frac{f_\beta pq(lp+mq)+(lq-mp)(q^2X+p^2Y)\sh_\beta^2}{f_\beta(p^2+q^2)}\sh_{2\alpha}d\sigma \Bigg]\wedge dz,\nn
f_\alpha&=& (p^2+q^2)\left[1+\frac{X-Y+p^2+q^2}{p^2+q^2}\sh_\alpha^2 \right].\nonumber
\eea
Writing the HJ equation for the metric \eqref{F1NS5Pleb} and multiplying it  by $f_\alpha$, we extract the Killing tensor from separation of variables as in the previous subsections
\bea
K^{MN}\d_M\d_N&=&-\bar{p}_\alpha\bar{p}_\beta\d_y^2+4mp\sh_\alpha\sh_\beta \d_y \d_z-p^2\d_z^2-X\d_p^2-\frac{1}{X}\d_\phi^2-\frac{2p^2}{X}\ch_\alpha \ch_\beta \d_\tau\d_\phi\nn
&&-\Big[p^2\sh_\beta^2+\frac{p^4\ch_\beta^2}{X}+\frac{p^4\ch_\beta^2\sh_\alpha^2}{X}+p\sh_\alpha^2(p+(2l+2p-\epsilon p)\sh^2_\beta)\Big]\d_\tau^2,\nn
&&+p\bar{p}_\alpha g^{MN}\d_M\d_N,
\eea
where we defined
\bea
\bar{p}_\alpha=p\ch_\alpha^2+(2l-\epsilon p)\sh_\alpha^2.
\eea
Note that setting the NUT charge to zero and choosing $\epsilon=1$ gives
\bea
\bar{p}|_{l=0,\epsilon=1}=p.
\eea
This example shows that the NUT charge does not spoil separability and consistent with results from Appendix \ref{App4DKerr}.


\section{Double Field Theory}\label{AppDFT}
\renewcommand{\theequation}{H.\arabic{equation}}
\setcounter{equation}{0}

In this appendix we review the Double Field Theory (DFT) \cite{DFT} and use rewrite the action of T duality on Killing vectors in a more symmetric form.

Double Field Theory is an elegant way of incorporating T duality as a symmetry of field theory. This is accomplished by extending the standard $D$ coordinates $x^m$ into a larder $2D$--dimensional space $x^M=(\tilde x_m, x^m)$. In this appendix we deviate from the notation used throughout this paper and denote the spacetime indices by lower--case letters, while reserving the capital ones to label the ``double space" spanning over regular and barred indices $N=(n,\bar{n})$. This notation is standard in the DFT literature. The theory is formulated with full duality group $O(D,D)$. 

Recall that the T duality group is associated to string compactifications on $T^n$ is $O(n,n)$, so we see that DFT gives a geometric interpretation to  the T duality transformation.

The next step in constructing DFT is defining the fields. One is looking for $O(D,D)$ invariant tensors. It turns out that the metric $g_{mn}$ and the $B_{mn}$ field can be unified into such kind of tensor called the generalized metric \cite{Odd, GenMetric}
\bea\label{DefDFT}
\cH_{MN}=\begin{pmatrix}g^{mn}& -g^{mk}B_{kn}\\B_{mk}g^{kn}&g_{mn}-B_{mk}g^{kl}B_{ln} \end{pmatrix}.
\eea 

Note that the generalized metric does not play the same role as the regular metric in general relativity: the indices are raised and lowered with the constant $O(D,D)$ invariant metric $\eta_{MN}$ rather than $\cH_{MN}$, where
\be
\eta_{MN}=\begin{pmatrix} 0& \delta^m{}_n\\ \delta_m{}^n&0\end{pmatrix}.
\ee
To define diffeomorphisms in DFT theory one needs to introduce the generalized Lie derivative \cite{DFTGaugeTransf} of the generalized metric
\be\label{AppGenLie}
{L}_{\xi}\mathcal{H}_{MN}=\xi^P\d_P\cH_{MN}+(\d_M\xi^P-\d^P\xi_M)\cH_{PN}+(\d_N\xi^P-\d^P\xi_N)\cH_{MP}.
\ee
where $\xi^I=(\tilde\lambda_i,\lambda^i), \xi_I=(\lambda^i,\tilde\lambda_i)$ is the generalized gauge parameter. Here $\tilde\lambda_i$ corresponds to the gauge transformation of the Kalb--Ramond field $B_{ij}$ and $\lambda^i$ is a usual diffeomorphism. 

Transformation (\ref{AppGenLie}) differs from the standard diffeomorphisms in $2D$ dimensions since the following condition must be preserved
\bea\label{AppchSquare}
\cH_{MA}\eta^{AB}\cH_{BN}=\eta_{MN}.
\eea
To demonstrate that (\ref{AppGenLie}) accomplishes this task, one begins with observing that 
\bea
{L}_{\xi}\eta_{MN}=(\d_M\xi^P-\d^P\xi_M)\eta_{PN}+(\d_N\xi^P-\d^P\xi_N)\eta_{MP}=0.
\eea
Then
\bea
&&{L}_{\xi}(\cH_{MA}\eta^{AB}\cH_{BN})\nonumber\\
&&=\left[\xi^P\d_P\cH_{MA}+(\d_M\xi^P-\d^P\xi_M)\cH_{PA}+(\d_A\xi^P-\d^P\xi_A)\cH_{MP}\right]\eta^{AB}\cH_{BN}
+(M\leftrightarrow N)
\nonumber\\
&&=\left[(\d_M\xi^P-\d^P\xi_M)\eta_{PN}+(\d_A\xi^P-\d^P\xi_A)\cH_{MP}\eta^{AB}\cH_{BN}\right]
+(M\leftrightarrow N)
\nonumber\\
&&=(\d_A\xi_Q-\d_Q\xi_A)\eta^{PQ}\eta^{AB}(\cH_{MP}\cH_{BN}+\cH_{NP}\cH_{BM})=0
\eea
This leads to the conclusion that the condition (\ref{AppchSquare}) 
is preserved by the modified diffeomorphism (\ref{AppGenLie}).

\addtocontents{toc}{\protect\setcounter{tocdepth}{0}}
\subsection{Killing vectors in DFT}
\addtocontents{toc}{\protect\setcounter{tocdepth}{6}}

\label{AppKVfromDFT}

To incorporate Killing vectors in the DFT framework, we recall that in the Riemannian geometry the Lie derivative of the metric $g_{mn}$ along a Killing vector $\la$ vanishes
\bea\label{DFTstandKV}
{\cal L}_\la g_{mn}=\nabla_m \lambda_n+\nabla_n \lambda_n=0.
\eea 
So to define the ``double Killing vector'' $\xi^M=({\tilde\la}_m,\la^m)$ we require vanishing of the generalized Lie derivative (\ref{AppGenLie})
\bea\label{AppDFTGenKV}
L_\xi \mathcal{H}_{MN}=\xi^P\d_P\cH_{MN}+(\d_M\xi^P-\d^P\xi_M)\cH_{PN}+(\d_N\xi^P-\d^P\xi_N)\cH_{MP}=0.
\eea
Next we will demonstrate that this equation incorporates both gauge transformation of $B$ field and usual diffeomorphism of the metric\footnote{Appearance of both ingredients in the generalized Lie derivative has been discussed in \cite{DFTGaugeTransf}.}. 

Let us begin with $\bar m\bar n$ components of equation \eqref{AppDFTGenKV}\footnote{In the following calculations we use the strong constraint $\tilde\d=0$ \cite{StrongConstraint}.} with $\cH_{MN}$ from (\ref{DefDFT})
\bea\label{LieDerMetr}
{L}_{\xi}\mathcal{H}_{\bar m\bar n}&=&\xi^P\d_P\cH_{\bar m\bar n}+\d_{\bar m}\xi^P\cH_{P\bar n}-\d^P\xi_{\bar m}\cH_{P\bar n}+\d_{\bar n}\xi^P\cH_{\bar mP}-\d^P\xi_{\bar n}\cH_{\bar mP}\nonumber\\
&=&\xi^p\d_p\cH_{\bar m\bar n}-\d_p\xi_{\bar m}\cH^{p}{}_{\bar n}-\d_p\xi_{\bar n}\cH_{\bar m}{}^p=\xi^p\d_pg^{mn}-\d_p\xi_{\bar m}g^{pn}-\d_p\xi_{\bar n}g^{mp}\nonumber\\&=&\lambda^p\d_pg^{mn}-\d_p\lambda^{ m}g^{pn}-\d_p\lambda^{ n}g^{mp}={\cal L}_{\xi}(g^{mn})=0.
\eea
This recovers the standard equation (\ref{DFTstandKV}) for the Killing vector. 
For the $\bar m n$ components of equation \eqref{AppDFTGenKV} we find
\bea\label{LieDerGau}
{L}_{\xi}\mathcal{H}_{\bar m n}&=&\xi^p\d_p\cH_{\bar mn}-\d_p\xi_{\bar m}\cH^p{}_{n}+\d_n\xi^P\cH_{\bar mP}-\d_p\xi_n\cH_{\bar m}{}^p\nonumber\\
&=&\lambda^p\d_p(-g^{mk}B_{kn})-\d_p\lambda^m(-g^{pk}B_{kn})+\d_n\lambda^p(-g^{mk}B_{kp})+\d_n\tilde\lambda_pg^{mp}-\d_p\tilde\lambda_ng^{mp}\nonumber\\
&=&\lambda^p\d_pB_{n}{}^m-\d_p\lambda^mB_{n}{}^p+\d_n\lambda^pB_{p}{}^m+(\d_n\tilde\lambda_p-\d_p\tilde\lambda_n)g^{mp}=0.
\eea
The first two terms give the regular Lie derivative of $B_n{}^m$ along the Killing vector $\la^m$, but this derivative does bot have to vanish since the Kalb--Ramond is defined only up to a gauge transformation. Equation (\ref{LieDerGau}) states that the Lie derivative of $B$ must be a pure gauge (with gauge parameter ${\tilde\la}_m$), which means that all physical effects from the Kalb--Ramond field are invariant under the diffeomorphisms generated by $\la^m$. The $mn$ components of (\ref{AppGenLie}) give nothing new due to the constraint (\ref{AppchSquare}). 

We conclude that the Lie derivative (\ref{AppGenLie}) can be used to formulate generalized Killing equation
\bea\label{AppGenKil}
\xi^P\d_P\cH_{MN}+(\d_M\xi^P-\d^P\xi_M)\cH_{PN}+(\d_N\xi^P-\d^P\xi_N)\cH_{MP}=0,
\eea
whose components give equation (\ref{LieDerMetr}) for the regular Killing vector and relation (\ref{LieDerGau}) for the Lie derivative of the $B$ field. 

For future reference we rewrite equations (\ref{LieDerMetr}) and (\ref{LieDerGau}) in terms of the covariant derivatives. For the first equation the transition is standard:
\be\label{AppGenLieKillEq1}
\mathcal{L}_\xi\cH_{\bar m\bar n}=0 \quad\Rightarrow\quad 
\nabla_m\lambda_n+\nabla_n\lambda_m=0,
\ee
and equation (\ref{AppchSquare}), 
\be\label{AppGenLieKillEq2}
\mathcal{L}_\xi\cH_{\bar m n}=0\quad \Rightarrow\quad \lambda^p\d_pB_n{}^m-\d_p\lambda^mB_n{}^p+\d_n\lambda^pB_p{}^m+(\d_n\tilde\lambda_p-\d_p\tilde\lambda_n)g^{mp}=0.	
\ee
requires additional work. Straightforward transformations lead to 
\bea\label{AppGenLieBCov}
\lambda^p\nabla_pB_n{}^m-\nabla_p\lambda^mB_n{}^p+\nabla_n\lambda^pB_p{}^m+
\nabla^m\tilde\lambda_n-\nabla_n\tilde\lambda^m=0,
\eea
and using the Killing equation (\ref{AppGenLieKillEq1}) the last relation can be rewritten in terms of the gauge--invariant field strength $H=dB$:
\bea
{H_{mnp}\lambda^p=\nabla_m\tilde\lambda'_n-\nabla_n\tilde\lambda'_m.}
\eea
where we defined
\bea
\tilde\lambda'_m=\tilde\lambda_m+\lambda_pb_m{}^p\,.
\eea
Notice that under the $O(D,D)$ transformations act as a rotation between ${\tilde\la}_m$ and $\lambda^m$, and 
$\tilde\lambda'_m$ transforms in a more complicated way.


\section{Complex structures}\label{AppComplexStructure}
\renewcommand{\theequation}{I.\arabic{equation}}
\setcounter{equation}{0}

Killing--Yano tensors are closely related to K{\"a}hler forms on complex manifolds, and in this appendix we will apply the reduction used for the KYT to arrive at the modified K{\"a}hler condition on manifolds with torsion to recover the well--known results \cite{Strominger,HullCS}. 
We begin with an arbitrary anti--symmetric tensor $J$ and define
\be\label{AppCSEq}
T_{PMN}=\nabla_P J_{MN}.
\ee
The Killing--Yano equation for $J$ can be written as
\bea
T_{(PM)N}=0,
\eea
and the K{\"a}hler condition, $dJ=0$, is
\bea
T_{[PMN]}=0.
\eea
Combination of the K{\"a}hler condition with integrability of the complex structure is equivalent to a simple constraint \cite{Zumino}
\bea\label{NewKahler}
T_{PMN}=0,
\eea
and we will now analyze its transformation under T duality.

Starting with a pure metric (\ref{DimRedSetup}) with $B=0$ 
and performing the dimensional reduction of (\ref{AppCSEq}) using (\ref{DimRedL}), we find
\bea\label{CStemp}
T_{zz}{}^n&=&\frac{1}{2}{J^{an}}\d_a e^C+\frac{1}{2}{\hat g}^{nb}e^C F_{ab}{J_z}^a,\nonumber\\
T_z{}^{mn}&=&\frac{1}{2}[{\hat g}^{mb}J^{an}-{\hat g}^{nb}J^{am}]e^C F_{ab}-
\frac{1}{2}[{\hat g}^{ma}{J_z}^n-{\hat g}^{na}{J_z}^m]\d_a C\\
T^{p}{}_z{}^n&=&{\hat\nabla}^p {J_z}^n-\frac{1}{2}g^{pa}{J_z}^n \d_a C-\frac{1}{2}g^{pb}e^C F_{ba}J^{an}
\nonumber\\
T^{pmn}&=&{\hat \nabla}^p J^{mn}
+\frac{1}{2}g^{pa}g^{mb} F_{ab}{J_z}^n
-\frac{1}{2}g^{pa}g^{nb} F_{ab}{J_z}^m.\nonumber
\eea
Introducing rescaled quantities 
\bea
{\tilde J}_z^{~m}=e^{-C}  J_z^{~m},\qquad {\tilde J}^{mn}= J^{mn},
\eea 
we can rewrite these relations as
\bea
T_{zz}{}^n&=&\frac{1}{2}{{\tilde J}^{an}}\d_a e^C+\frac{1}{2}{\hat g}^{nb}e^{2C} {\tilde H}_{abz}{{\tilde J}_z}{}^a,\nonumber\\
T_z{}^{mn}&=&\frac{1}{2}[{\hat g}^{mb}J^{an}-{\hat g}^{nb}J^{am}]e^C {\tilde H}_{abz}+e^C
\frac{1}{2}[{\hat g}^{ma}{\tilde J_z}{}^n-{\hat g}^{na}{\tilde J_z}{}^m]\d_a {\tilde C},\nn
T^{p}{}_z{}^n&=&e^C{\hat\nabla}^p {\tilde J_z}{}^n-\frac{1}{2}e^Cg^{pa}{J_z}^n \d_a {\tilde C}-\frac{1}{2}g^{pb}e^C {\tilde H}_{baz}J^{an},\\
T^{pmn}&=&{\hat \nabla}^p J^{mn}
+\frac{e^C}{2}g^{pa}g^{mb} {\tilde H}_{abz}{J_z}^n
-\frac{e^C}{2}g^{pa}g^{nb} {\tilde H}_{abz}{J_z}^m,\nonumber
\eea
where tildes refer to expressions after the T duality. If we define a tensor
\bea\label{ModifComplStr}
{\tilde T}_{PMN}\equiv {\nabla_P {\tilde J}_{MN}+\frac{1}{2}{\tilde H}_{PNA}{\tilde g}^{AB}{\tilde J}_{MB}-
\frac{1}{2}{\tilde H}_{PMA}{\tilde g}^{AB}{\tilde J}_{NB}}
\eea
after duality, then
\bea
{\tilde T}_{zz}{}^n=-e^{-2C}{T}_{zz}{}^n,\quad
{\tilde T}_z{}^{mn}=-e^{-C}T_z{}^{mn},\quad 
{\tilde T}^{p}{}_z{}^n=e^{-C}T^{p}{}_z{}^n,\quad
{\tilde T}^{mnp}=T^{mnp}.
\eea
In particular we observe that the K{\"a}hler condition (\ref{NewKahler}) is preserved by the T duality, as long as one uses the modified expression (\ref{ModifComplStr}) for ${\tilde T}_{PMN}$ in the presence of the $B$ field. Expression (\ref{ModifComplStr}) can be interpreted as a covariant derivative on a manifold with torsion, and equation 
${\tilde T}_{PMN}=0$ coincides with well--known requirement of supersymmetry for geometries supported by the Kalb--Ramond field \cite{Strominger}.

\end{document}